\documentclass[12pt,preprint]{aastex}

\def\linebreak{\hfil\break}

\def\
{\hfil\linebreak}

\def\degree{\ifmmode {^\circ}\else {$^\circ$}\fi}
\def\mum{\ifmmode {\rm \mu {\rm m}}\else $\rm \mu {\rm m}$\fi}
\def\arcsec{\ifmmode ^{\prime \prime}\else $^{\prime \prime}$\fi}

\def\inch{\ifmmode ^{\prime \prime}\else $^{\prime \prime}$\fi}
\def\arcmin{\ifmmode ^{\prime}\else $^{\prime}$\fi}

\def\mearth{M$_\oplus$}
\def\msun{M$_\odot$}

\def\ldlstar{$L_d / L_\star$}
\def\lstar{$L_\star$}
\def\mstar{$M_\star$}
\def\tstar{$T_\star$}
\def\qdstar{$Q_d^\star$}
\def\2470{[24]--[70]}

\def\ms{m~s$^{-1}$}
\def\gyr{g~yr$^{-1}$}

\newbox\grsign \setbox\grsign=\hbox{$>$} \newdimen\grdimen \grdimen=\ht\grsign
\newbox\simlessbox \newbox\simgreatbox
\setbox\simgreatbox=\hbox{\raise.5ex\hbox{$>$}\llap
     {\lower.5ex\hbox{$\sim$}}}\ht1=\grdimen\dp1=0pt
\setbox\simlessbox=\hbox{\raise.5ex\hbox{$<$}\llap
     {\lower.5ex\hbox{$\sim$}}}\ht2=\grdimen\dp2=0pt

\begin{document}

\title{Variations on Debris Disks II. Icy Planet Formation as a 
Function of the Bulk Properties and Initial Sizes of Planetesimals}
\vskip 7ex
\author{Scott J. Kenyon}
\affil{Smithsonian Astrophysical Observatory,
60 Garden Street, Cambridge, MA 02138} 
\email{e-mail: skenyon@cfa.harvard.edu}

\author{Benjamin C. Bromley}
\affil{Department of Physics, University of Utah, 
201 JFB, Salt Lake City, UT 84112} 
\email{e-mail: bromley@physics.utah.edu}
%
%-------------------------- ABSTRACT ----------------------------------
%
%\doublespace

\begin{abstract}
We describe comprehensive calculations of the formation of icy planets and debris disks at 30--150~AU 
around 1--3 \msun\ stars. Disks composed of large, strong planetesimals produce more massive planets 
than disks composed of small, weak planetesimals. The maximum radius of icy planets ranges from $\sim$ 
1500~km to 11,500~km.  The formation rate of 1000~km objects -- `Plutos' -- is a useful proxy for the 
efficiency of icy planet formation. Plutos form more efficiently in massive disks, in disks with small 
planetesimals, and in disks with a range of planetesimal sizes. Although Plutos form throughout massive 
disks, Pluto production is usually concentrated in the inner disk. Despite the large number of Plutos 
produced in many calculations, icy planet formation is inefficient.  At the end of the main sequence 
lifetime of the central star, Plutos contain less than 10\% of the initial mass in solid material.  
This conclusion is independent of the initial mass in the disk or the properties of the planetesimals. 
Debris disk formation coincides with the formation of planetary systems containing Plutos.  As 
Plutos form, they stir leftover planetesimals to large velocities. A cascade of collisions then 
grinds the leftovers to dust, forming an observable debris disk. In disks with small 
($\lesssim$ 1--10~km) planetesimals, collisional cascades produce luminous debris disks 
with maximum luminosity $\sim 10^{-2}$ times the stellar luminosity. Disks with larger 
planetesimals produce debris disks with maximum luminosity $\sim 5 \times 10^{-4}$ (10~km) to 
$5 \times 10^{-5}$ (100~km) times the stellar luminosity.  Following peak luminosity, the evolution 
of the debris disk emission is roughly a power law, $ f \propto t^{-n}$ with $n \approx$ 0.6--0.8. 
Observations of debris disks around A-type and G-type stars strongly favor models with small planetesimals. 
In these models, our predictions for the time evolution and detection frequency of debris disks agree 
with published observations. We suggest several critical observations that can test key features of 
our calculations.
\end{abstract}


\section{INTRODUCTION}

Dusty disks of debris surround many main sequence stars \citep{bac93,che05,rie05,moor06,rhe07a,wya08}. 
Among young stars, the frequency of debris disks ranges from $\sim$ 50\% for B-type and 
A-type stars \citep[][2009]{su06,cur08b} to $\sim$ 10\% to 20\% for solar-type stars 
\citep{tri08,mey08} to $\sim$ 5\% for M-type stars \citep{pla09,les09}. Binary stars and single
stars are equally likely to have debris disks \citep{stau05,su06,bry06,gor07,sie07,tri07}.
Among stars with masses \mstar\ $\gtrsim$ 0.8--1~\msun, the debris disk frequency declines 
with stellar age \citep{rie05,cur08,car09b}. 

Debris disks are signposts of planet formation. Infrared (IR) and radio observations indicate 
that grains with typical sizes of 1--100 $\mu$m produce most of the dust emission. Radiation 
processes remove these grains on timescales much shorter than the age of the central star 
\citep{bac93,wya08}.  To maintain 
emission from small grains throughout the main sequence lifetime, some process must replenish the 
dust. The simplest replenishment mechanism invokes a 10--100 \mearth\ reservoir of $\sim$ 1~km 
solid objects which continuously collide at high velocities and fragment into smaller objects 
\citep{bac93,hab01,kb04b,wya08,heng10}. Although this mass is plausible \citep[e.g., ][2007b]{and05},
collisional damping among an ensemble of 1~$\mu$m to 1 m objects rapidly reduces collision 
velocities to low values \citep[e.g.,][]{kb01}. Thus, debris disks require a mechanism to 
maintain the high velocities of the solids. Gravitational stirring by massive planets is the 
most successful mechanism \citep{wya08}. Thus, in the current picture, maintenance of debris 
disks requires massive planets.

Two broad classes of planetary systems can explain the general observations of debris disks.
In the Solar System, external perturbations power the local debris disk. Jupiter's gravity
excites the orbits of the asteroids and produces the Jupiter family comets; collisions among 
asteroids and mass loss from comets produce the Zodiacal light \citep[e.g.,][2008, 2009]{nes06}.  
Beyond 30~AU, Neptune's gravity plays a similar role for the 
Kuiper belt and the scattered disk \citep[e.g.,][]{mor04,cha07}.  The recent discovery of gas
giant planets associated with the debris disks in HR 8799 and Fomalhaut \citep{kal08,mar08,che09,su09} 
suggests similar processes occur in other planetary systems \citep[see also][and references 
therein]{wil02,moran04,del05,mor05,qui06,wya06,fab07}. In particular, \citet{must09} show that the 
gravitational perturbations of newly-formed planets at 5--10~AU can produce dusty debris at
30--100~AU. Their results suggest that these perturbation may yield the most luminous debris
disks.

Smaller planets can also produce debris disks. In a series of papers, we show that the collisional 
evolution of solid material in a gaseous protostellar disk naturally leads to the formation of planets 
along with copious amounts of dusty debris \citep[e.g.,][2004b, 2005]{kb02b}. In our picture, 
mergers of~km-sized objects first produce 500--1000~km protoplanets. These protoplanets stir up 
leftover planetesimals along their orbits. Destructive collisions among the leftovers initiate a 
collisional cascade, which produces a dusty debris disk 
\citep[for different approaches to this problem, see also][]{kri08,the08,wya08,heng10,kenn10}. 
In the terrestrial zone at a few~AU 
from the central star, rocky protoplanets reach masses of 0.5--2~\mearth\ and rapidly remove the 
debris \citep[][2005, 2006; see also Bottke et al 2007]{kb04a}. At 5--20~AU, the gas entrains 
the debris, allowing icy protoplanets 
to grow rapidly into the cores of gas giant planets \citep{kb09}. At 30--150~AU, icy protoplanets 
reach maximum sizes of only $\sim$ 1500--2000~km \citep[][hereafter KB08]{kb08}. The collisional
cascade among the leftover planetesimals produces a luminous debris disk; the maximum brightness
and evolution of the debris matches the observations of known debris disks reasonably well (KB08).

Here, we continue our exploration of the formation and evolution of icy planets and debris disks.
In KB08, we described a suite of calculations for disks at 30--150~AU around 1--3~\msun\ stars.
We considered disks with a single surface density law, a single initial size distribution for
planetesimals, and a single set of fragmentation parameters for solid objects. For the calculations
discussed here, we examine planet formation in disks with an expanded set of initial conditions
for the initial surface density of the disk and for the initial properties of the planetesimals. 
These results yield new predictions for the maximum sizes of icy planets as a function of initial 
planetesimal size. Results for the long-term evolution of debris disks continue to account for 
many fundamental aspects of the data.  Our new analysis demonstrates that luminous debris disks 
at 30--150~AU require planetesimals with initial sizes of 1--10~km instead of 100--1000~km. 

Our calculations suggest that the minimum stable grain size, the slope of the size distribution 
for small grains, and the slope of the IR emissivity law 
are also critical parameters.  Spatially resolved images of debris disks around A-type and solar-type 
stars can improve our understanding of the minimum stable grain size. Larger samples of debris 
disks with high quality submm data from the Atacama Large Millimeter/Submillimeter Array (ALMA), 
the {\it Herschel Space Observatory}, and the Stratospheric Observatory for Infrared Astronomy 
(SOFIA) can place better constraints on the size distribution for small objects and the slope of 
the emissivity law.  Together, these data can 
test our predictions for the time evolution of debris disk emission around 1--3~\msun\ stars and 
can provide input for more complete calculations that include the formation and dynamical evolution of 
giant planets.

We outline the numerical model in \S2. We describe the formation of icy planets in \S3 and the 
evolution of debris disks in \S4; these sections include `highlights' (\S3.3 and \S4.3) which
summarize the main results and conclusions. In \S5, we consider applications of our results to 
the observed time evolution (\S5.1) and frequency (\S5.2) of debris disks around A-type and 
solar-type stars.  We conclude with a brief summary in \S6.

\section{CALCULATIONS}

To calculate the formation and evolution of debris disks, we use a hybrid 
multiannulus numerical code for planet formation.  We compute the collisional 
evolution of an ensemble of planetesimals in a circumstellar disk orbiting 
a star of mass \mstar. The code uses statistical algorithms to evolve the 
mass and velocity distributions of low mass objects with time and an $n$-body 
algorithm to follow the individual trajectories of massive objects. KB08
describe the statistical (coagulation) code; \citet{bk06} describe the $n$-body 
code. Here, we briefly summarize the basic aspects of our approach.

We perform calculations on a cylindrical grid with inner radius $a_{in}$ and outer
radius $a_{out}$. The model grid contains $N$ concentric annuli with widths 
$\delta a_i$ centered at semimajor axes $a_i$. Calculations begin with a mass 
distribution $n(m_{ik}$) of planetesimals with horizontal and vertical velocities 
$h_{ik}(t)$ and $v_{ik}(t)$ relative to a circular orbit.  The horizontal velocity 
is related to the orbital eccentricity, $e_{ik}^2(t)$ = 1.6 $(h_{ik}(t)/V_{K,i})^2$, 
where $V_{K,i}$ is the circular orbital velocity in annulus $i$.  The orbital 
inclination depends on the vertical velocity, 
$i_{ik}^2(t)$ = sin$^{-1}(2(v_{ik}(t)/V_{K,i})^2)$.

In the coagulation code, the mass and velocity distributions evolve in time due 
to inelastic collisions, drag forces, and long-range gravitational forces. The 
collision rate is $n \sigma v f_g$, where $n$ is the number density of objects,
$\sigma$ is the geometric cross-section, $v$ is the relative velocity, and $f_g$ is 
the gravitational focusing factor \citep[][KB08]{saf69,lis87,spa91,wet93,wei97,kl98,kri06,the07,loh08}. 
The collision
outcome depends on the ratio of the collision energy needed to eject half the mass 
of a pair of colliding planetesimals \qdstar\ to the center of mass collision energy 
$Q_c$. If $m_1$ and $m_2$ are the masses of two colliding planetesimals, the mass of 
the merged planetesimal is 
\begin{equation}
m = m_1 + m_2 - m_{ej} ~ ,
\label{eq:msum}
\end{equation}
where the mass of debris ejected in a collision is
\begin{equation}
m_{ej} = 0.5 ~ (m_1 + m_2) \left ( \frac{Q_c}{Q_d^*} \right)^{9/8} ~ .
\label{eq:mej}
\end{equation}
This approach allows us to derive ejected masses for catastrophic collisions 
with $Q_c \sim Q_d^*$ and for cratering collisions with $Q_c \ll Q_d^*$ 
\citep[see also][]{wet93,wil94,tan96,st97,kl99,obr03,kob10}. Consistent with N-body 
simulations of collision outcomes \citep[e.g.,][]{ben99,lein08,lei09}, we set 
\begin{equation}
Q_d^* = Q_b r^{\beta_b} + Q_g \rho_g r^{\beta_g}
\label{eq:Qd}
\end{equation}
where $Q_b r^{\beta_b}$ is the bulk component of the binding energy,
$Q_g \rho_g r^{\beta_g}$ is the gravity component of the binding energy,
$r$ is the radius of a planetesimal, and $\rho_g$ is the mass density of 
a planetesimal.

To compute the evolution of the velocity distribution, we include collisional
damping from inelastic collisions, gas drag, and gravitational interactions.
For inelastic and elastic collisions, we follow the statistical, Fokker-Planck
approaches of \citet{oht92} and \citet{oht02}, which treat pairwise interactions 
(e.g., dynamical friction and viscous stirring) between all objects in all annuli. 
As in \citet{kb01}, we add terms to treat the probability that objects in annulus 
$i$ interact with objects in annulus $j$ \citep[see also][KB08]{kb04b}. We also 
compute long-range stirring from distant oligarchs \citep{wei89}. For gas drag, 
we follow \citet{wet93}, who calculate drag in the quadratic limit \citep[see also][]
{ada76, wei77b}.

To evolve the gas in time, we consider a simple nebular model for the gas density.
We adopt a scale height $H_{gas}(a) = H_{gas,0} (a/a_0)^{1.125}$ \citep{kh87}
and assume that the gas surface density declines exponentially with time
\begin{equation}
\Sigma_{gas}(a,t) = \Sigma_{gas,0} ~ x_m ~ a^{-n} ~ e^{-t/t_{gas}}
\label{eq:sigma-gas}
\end{equation}
where $\Sigma_{gas,0}$ and $x_m$ are scaling factors and $t_{gas}$ is the gas 
depletion time. To enable comparisons with results in KB08, we adopt $t_{gas}$ = 
10~Myr. Although longer than the 2--5~Myr timescale estimated from observations of 
the lifetimes of accretion disks in pre-main sequence stars \citep{cur09,kenn09,mam09}, 
shorter gas depletion times have little impact on our results.

In the $n$-body code, we directly integrate the orbits of objects with masses larger 
than a pre-set `promotion mass' $m_{pro}$. The calculations allow for mergers among
the $n$-bodies. Additional algorithms treat mass accretion from the coagulation 
grid and mutual gravitational stirring of $n$-bodies and mass batches in the
coagulation grid. For the applications in this paper, the few large objects capable 
of promotion into the $n$-body code never contain a significant fraction of the mass 
in an annulus and never contribute significantly to the local stirring. To treat 
situations where a few large objects might impact the evolution, we set 
$m_{pro} = 10^{26}$ g. However, our calculations never produced more than a few 
$n$-bodies. These remained on circular orbits throughout their evolution.

The initial conditions for these calculations are appropriate for a disk
with an age of $\lesssim$ 1--2~Myr \citep[e.g.][]{dul05,nom06,cie07,gar07}.
We consider systems of $N$ annuli in disks with $a_i$ = 30--150~AU and 
$\delta a_i/a_i$ = 0.025. We assume a power law variation of the initial surface 
density of solid material with semimajor axis,
\begin{equation}
\Sigma_{d,i} = \Sigma_{d,0}(M_{\star}) ~ x_m ~ a_i^{-n} ~ , 
\label{eq:sigma-dust}
\end{equation}
where $a_i$ is the central radius of the annulus in~AU, $n$ = 1 or 3/2, and
$x_m$ is a scaling factor.  Consistent with observations of disks surrounding 
pre-main sequence stars \citep[e.g.,][]{sch2006,sch09}, we scale the reference 
surface density of solids with the stellar mass, $\Sigma_{d,0} (M_{\star})$ = 
30 $(M_{\star} / M_{\odot}$) g cm$^{-2}$ \citep[see also][]{wei77a,hay81}.  For a 
standard gas to dust ratio of 100:1, $\Sigma_{gas,0} = 100 ~ \Sigma_{d,0} (M_{\star})$. 
To explore a range of disk masses similar to the observed range among the youngest stars, 
we consider $x_m$ = 0.01--3; disks with $x_m \approx$ 0.1 have masses similar to the
median disk masses observed around young stars in nearby dark clouds 
\citep[][2007b]{ost95,mot01,and05}. 

As a baseline model, we consider disks 
composed of small planetesimals with initial radii of 1--1000 m and an initial mass 
distribution $n_i(m_{ik})$ in each annulus. The mass ratio between adjacent bins 
is $\delta = m_{ik+1}/m_{ik}$ = 1.4--2. At the start of the calculations, each 
bin has the same total mass, eccentricity $e_0 = 1-3~\times~10^{-4}$, and 
inclination $i_0 = e_0/2$.  To examine the dependence of our results on the initial 
radii of planetesimals, we also perform calculations where all of the initial mass 
is in planetesimals with radii of 1~km, 10~km, or 100~km. 

Although our adopted range of disk masses and surface density gradients is 
consistent with observations of the youngest stars \citep[][2007b]{and05}, our 
adopted outer radius of 150~AU is probably larger than the typical outer disk 
radius. Analyses of high quality submm observations of nearby young stars yield 
a broad range of outer disk radii, $\sim$ 20~AU to $\gtrsim$ 200~AU 
\citep[e.g.,][]{and07a}. However, recent studies suggest the typical outer radius 
is $\sim$ 75--100~AU \citep{hug08,and09,ise09}. In KB08, we showed that smaller
disks produce roughly comparable amounts of debris disk emission at 
$\lambda \lesssim$ 70 $\mu$m and much less emission at longer wavelengths. Here,
we concentrate on understanding the evolution of larger disks and use the results
from KB08 to examine the impact of smaller disks on our results.

To explore the sensitivity of our results to the fragmentation algorithm,
we consider two sets of fragmentation parameters $f_i$. Strong planetesimals 
have $f_S$ = \citep[$Q_b$ = $10^1$, $10^3$, or $10^5$ erg g$^{-1}$, $\beta_b$ = 0,
$Q_g$ = 2.25 erg g$^{-2}$ cm$^{1.75}$, $\beta_g$ = 1.25; KB08, ][]{ben99}.
Weaker planetesimals have $f_W$ =
\citep[$Q_b$ = $2 \times 10^5$ erg g$^{-1}$ cm$^{0.4}$, $\beta_b = -0.4$,
$Q_g$ = 0.33 erg g$^{-2}$ cm$^{1.7}$, $\beta_g$ = 1.3;][]{lei09}.

Tables 1--2 summarize the model grids.  Table \ref{tab:massgrid} lists the 
initial disk masses for the ranges in $M_{\star}$, $x_m$ we consider and the
adopted main sequence lifetimes for the central stars. Because the growth of 
planets has large stochastic variations, we repeated the calculations 5--15
times for each set of starting conditions, $M_{\star}$, $n$, $x_m$, and $Q_b$.  
Table \ref{tab:modgrid} lists the number of calculations for each 
($M_{\star}$, $x_m$) pair. For calculations with strong planetesimals, the
number of repetitions for each $Q_b$ ($10^1$, $10^3$, or $10^5$ erg g$^{-1}$)
is approximately 1/3 the number listed in the Table.

Our calculations follow the time evolution of the mass and velocity
distributions of objects with a range of radii, $r_{ik} = r_{min}$
to $r_{ik} = r_{max}$.  The upper limit $r_{max}$ is always larger
than the largest object in each annulus.  To save computer time in
our main calculation, we do not consider small objects which do not
significantly affect the dynamics and growth of larger objects,
$r_{min}$ = 100 cm.
Erosive collisions produce objects with $r_{ik}$ $< r_{min}$ which
are `lost' to the model grid. Lost objects are more likely to be
ground down into smaller objects than to collide with larger objects
in the grid \citep[see][2004a]{kb02a}.

To derive the amount of dusty debris as a function of time, we follow the 
evolution of the `lost' objects using a simple collision algorithm that 
includes Poynting-Robertson drag and radiation pressure \citep[see][KB08]{kb04a}.
Every timestep, we add new debris to each annulus. The new debris has a 
fixed size distribution, $n \propto r^{-s}$ with $s$ = 3.5, between $r_{min}$
and the minimum stable grain size $r_2$. In parallel, we derive collision rates 
and Poynting-Robertson drag rates for the old debris. As long as the particle 
velocities are not size dependent, this routine yields reasonably accurate 
results for the evolution of the size distribution.

Our approach to the evolution of dusty debris ignores collisions between 
very small grains ejected by radiation pressure and larger grains on bound 
orbits \citep[e.g.,][]{grig07}. These collisions provide an extra source of 
small grains, potentially enhancing the optical depth throughout the disk. 
To estimate the potential impact of these collisions in our calculations,
we compare (i) the rate very small grains produce debris through collisions
with larger grains and (ii) the rate oligarchs and other large objects immune
to disruption sweep up the extra debris. Our estimates suggest that these 
two processes approximately balance. Thus, the collisional avalanches
described by \citet{grig07} probably have little impact on our results.

To set the main parameters -- $r_2$ and $s$ -- in this approach, we follow 
previously published results (see also KB08). Theoretical estimates for the
minimum stable grain size yield $r_2 \approx$ 0.5--2~$M_{\star}^3$ 
\citep{bur79, art88, bac93, kim02}. Because the coefficient is sensitive to
the composition, internal structure, and radiative properties of the grains,
we adopt $r_2$ = 1 $\mu$m. Thus, we probably overestimate the number of small 
grains and the amount of infrared excesses for stars more massive than 1~\msun.
For small particles with $Q_d^* \propto r^{\beta_b}$, 
$s$ = $(21 + \beta_b) / (6 + \beta_b)$ \citep{obr03,kob10}.
Thus, $s$ = 3.5 (3.67) for strong (weak) planetesimals.
Our adopted $s$ = 3.5 underestimates infrared excesses for systems with weak
planetesimals. In \S5, we consider the impact of these choices in more detail.  

As in KB08, we use simple scaling relations to show how our results depend on 
initial conditions and the properties of the grid.  For each set of 
calculations (Table 2), we derive median results for the size distribution, 
the size of the largest object as a function of $a$ and $t$, and other physical 
variables. Substituting the inter-quartile range for the dispersion, we then 
perform least-squares fits to relate these median results to input parameters 
(e.g., $x_m$) and the properties of the grid (e.g., $a$). For parameters where 
analytic theory predicts a relation (e.g., the growth time as a function of $a$), 
we derive the best-fitting coefficient, test whether different fitting functions 
provide better fits to our results, and keep the result that minimizes $\chi^2$
per degree of freedom. When analytic theory provides no guidance, we derive 
fitting functions that yield the sensitivity of our results to
all important physical variables. Thus, our fits test some aspects of
analytic theory and guide other aspects.

\section{PLANET FORMATION CALCULATIONS}

\subsection{Icy Planet Formation in Disks Around 1 \msun\ Stars}

We start with a description of icy planet formation in disks at 30--150~AU around
a 1~\msun\ star. For disks with masses $M_d \gtrsim$ 0.003~\msun, the timescale 
to form large objects at the outer edge of the disk (150~AU) is shorter than the 
main sequence lifetime of the central star. In these disks, the outcome of icy 
planet formation depends on the physics of icy solids rather than on stellar 
physics. For lower mass disks, the central star evolves off the main sequence
before icy planet formation reaches a `standard' outcome throughout the disk.
Because post-main sequence evolution changes the structure in the disk considerably 
(see the discussion in KB08), we do not follow the growth of icy planets after 
the central star evolves off the main sequence. Thus, icy planet formation is
truncated in the lowest mass disks.

In this section, we review the stages of icy planet formation and describe the
outcome of collisional evolution throughout the disk. For our adopted grids of
initial disk masses (Table \ref{tab:modgrid}) and a range of initial planetesimal
sizes and fragmentation parameters, we derive relations for the time scale and 
maximum radius of planets as a function of initial conditions. We also demonstrate
how the dust production rate and the mass in dust grains depend on initial conditions 
and time. 

In \S3.2, we compare these results with calculations for 1.5--3~\msun\ stars. For
disks around more massive stars, the growth time is a smaller and smaller fraction
of the main sequence lifetime. Thus, icy planet formation is truncated at smaller
and smaller disk radii for more and more massive stars. Here, we show how larger
disk masses and shorter main sequence lifetimes change the results derived for 
1~\msun\ stars.

\subsubsection{Growth of Large Objects}

At the start of our simulations, planetesimals have random velocities comparable to 
their escape velocities, $\sim$ 1 \ms. Both velocities are smaller than the velocities, 
$\gtrsim$ 10 \ms, required to disrupt a colliding pair of planetesimals. Thus, 
collisions produce mergers instead of debris. Because the growth of planetesimals
depends on the initial size distribution, we first discuss the evolution of an ensemble 
of 1~m to 1~km planetesimals. We then consider the evolution of a swarm of planetesimals
of a single size.

For an ensemble of 1~m to 1~km planetesimals, icy planet formation in the outer 
regions of a disk surrounding a low mass star has three distinct stages 
\citep{kl99, kb04a, kbod08}. Initially, planetesimals grow slowly. As they grow,
dynamical friction damps $e$ and $i$ for the largest objects; dynamical friction and 
viscous stirring raise $e$ and $i$ for the smallest objects \citep{gre84,wet93,gol04}.
Stirring creates a rough equipartition in kinetic energy between large and small objects;
thus, gravitational focusing increases collisional cross-sections by factors of 10--100. 
Slow, orderly growth ends. Runaway growth begins. During runaway growth, a few large 
objects grow much faster than smaller objects and `run away' from the rest of the 
planetesimal swarm. Gravitational stirring continues to raise $e$ and $i$ for the 
smallest objects.  As viscous stirring raises $e$ and $i$ for the runaways, 
gravitational focusing factors decline; growth of the largest objects slows. Runaway
growth ends. Oligarchic growth -- where the largest objects still grow more rapidly
than smaller planetesimals -- begins.  As oligarchic growth proceeds, the oligarchs 
contain an ever increasing fraction of the mass in the disk 
\citep{ida93,wet93,kok98,raf03,cha06,naga07}. 

Throughout runaway and oligarchic growth, other physical processes modify the growth 
rates of massive oligarchs. At the start of our simulations, gas drag damps the velocities 
of small objects and transports small particles radially inward. Although the net transport 
towards the central star is small, damping increases gravitational focusing factors and 
enhances runaway growth \citep[][see also the Appendix of KB08]{raf04}. In the inner disk,
the timescale for gas dissipation is comparable to the timescale for runaway growth. 
Thus, gas drag produces shorter growth times for large objects \citep[][KB08]{raf04}. 

During the transition from runaway to oligarchic growth, collisions start to produce 
copious amounts of dust. As oligarchs grow from 100~km to 1000~km, they stir leftover
1--10~km planetesimals to large collision velocities. Collisions between small objects
then produce debris instead of mergers \citep[see][and references therein]{kbod08}. 
Once debris production begins, continued gravitational stirring leads to a collisional
cascade, where leftover planetesimals are ground down into smaller and smaller objects.
Destructive collisions among the leftovers are much more likely than mergers of 
leftovers with much larger oligarchs. Thus, the collisional cascade grinds 1--10~km 
planetesimals into small dust grains.

Collisions and radiative processes rapidly remove dust grains from the disk.  For 
grains with $r \lesssim$ 10--100 $\mu$m, the collision time is much shorter than the
timescales to remove particles by gas drag \citep{ada76} or by Poynting-Robertson
drag \citep{bur79}. Thus, the collisional cascade continues until particles reach
sizes of a few microns. At these sizes, the orbital time is usually shorter than 
the collision time. Radiation pressure then rapidly removes small particles from the
cascade \citep{bur79,art88}. This ejection produces a disk-shaped `wind' of small 
particles \citep[e.g.,][]{lec98,tak01,kb04a,su05}.

Because runaway growth leaves most of the mass in 1--10~km objects, the collisional
cascade removes a significant fraction of the solid material in the disk. Thus, the
maximum sizes of oligarchs are small, $\sim$ 1500--2500~km (KB08). Although the
gaseous disk can entrain small particles and halt the cascade in the inner disk
\citep[a $\sim$ 5--10~AU;][]{kb09}, the gaseous disk dissipates during runaway growth
at $a \gtrsim$ 30~AU. Thus, interactions with the gas cannot halt the cascade during 
oligarchic growth at $a \gtrsim$ 30~AU.

\begin{figure} 
\includegraphics[width=6.5in]{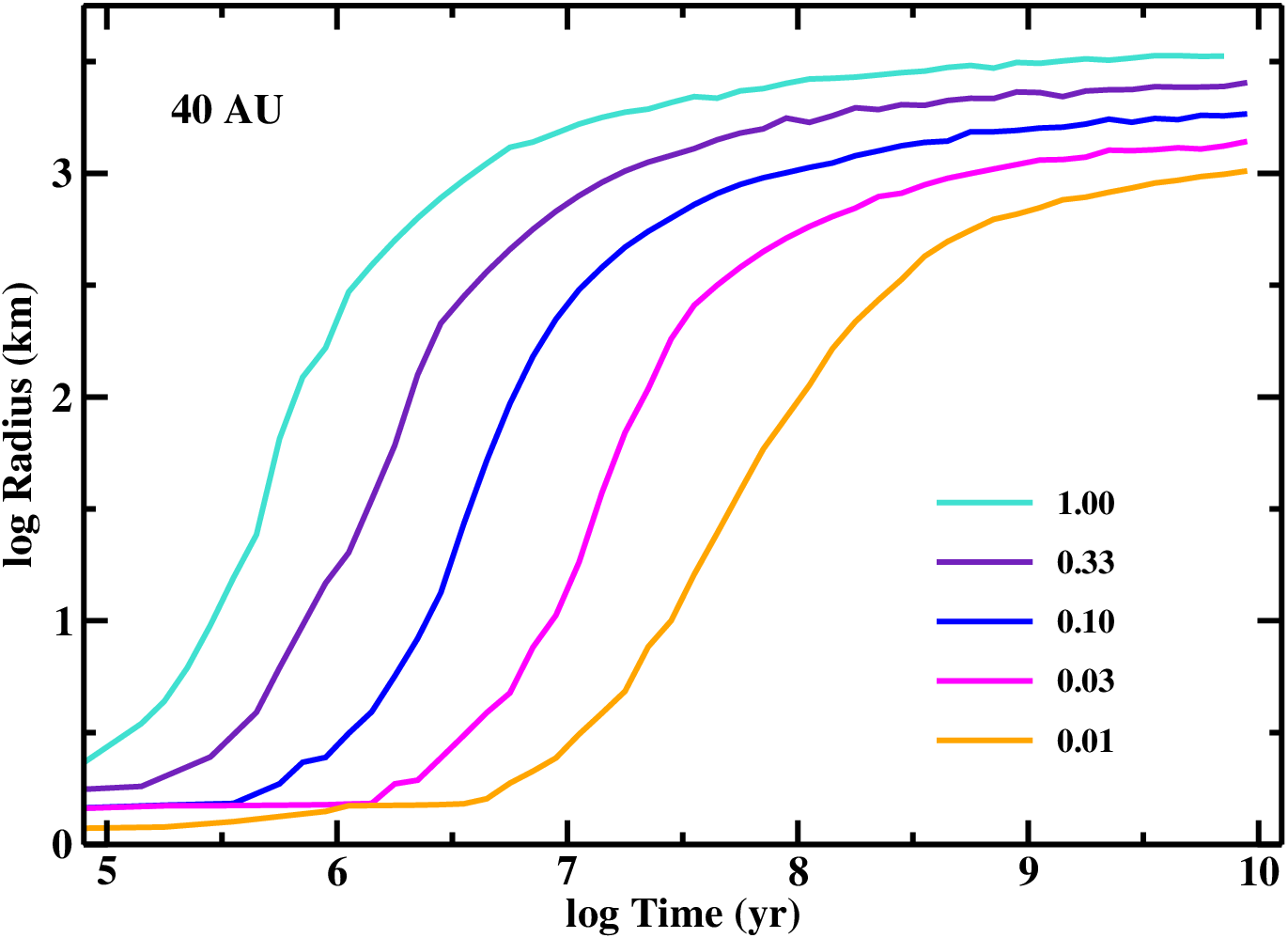}
\vskip 3ex
\caption{
Time evolution of the median radius of the largest object
at 40~AU for disks with initial dust surface density 
$\Sigma_d = 30 ~ x_m ~ a^{-1}$ surrounding a 1 \msun\ star. 
The legend indicates the scaling factor $x_m$ for each curve. 
More massive disks produce larger objects faster than less
massive disks.
\label{fig: rad40-1}
}
\end{figure}

Figure \ref{fig: rad40-1} shows the growth of the largest object at 40~AU in disks 
with initial surface density $\Sigma_d = 30 ~ x_m ~ a^{-1}$ g~cm$^{-2}$ surrounding 
a 1~\msun\ star.  In the lowest mass disks 
($x_m \sim$ 0.01--0.03), slow, orderly growth lasts 1--10~Myr. During runaway growth,
the largest objects reach sizes of $\sim$ 300~km in 10--100~Myr. As the swarm makes
the transition to oligarchic growth, the largest objects reach sizes of 500--1000~km.
The largest objects then grow very slowly to sizes of 1000--1250~km as the central star 
evolves off the main sequence. In more massive disks, growth is much faster. For 
$x_m \gtrsim$ 0.1, slow growth and runaway growth produce 300~km objects in 
$\lesssim$ 10~Myr. Once these systems make the transition to the oligarchic phase,
growth slows significantly. Despite very rapid early growth, the largest objects in
the most massive disks are still relatively small, $r \sim$ 3000~km, at the end of 
the main sequence lifetime of the central star.

\begin{figure} 
\includegraphics[width=6.5in]{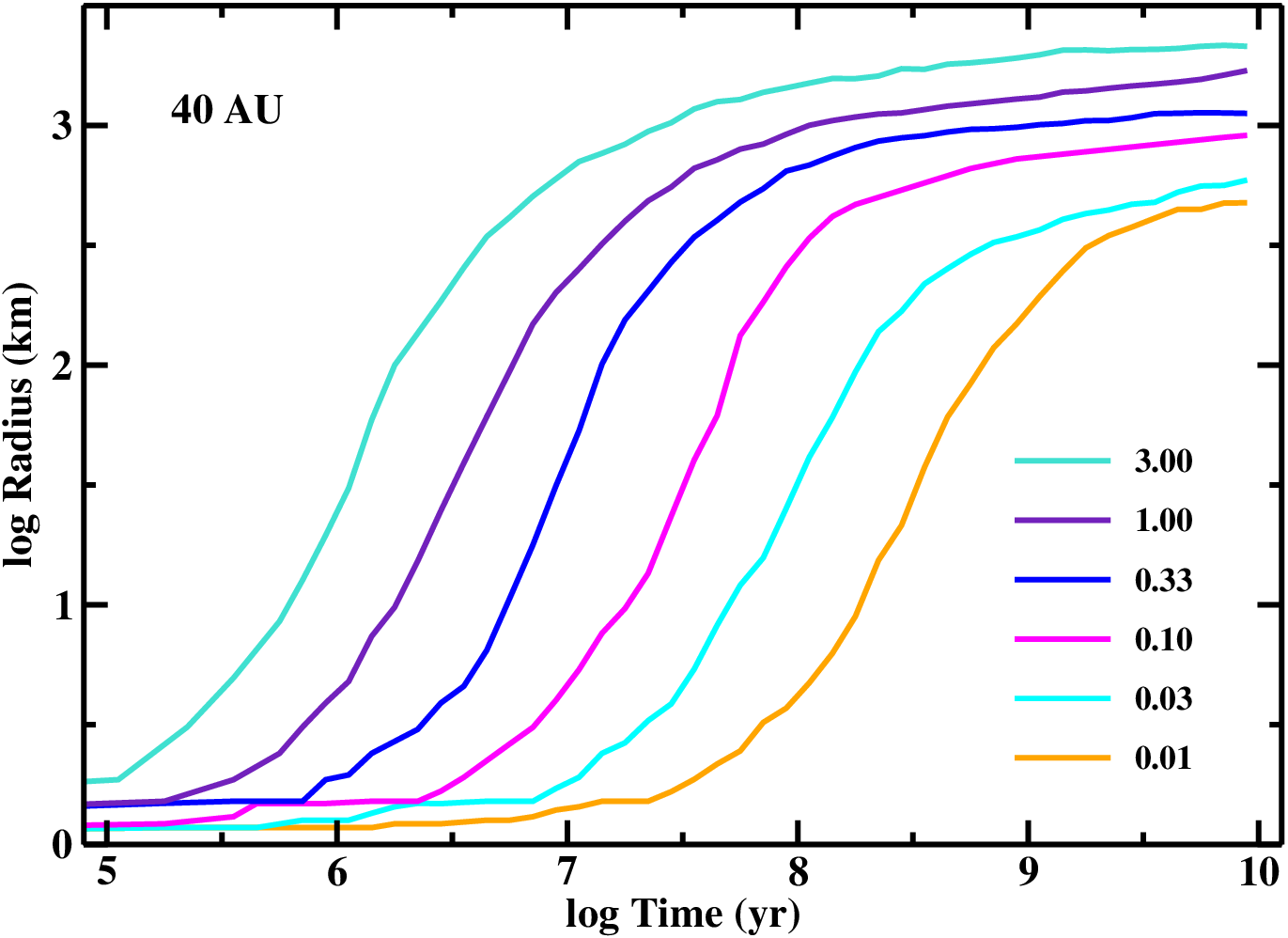}
\vskip 3ex
\caption{
As in Figure \ref{fig: rad40-1} for disks with initial dust 
surface density $\Sigma_d = 30 ~ x_m ~ a^{-3/2}$.
\label{fig: rad40-2}
}
\end{figure}

Disks with steeper surface density gradients follow the same evolutionary path (KB08; 
Figure \ref{fig: rad40-2}). For $\Sigma_d = 30 ~ x_m ~ a^{-3/2}$ g~cm$^{-2}$, disks 
with identical $x_m$ have a factor of $\sim$ 6 less material at 30--40~AU than disks 
with $\Sigma_d = 30 ~ x_m ~ a^{-1}$.  Because the collision rate is proportional to 
the total mass in solids, these disks take a factor of 6 longer to reach runaway and 
oligarchic growth. Once the runaway begins, growth still produces 300~km objects fairly 
rapidly. During oligarchic growth, the largest objects at 30--40~AU approach sizes of 
$\sim$ 1000~km. As oligarchic growth proceeds, the collisional cascade removes most of 
the leftover planetesimals. Thus, the largest objects reach maximum sizes of $\sim$ 
1000--1500~km.

Combined with results from KB08 (e.g., Figure 8), Figures \ref{fig: rad40-1}--\ref{fig: rad40-2} 
also illustrate the impact of the bulk properties of planetesimals on growth rates. 
In calculations with the strong fragmentation parameters ($f_S$), massive objects have 
$r_{max} \sim$ 3000~km in massive disks and $r_{max} \sim$ 1000~km in the low mass disks 
(Figure \ref{fig: rad40-1}; see also Figures 2--3 of KB08). Calculations with 
the $f_W$ parameters yield $r_{max} \sim$ 2000~km (500~km) in massive (low mass) disks. 
The collisional cascade is responsible for this difference.  During oligarchic growth, 
the collision velocities scale with the mass of the largest object.  When planetesimals 
are weaker, they fragment at smaller collisional velocities.  Thus, weaker planetesimals 
begin to fragment when the largest oligarchs are smaller. Because the collisional cascade 
robs growing oligarchs of leftover planetesimals, oligarchs cannot grow as large when 
planetesimals are weaker. 

\begin{figure} 
\includegraphics[width=6.5in]{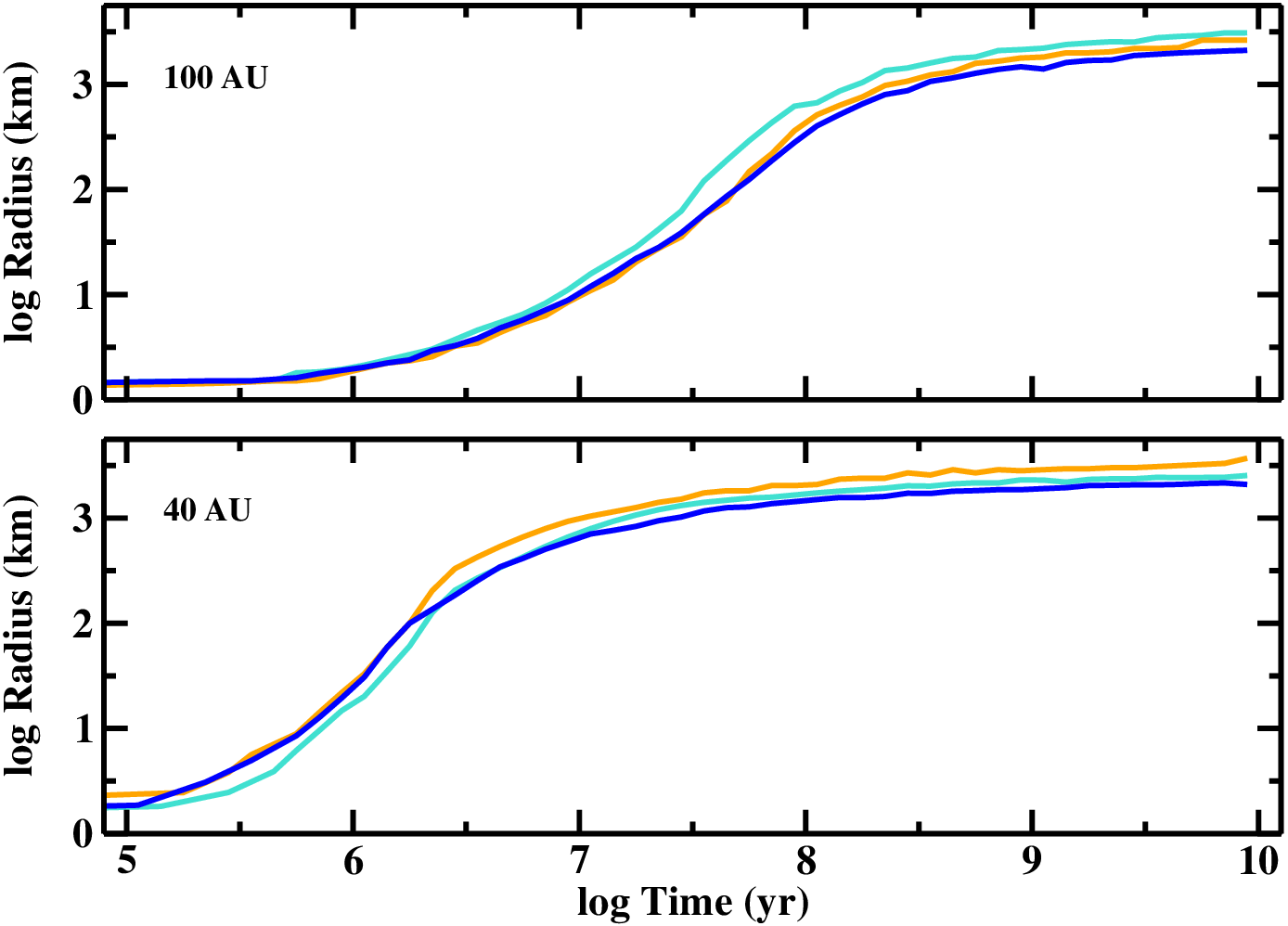}
\vskip 3ex
\caption{
Growth of the largest object at 40~AU (lower panel) and 100~AU 
(upper panel) in disks with similar initial masses but different 
starting conditions. 
Turquoise lines: initial $\Sigma_d = 30 ~ x_m ~ a^{-1}$,
$x_m$ = 1/3, and the $f_S$ fragmentation parameters;
blue lines: disks with initial $\Sigma_d = 30 ~ x_m ~ a^{-3/2}$,
$x_m$ = 3, and the $f_W$ fragmentation parameters;
orange lines: disks with initial $\Sigma_d = 30 ~ x_m ~ a^{-3/2}$,
$x_m$ = 3, and the $f_S$ fragmentation parameters.
Calculations with the $f_W$ parameters yield smaller objects.
In shallower density laws, planets form faster at larger $a$.
\label{fig: rad40-100-1}
}
\end{figure}

Figure \ref{fig: rad40-100-1} compares the variation of $r_{max}$ with our input 
parameters in more detail.  For both values of $a$ in the plots, calculations with
the $f_W$ parameters produce smaller objects than those with the $f_S$ parameters.
In both panels, the curves diverge near the end of runaway growth when 
$r_{max} \approx$ 100~km. When objects reach this radius, collisions destroy weak 
planetesimals. Stronger planetesimals survive until oligarchs reach sizes of $\sim$
300~km. Because the collisional cascade begins earlier when planetesimals are weak,
growing oligarchs are smaller and have a smaller reservoir of leftover planetesimals
to accrete. Thus, the largest oligarchs are always smaller in calculations with weak
planetesimals.

Figure \ref{fig: rad40-100-1} also illustrates how the growth time depends on $n$,
the initial gradient in our adopted relation for the radial surface density 
(equation (\ref{eq:sigma-dust})). 
For disks with identical total masses inside 150~AU, objects at 40~AU (100~AU) grow 
faster in disks with larger (smaller) $n$. This difference persists throughout the
main sequence lifetime of the central star. Objects at 40~AU (100~AU) are larger in 
disks with larger (smaller) $n$. 

To combine these general conclusions into a robust relationship between the growth 
time and the input parameters,
we consider standard coagulation theory. For oligarchs embedded in a swarm of
planetesimals, the accretion rate is roughly the ratio of the mass in an annulus
to the orbital period. Thus, the growth time is $t_{gro} \propto P / \Sigma_d$, where 
$P$ is the orbital period \citep[e.g.,][KB08]{lis87,gol04}. For $P \propto a^{3/2}$ 
and $\Sigma_d \propto x_m ~ a^{-n}$ (equation (\ref{eq:sigma-dust})), the growth time 
is $t_{gro} \propto x_m^{-1} ~ a^{n-3/2}$. Gas drag enhances dynamical friction during 
runaway growth, reducing the growth time \citep{raf04}. Including this process yields 
\begin{equation}
t_{gro} \propto x_m^{-\gamma-1} ~ a^{n-3/2} ~ .
\label{eq: tgrow}
\end{equation}
For disks with a long-lived gaseous component, \citet{raf04} derived $\gamma \approx$ 
0.3 to 0.4. In our calculations, the lifetime of the gaseous disk is shorter than 
the runaway growth time. Thus, we expect $\gamma \approx$ 0.1--0.2 (KB08).

To apply this prediction to our simulations, we measure the median time required for 
objects to reach sizes of 1000~km, $t_{1000}$, as a function of the input parameters\footnote{
Planets with radii of 1000~km are a natural choice for this comparison. Calculations with
weak planetesimals do not reach significantly larger radii (equation \ref{eq: rmax}). 
Timescales to reach a 500~km radius have more scatter.}.
To derive the best exponents in equation (\ref{eq: tgrow}), we use a least-squares fit. 
For our calculations, we derive:
\begin{equation}
t_{1000}~(\rm Myr) = \left\{ \begin{array}{rll}
     30 ~ x_m^{-1.15} ~ a_{80}^{2.5}, &  &\Sigma = 30 ~ x_m ~ a^{-1}, ~ f_i = f_S \\
\\
    400 ~ x_m^{-1.15} ~ a_{80}^{3}, & \hspace{5mm} &\Sigma = 30 ~ x_m ~ a^{-3/2}, ~ f_i = f_S  \\
\\
   1100 ~ x_m^{-1.1} ~ a_{80}^{3}, &  &\Sigma = 30 ~ x_m ~ a^{-3/2}, ~ f_i = f_W 
   \end{array} \right.
\label{eq: t1000}
\end{equation}
where $a_{80} = a$/80~AU. For each set of calculations, the inter-quartile range in the 
median $t_{1000}$ is 15\% to 20\%.

These results agree with the analytic predictions. For each surface density law, the 
growth time scales with $a$ and $x_m$ as expected.  When planetesimals are weak, the 
collisional cascade should remove more material from the disk. Thus, growth times for 
calculations with the $f_W$ parameters are longer than those with the $f_S$ parameters. 
The difference in coefficients for the first two relations of equation (\ref{eq: t1000}) 
is a result of the different initial surface density at 80~AU when $x_m$ = 1. In our formalism,
$\Sigma (n = 1, a = 80~{\rm~AU}) \approx 9 ~ \Sigma (n = 3/2, a = 80~{\rm~AU})$. Thus,
the scaling law in equation (\ref{eq: t1000}) yields expected formation times roughly 
$9^{1.15} \approx$ 12.5 longer in disks with $n$ = 3/2. This result is close to the 
ratio of the coefficients, $400 / 30 \approx$ 13.3, in our expression for the growth time.

To derive scaling relations for the maximum radius $r_{max}$ of icy planets as a function of 
input parameters, we examine results near the end of the main sequence lifetime of the central 
star.  For each calculation, we derive $r_l(a,t)$ the radius of the largest object as a 
function of $a$ and $t$. When the collisional cascade has destroyed nearly all leftover 
planetesimals throughout the disk, $r_l(a,t)$ is roughly constant with $a$. Thus, we adopt 
the median value of $r_l(a,t)$ as $r_{max}$. In massive disks with $x_m \gtrsim 0.1$, the 
collisional cascade rapidly removes leftover planetesimals; icy planets reach 
$r \approx r_{max}$ for $t < t_{ms}$ (e.g., Figure \ref{fig: rad40-1}). In low mass disks,
icy planets reach their limiting radii well after $t_{ms}$. 

For this ensemble of calculations, we derive
\begin{equation}
r_{max} (\rm~km) = \left\{ \begin{array}{rll}
   3500 ~ x_m^{0.22}, ~ &  &\Sigma = 30 ~ x_m ~ a^{-1}, ~ f_i = f_S \\
\\
   2000 ~ x_m^{0.25}, ~ & \hspace{5mm} &\Sigma = 30 ~ x_m ~ a^{-3/2}, ~ f_i = f_S  \\
\\
   1250 ~ x_m^{0.22}, ~ &  &\Sigma = 30 ~ x_m ~ a^{-3/2}, ~ f_i = f_W 
   \end{array} \right.
\label{eq: rmax}
\end{equation}
The inter-quartile range in $r_{max}$ is 10\%.

These results also agree with expectations. Because the collisional cascade removes more 
material from the disk when planetesimals are weak, the largest objects are much smaller. 
In disks with strong planetesimals, we expect regions with similar surface density to 
produce objects with similar sizes.  With 
$\Sigma (n = 1, a = 80~{\rm~AU}) \approx 9 ~ \Sigma (n = 3/2, a = 80~{\rm~AU})$, 
our scaling laws yield $r_{max} \approx$ 3500~km for ($n, x_m$) = (1, 1) and 
$r_{max} \approx$ 3450~km for ($n, x_m$) = (3/2, 9).  For a broad range of disk masses, 
the two surface density laws yield similar masses for icy planets at $a \gtrsim$ 30~AU.
 
To explore the growth of icy planets as a function of initial conditions in more detail, we 
now consider calculations when planetesimals of one size contain all of the initial mass. When 
the range of initial planetesimal sizes is small, dynamical friction cannot damp $e$ and $i$
for the largest objects. Viscous stirring raises $e$ and $i$ for all planetesimals. 
Gravitational focusing factors remain small; growth is very slow. As the largest objects 
grow to sizes roughly a factor of ten larger than the initial size, dynamical friction becomes 
more effective. Gravitational focusing factors increase; runaway growth begins.  Compared to 
calculations that start with a broad range of planetesimal sizes, gravitational focusing 
is weaker and viscous stirring is stronger. Thus, the evolution makes the transition from
runaway to oligarchic growth sooner. 

Although growth is slow when planetesimals are larger, objects grow to larger sizes. When
runaway and oligarchic growth begin, the average size of a planetesimal is roughly proportional 
to the initial size. Larger planetesimals are harder to fragment. As the initial size of
planetesimals increases, the collisional cascade begins later and later relative to the 
onset of oligarchic growth. Because the collisional cascade removes planetesimals faster 
than oligarchs can accrete them, a delayed collisional cascade allows oligarchs to grow
to larger sizes. 

\begin{figure} 
\includegraphics[width=6.5in]{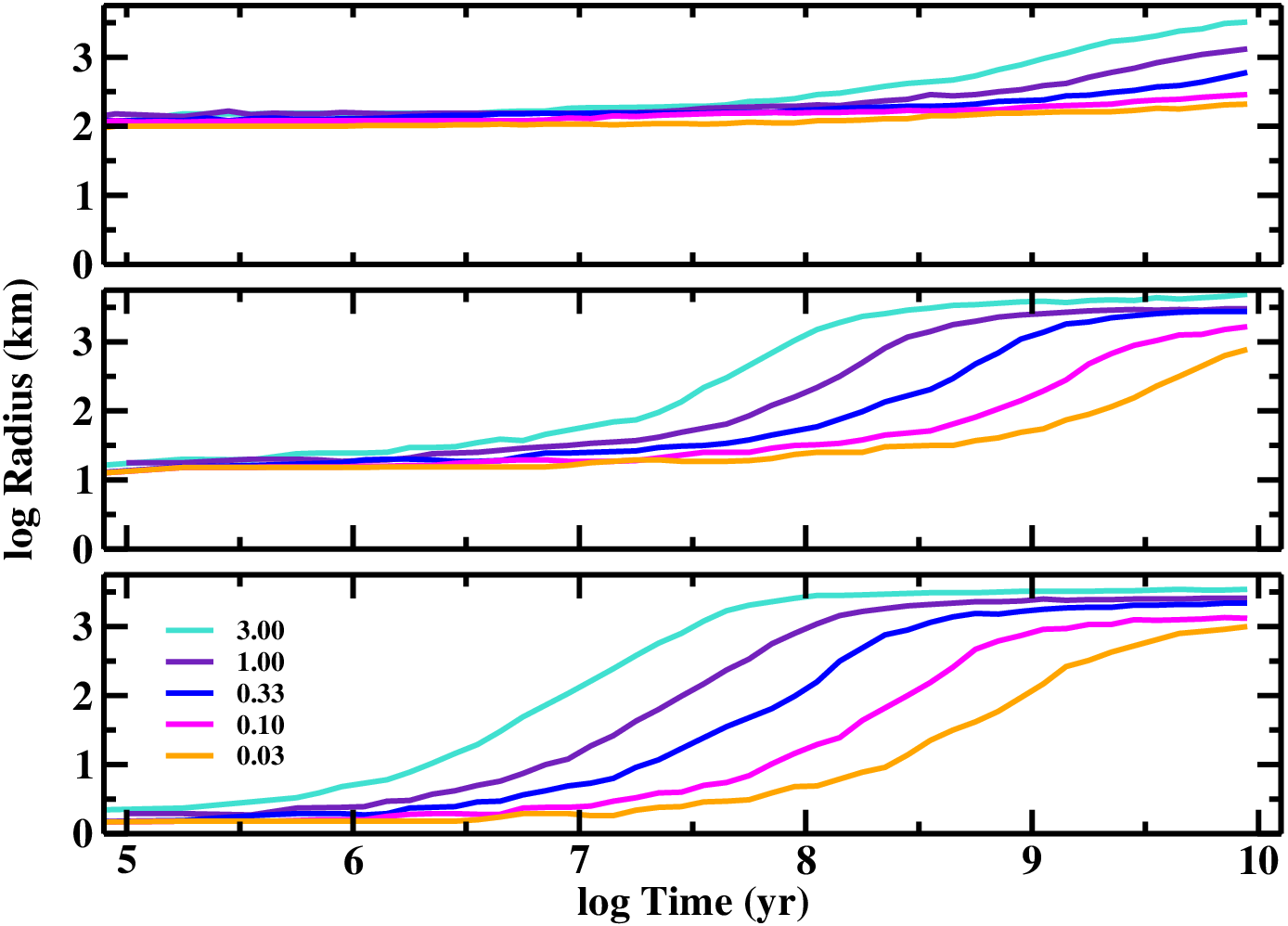}
\vskip 3ex
\caption{
As in Figure \ref{fig: rad40-2} for disks with different
initial planetesimal sizes. The legend in the lower panel
indicates the scaling factor $x_m$ for each curve. 
Lower panel: all planetesimals have $r_0$ = 1~km;
middle panel: all planetesimals have $r_0$ = 10~km;
upper panel: all planetesimals have $r_0$ = 100~km.
Objects with $r >$ 1000~km form more rapidly in disks
with smaller planetesimals.
\label{fig: rad40-rall}
}
\end{figure}

Figure \ref{fig: rad40-rall} shows the growth of the largest object at 40~AU in disks with 
$\Sigma = 30~x_m~a^{-3/2}$ and initial planetesimal sizes of 1~km (lower panel), 10~km (middle 
panel), and 100~km (upper panel). In low mass disks with 1~km planetesimals, slow growth lasts 
$\sim$ 100~Myr, 5--10 times longer than in calculations with an ensemble of 1~m to 1~km
planetesimals.  During a 1~Gyr period of runaway growth, icy planets grow from $\sim$ 10~km to 
$\sim$ 300~km. These disks then make the transition to oligarchic growth, when icy planets reach 
maximum sizes of $\sim$ 1000~km. Although planets grow more rapidly in the most massive disks, the 
time to reach runaway growth is still long. Comparing with results in Figure \ref{fig: rad40-100-1},
icy planets take $\sim$ 5 times longer to reach sizes of 100~km and $\sim$ 3 times longer to reach
sizes of 1000~km.

For calculations with larger planetesimals, growth is even slower. As the initial planetesimal
radius increases from 1~km to 10~km, the rate of growth slows by a factor of roughly 4. In
calculations with 100~km planetesimals, icy planets in the most massive disks reach their
maximum sizes near the end of the main sequence lifetime of the 1~\msun\ central star, roughly
10 times later than calculations with an ensemble of 1~km planetesimals. In the lowest mass
disks, the largest objects have barely grown by a factor of two as the central star evolves
off the main sequence.

Although the faster growth of smaller planetesimals may seem counterintuitive, collision physics 
provides a simple explanation. For an individual oligarch, the growth rate is 
$\dot{M} = \Sigma \Omega \sigma f_g$, where $\Omega$ is the angular frequency, $\sigma$ is the 
collision cross-section, and $f_g$ is the gravitational focusing factor (see the Appendix of 
KB08 and references therein). For calculations with $r_0$ = 10 km and $r_0$ = 100 km, the 
ratio of growth rates in the dispersion-dominated regime is
\begin{equation}
\xi_{10,100} \equiv \frac{\dot{M}_{r_0 = 10~{\rm km}}}{\dot{M}_{r_0 = 100~{\rm km}}} = 
\frac{\sigma_{10} f_{g,10}}{\sigma_{100} f_{g,100}} ~ .
\end{equation}
For oligarchs with $r$ = 200 km, the ratio of cross-sections is 
$\sigma_{10} / \sigma_{100} \approx$ 0.8. However, dynamical friction
among growing oligarchs produces a broad range in $f_g$. For this example,
the ratio of gravitational focusing factors from our calculations is 
$f_{g,10} / f_{g,100} \approx$ 2; $\xi \approx$ 1.5. When $r_0$ = 1 km, the ratio of growth 
rates is  $\xi_{1,100} \approx$ 8. Thus, 200~km oligarchs grow faster in a sea of smaller 
planetesimals.

To quantify the relative growth rates in our calculations, we derive the time 
for objects to grow to 1000~km as a function of the initial planetesimal size. 
We infer
\begin{equation}
t_{1000} (\rm Myr) = \left\{ \begin{array}{rll}
    925 ~ x_m^{-1.05} ~ a_{80}^{3.3}, &  &r_0 = {\rm 1 ~ km}, ~ f_i = f_S \\
\\
   3400 ~ x_m^{-1.05} ~ a_{80}^{3.3}, &  &r_0 = {\rm 10 ~ km}, ~ f_i = f_S \\
\\
   9500 ~ x_m^{-1.00} ~ a_{80}^{3.3}, &  &r_0 = {\rm 100 ~ km}, ~ f_i = f_S \\
   \end{array} \right.
\label{eq: t1000-rall}
\end{equation}
For these calculations, the inter-quartile range in the median $t_{1000}$ is 15\%.

Compared to our calculations with an ensemble of 1~m to 1~km planetesimals, these growth times
have two major differences. Because the slow growth phase lasts much longer, damping by gas
drag is not important. Thus, the time to form 1000~km objects scales more weakly with $x_m$. 
Because all growth phases are slow, stirring by long-range perturbations is more important 
in the outer disk.  Larger stirring rates in the outer disk slows growth relative to the inner 
disk. Thus, $t_{1000}$ scales more strongly with $a$.

Our results demonstrate that larger planetesimals produce larger oligarchs. For disks around
1~\msun\ stars, the largest objects have radii
\begin{equation}
r_{max} (\rm km) = \left\{ \begin{array}{rll}
   3000 ~ x_m^{0.30} ~ &  &r_0 = {1 ~ km}, ~ f_i = f_S \\
\\
   4000 ~ x_m^{0.25} ~ &  &r_0 = {10 ~ km}, ~ f_i = f_S \\
\\
   6000 ~ x_m^{0.25} ~ &  &r_0 = {100 ~ km}, ~ f_i = f_S \\
   \end{array} \right.
\label{eq: rmax-rall}
\end{equation}
with a typical inter-quartile range of 10\%. 
Compared to our other set of calculations, these objects are 50\% to 3 times larger
and 3--30 times more massive. Thus delaying the collisional cascade has a significant
impact on the sizes of the largest objects.

To conclude this discussion of the formation of icy planets, we consider the fraction
of initial disk mass left in solid material of various sizes. For calculations with
an initial ensemble of 1~m to 1~km planetesimals, the median fraction of solids remaining 
in the disk ranges from $\sim$ 10\% to 20\% in massive disks to more than 95\% in the 
lowest mass disks. This depletion varies by a factor of 2--3 from the inner disk to 
the outer disk.  The median fraction of initial mass in 1000~km and larger objects is 
roughly 1\% to 2\% in massive disks and less than 0.1\% in the lowest mass disks. The 
median fraction of material in 100~km and larger objects is roughly a factor of two 
larger. 

Results for calculations with a single planetesimal size are similar. The collisional
cascade removes a much smaller fraction of the initial disk mass, ranging from $\sim$
80\% in the inner disks of the most massive disks composed of 1~km planetesimals to
$<$ 1\% throughout the disks composed of 100~km planetesimals. Despite the ability
to produce larger objects overall, very little mass ends up in 1000~km or larger 
planetesimals. Over the lifetime of the central star, the lowest mass disks never 
produce 1000~km objects. For the most massive disks, we derive $<$ 2\% (1~km), $<$ 5\% 
(10~km), and $<$ 10\% (100~km) of the initial mass in 1000~km or larger objects 
at the end of the main sequence lifetime of the central star.

These results show that planet formation at 30--150~AU is very inefficient.  Most of 
the initial mass in solids is either removed by the collisional cascade or remains in 
small planetesimals with sizes comparable to their initial size. Very little of the 
initial mass is incorporated into much larger objects. Although more mass ends up in 
much larger objects when planetesimals are initially large, 1000~km or larger objects 
never contain more than 10\% of the initial solid mass in the disk.

\subsubsection{Evolution of Dust}

In our calculations, the collisional cascade converts $\sim$ 1--2\% to more than 95\% 
of the initial solid mass into dust. Because oligarchs and leftover planetesimals with 
$r \lesssim 10^4$~km are unobservable with current techniques, dust emission is the 
only observational diagnostic of icy planet formation at 30--150~AU around other stars. 
Here, we consider the time evolution of the dust around 1~\msun\ stars as a function 
of our input parameters.

To describe our results, we follow KB08 and divide the dust into large grains with 
$r$ = 1 mm to 1 m, small grains with $r$ = 1 $\mu$m to 1 mm, and very small grains
with $r <$ 1 $\mu$m. Although Poynting-Robertson drag removes some large grains in
the inner regions of low mass disks, the collisional cascade grinds nearly all large
grains into small grains on timescales of 10~Myr to 10~Gyr. Poynting-Robertson drag
removes from $<$1\% to 40\% of the mass in small grains; collisions grind the rest into
very small grains.  For most stars with \mstar\ $\gtrsim$ 1~\msun, radiation pressure 
rapidly ejects very small grains \citep[see also][]{kri00,wya05}. These grains then 
produce an outflowing wind of small particles in the disk midplane. This wind contains 
60\% to 100\% of the total amount of mass lost from the disk. More massive disks lose 
more mass in the wind; low mass disks lose less mass in the wind (see also KB08).

\begin{figure} 
\includegraphics[width=6.5in]{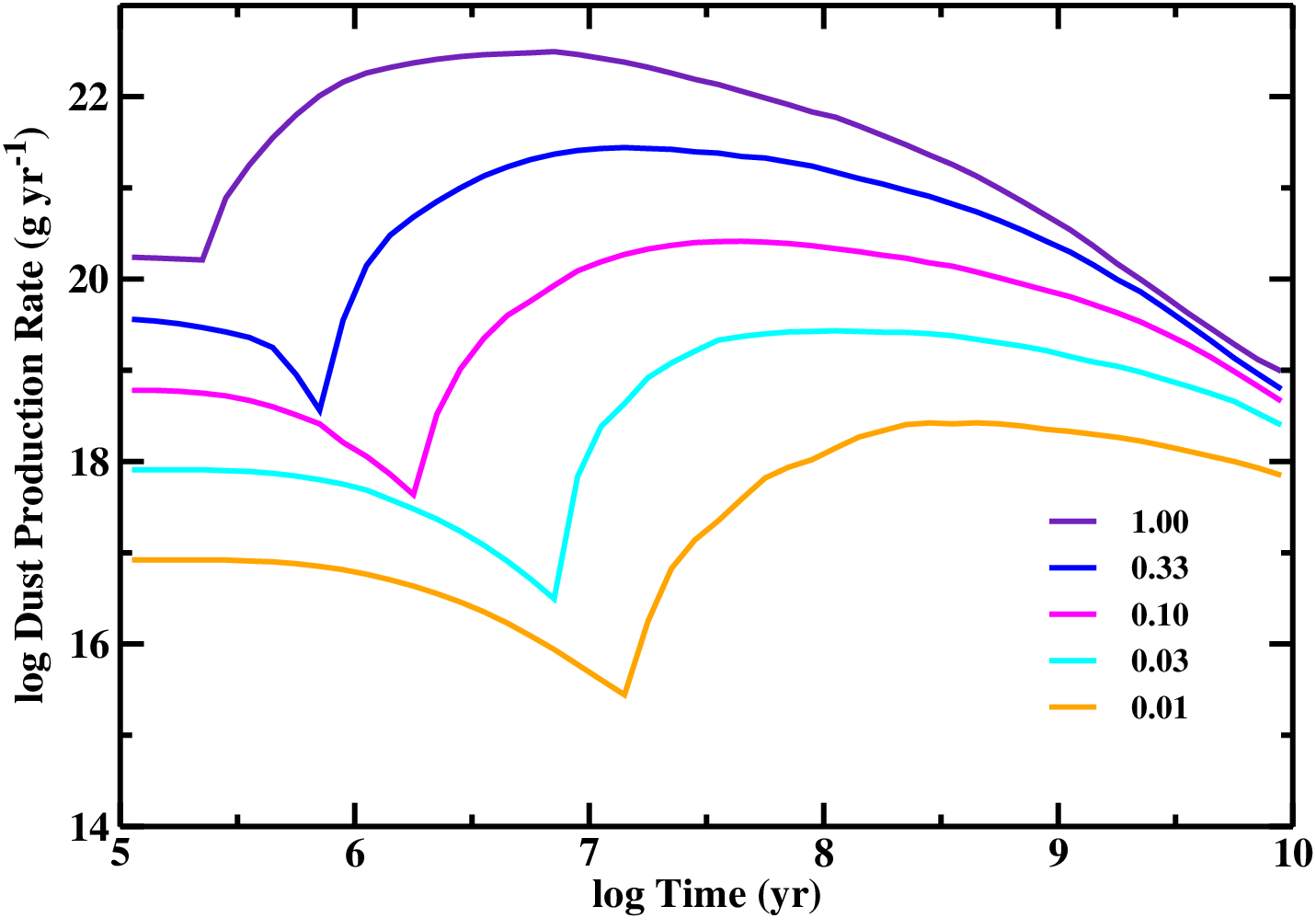}
\vskip 3ex
\caption{
Time evolution of the dust production rate for disks with initial dust 
surface density $\Sigma_d = 30 ~ x_m ~ a^{-1}$ surrounding a 1 \msun\ star. 
The legend indicates the scaling factor $x_m$ for each curve. More massive 
disks produce more dust faster than lower mass disks.
\label{fig: dust1-1msun}
}
\end{figure}

Figure \ref{fig: dust1-1msun} shows the time evolution of the production rate for 
very small grains in disks with $\Sigma \propto a^{-1}$ and a range of initial disk
masses.  When each calculation begins, collisional damping, dynamical friction, and
gas drag reduce collision velocities throughout the disk. Thus, collisions produce
less and less debris; the dust production rate slowly declines with time. During the 
late stages of runaway growth and the onset of oligarchic growth, the largest objects 
reach sizes of 300~km to 500~km. These oligarchs stir leftover planetesimals along
their orbits. Dust production rapidly increases. As oligarchs grow, continued stirring 
leads to a collisional cascade and a peak in the dust production rate. After the dust 
production rate peaks, the cascade removes more and more leftover planetesimals from 
the disk. Fewer planetesimals have less frequent collisions and produce less dust.
The dust production rate then slowly declines with time.

\begin{figure} 
\includegraphics[width=6.5in]{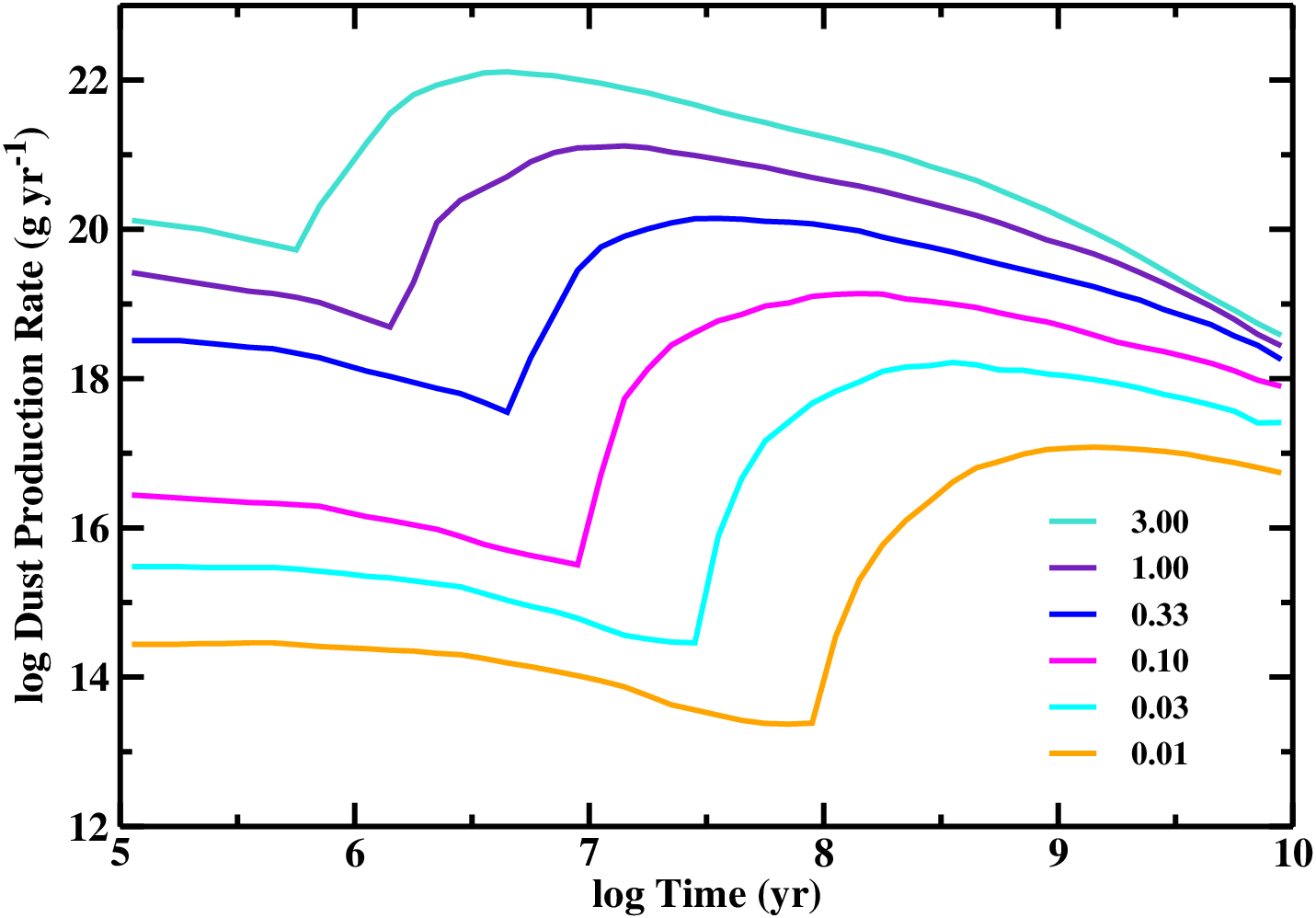}
\vskip 3ex
\caption{
As in Figure \ref{fig: dust1-1msun} for disks with initial dust surface density 
$\Sigma_d = 30 ~ x_m ~ a^{-3/2}$ surrounding a 1 \msun\ star. 
\label{fig: dust2-1msun}
}
\end{figure}

Figure \ref{fig: dust2-1msun} repeats the plot in Figure \ref{fig: dust1-1msun} for
disks with $\Sigma \propto a^{-3/2}$. In both cases, the maximum production rate of
very small grains is 
roughly $10^{22}$ g yr$^{-1}$, implying a maximum mass loss rate of 1.5 \mearth\ Myr$^{-1}$.
This maximum rate is roughly 5 orders of magnitude larger than the mass loss rates in the 
lowest mass disks, $\sim$ 1\% of a lunar mass every million years. The large range in 
local collision rates yields the large difference in dust production rates.  The number of
destructive collisions scales with the square of the local mass density of leftover 
planetesimals. Thus, the dust production rate scales with $x_m^2$. For these calculations, 
we derive maximum dust production rates of

\begin{equation}
\dot{M}_{max}~({\rm g~yr^{-1}}) = \left\{ \begin{array}{rll}
    3.0 \times 10^{22} ~ x_m^2, &  &\Sigma = 30 ~ x_m ~ a^{-1}, ~ f_i = f_S \\
\\
    6.2 \times 10^{20} ~ x_m^2, & \hspace{5mm} &\Sigma = 30 ~ x_m ~ a^{-3/2}, ~ f_i = f_S  \\
\\
    1.4 \times 10^{21} ~ x_m^2, &  &\Sigma = 30 ~ x_m ~ a^{-3/2}, ~ f_i = f_W 
   \end{array} \right.
\label{eq: mdot}
\end{equation}

The time evolution of the collision rate yields a simple relation between the time of maximum
dust production and the initial disk mass. For the calculations described here and in KB08, 
we infer:
\begin{equation}
t_{\dot{M}_{max}}~({\rm Myr}) = \left\{ \begin{array}{rll}
     5.5 ~ x_m^{-1}, &  &\Sigma = 30 ~ x_m ~ a^{-1}, ~ f_i = f_S \\
\\
    14 ~ x_m^{-1}, & \hspace{5mm} &\Sigma = 30 ~ x_m ~ a^{-3/2}, ~ f_i = f_S  \\
\\
   12.5 ~ x_m^{-1}, &  &\Sigma = 30 ~ x_m ~ a^{-3/2}, ~ f_i = f_W 
   \end{array} \right.
\label{eq: tmdot}
\end{equation}
In these expressions, the time of maximum dust production rate scales with the collision
time for leftover planetesimals in the inner disk. This timescale scales inversely with
the disk mass. In both sets of equations, the inter-quartile range is 5\% to 10\%.

These relations illustrate the impact of different surface density laws and fragmentation parameters 
on the production rate of very small grains. In disks with $\Sigma \propto a^{-3/2}$, calculations
with weaker planetesimals produce more dust sooner than calculations with stronger planetesimals.
For the two sets of fragmentation parameters we investigate, the difference in dust production
rate
is roughly a factor of two. Dust production peaks roughly 10\% earlier in time for calculations
with weaker planetesimals. For calculations with identical fragmentation parameters but different
surface density relations, disks with similar total masses yield similar maximum dust production
rates.

\begin{figure} 
\includegraphics[width=6.5in]{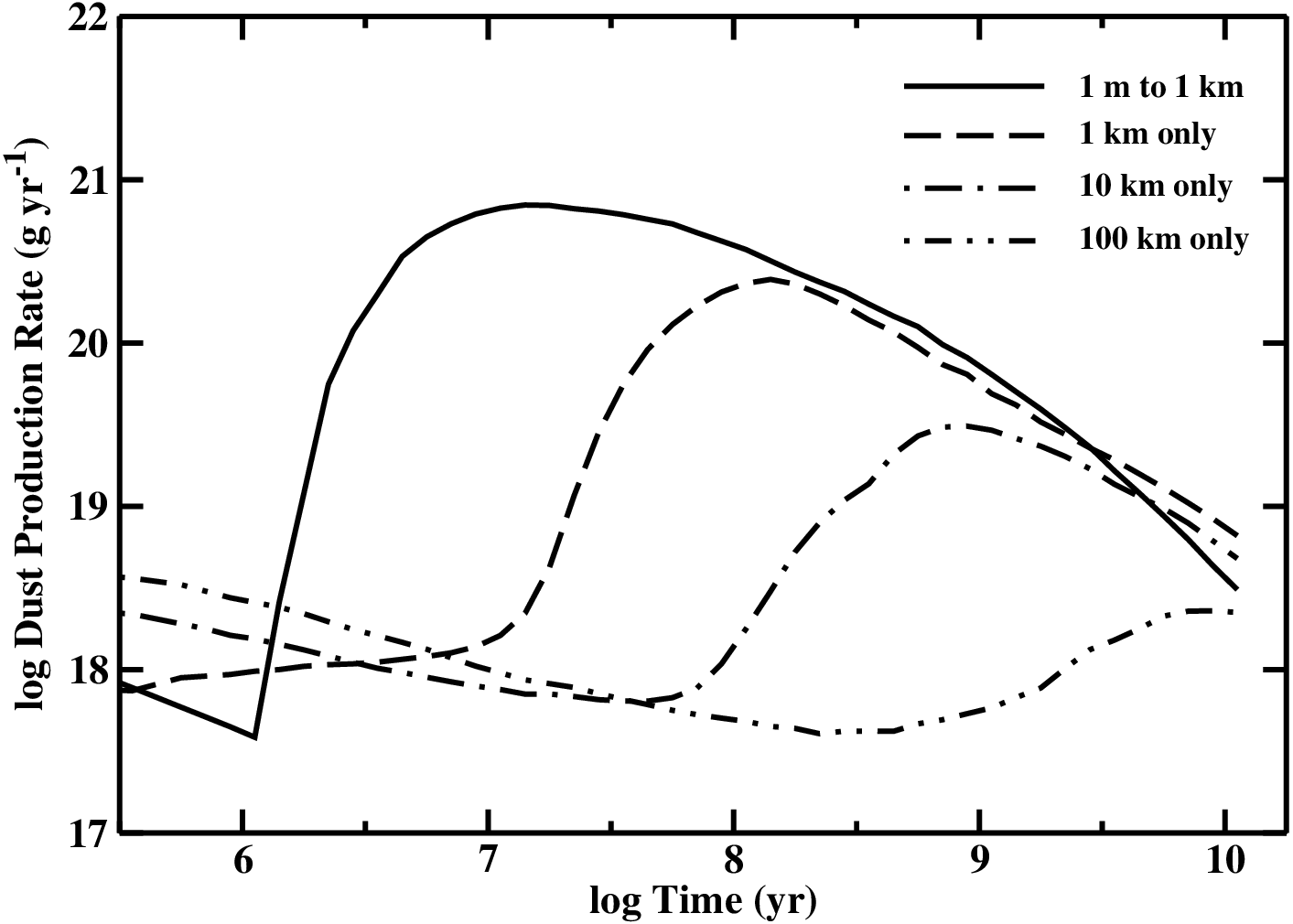}
\vskip 3ex
\caption{
Comparison of production rates of very small grains for calculations with various initial 
sizes of planetesimals as indicated in the legend. All disks have initial 
dust surface density $\Sigma_d = 30 ~ a^{-3/2}$ g cm$^{-2}$.
Disks with smaller planetesimals and a large range of planetesimal sizes 
produce more dust faster than disks with larger planetesimals and a small
range of initial planetesimals sizes.
\label{fig: dust3-1msun}
}
\end{figure}

Dust production is very sensitive to the initial size of planetesimals
(Figure \ref{fig: dust3-1msun}). For $\Sigma  = 30 ~ a^{-3/2}$ g cm$^{-2}$ and the $f_S$
fragmentation parameters, an ensemble of 1~m to 1~km planetesimals produces a maximum 
dust production rate of roughly $6 \times 10^{21}$ \gyr\ at 14~Myr. Collisional cascades in
an ensemble of 1~km planetesimals, however, yield a factor of 30 smaller dust production 
rate, $\approx 2 \times 10^{20}$ \gyr, almost an order of magnitude later in time,
$t_{\dot{M}_{max}} \approx$ 775~Myr.  Ensembles of larger, 10~km and 100~km, planetesimals
yield even smaller maximum dust production rates at very late times. For 1--100~km 
planetesimals, our results suggest that the maximum dust production rate declines by 
roughly a factor of ten for every factor of ten increase in the initial planetesimal size.
The time when this maximum occurs increases by a factor of 6--7 for each factor of ten
increase in the initial size of planetesimals.

Although the dust production rate and time of maximum dust production depend on the
initial planetesimal size, we derive similar scaling laws with initial disk mass.
Our calculations yield

\begin{equation}
\dot{M}_{max} ({\rm g~yr^{-1}}) = \left\{ \begin{array}{rll}
   1.8 \times 10^{20}~ x_m^2 ~ &  &r_0 = {1 ~ km}, ~ f_i = f_S \\
\\
   3.6 \times 10^{19} ~ x_m^2 ~ &  &r_0 = {10 ~ km}, ~ f_i = f_S \\
\\
   3.6 \times 10^{18} ~ x_m^2 ~ &  &r_0 = {100 ~ km}, ~ f_i = f_S \\
   \end{array} \right.
\label{eq: mdotmax-all}
\end{equation}
for the maximum dust production rates and 
\begin{equation}
t_{\dot{M}_{max}} ({\rm Myr}) = \left\{ \begin{array}{rll}
   125 ~ x_m^{-1} ~ &  &r_0 = {1 ~ km}, ~ f_i = f_S \\
\\
   775 ~ x_m^{-1} ~ &  &r_0 = {10 ~ km}, ~ f_i = f_S \\
\\
   5600 ~ x_m^{-1} ~ &  &r_0 = {100 ~ km}, ~ f_i = f_S \\
   \end{array} \right.
\label{eq: tmdot-all}
\end{equation}
for the time of maximum dust production. In both sets of equations, the 
inter-quartile range is 5\% to 10\%.

The large production rates of very small grains in our calculations require 
massive reservoirs of small and large dust grains. In each annulus, the 
evolution of the mass in small and large grains follows a standard pattern. 
During slow growth and most of runaway growth, dust production rates decline 
with time (Figure \ref{fig: dust1-1msun}). Despite this decline, collisions 
among planetesimals produce large and small grains faster than collisions 
or Poynting-Robertson drag remove them. The mass in small grains grows slowly 
with time. When oligarchic growth begins, dust production and the masses of 
small and large grains increase rapidly. The collisional cascade begins, 
converting km-sized planetesimals into dusty debris. Initially, the cascade 
rapidly converts planetesimals into very small grains which are ejected by 
radiation pressure. As the cascade proceeds, the collision rate declines. 
Poynting-Robertson drag removes more and more small grains from the disk. 
Eventually, Poynting-Robertson drag dominates collisions and the mass in
small grains in the annulus rapidly declines to zero.

\begin{figure} 
\includegraphics[width=6.5in]{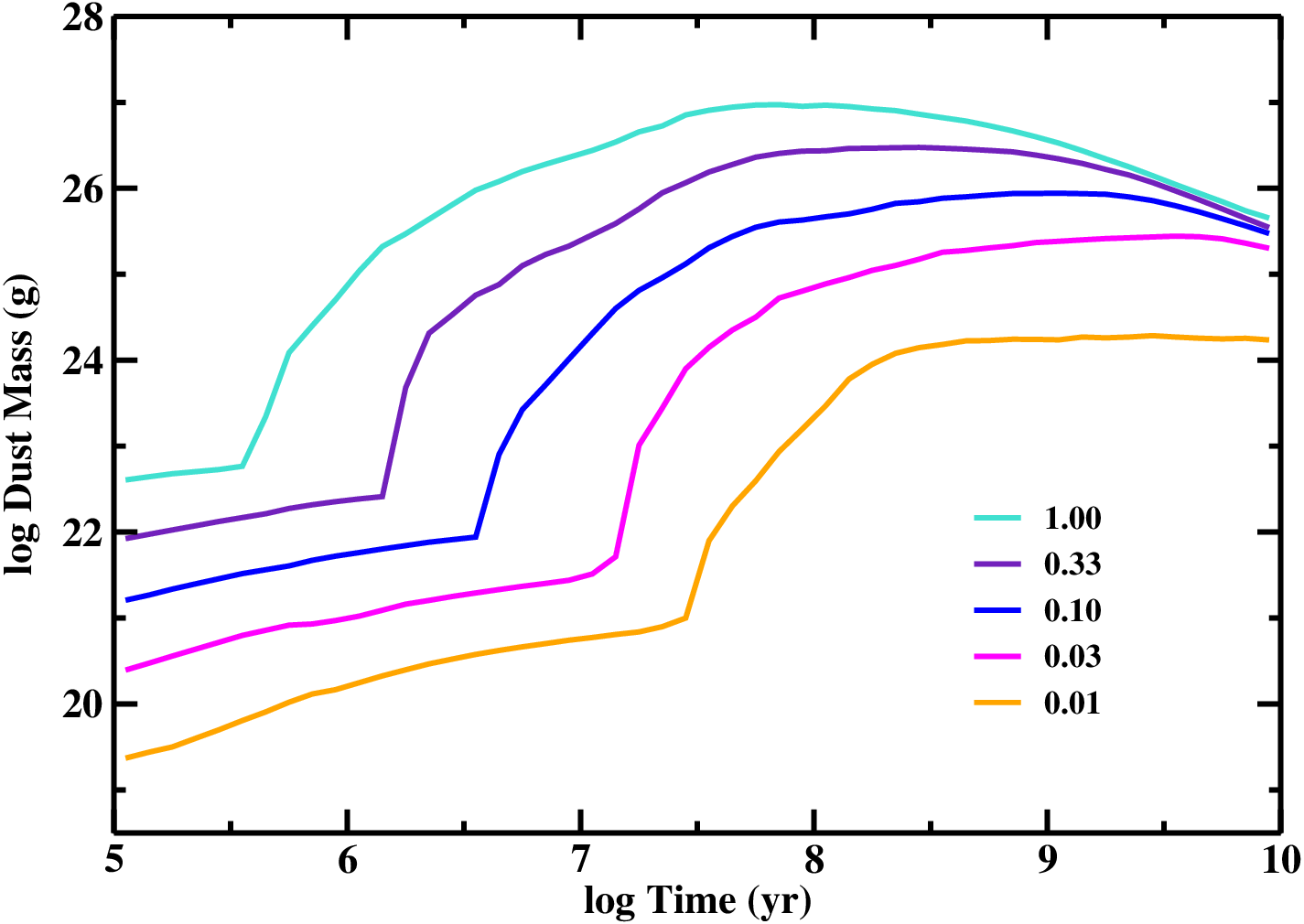}
\vskip 3ex
\caption{
Time evolution of the mass in 1 $\mu$m to 1 mm particles for disks 
with initial dust surface density $\Sigma_d = 30 ~ x_m ~ a^{-1}$ g cm$^{-2}$
surrounding a 1 \msun\ star. The legend indicates the scaling 
factor $x_m$ for each curve. More massive disks produce more 
dust faster than lower mass disks.
\label{fig: mdust1-1msun}
}
\end{figure}

Figure \ref{fig: mdust1-1msun} plots the time evolution of the total mass in 
small grains for disks with $\Sigma \propto a^{-1}$. Initially, all annuli 
slowly produce more and more dust grains. The mass in small grains throughout 
the disk grows slowly with time. When oligarchic growth begins in the inner 
disk, the mass in small grains rapidly grows by 1--2 orders of magnitude.  
As oligarchs form farther and farther out in the disk, the collisions among 
leftover planetesimals continue to produce small grains more rapidly than 
other processes remove them. The mass in grains throughout the disk continues 
to grow. Once the collisional cascade reaches the outer edge of the disk, the
overall dust production rate cannot keep up with removal of small grains by 
collisional erosion or Poynting-Robertson drag. The dust mass begins to decline.

\begin{figure} 
\includegraphics[width=6.5in]{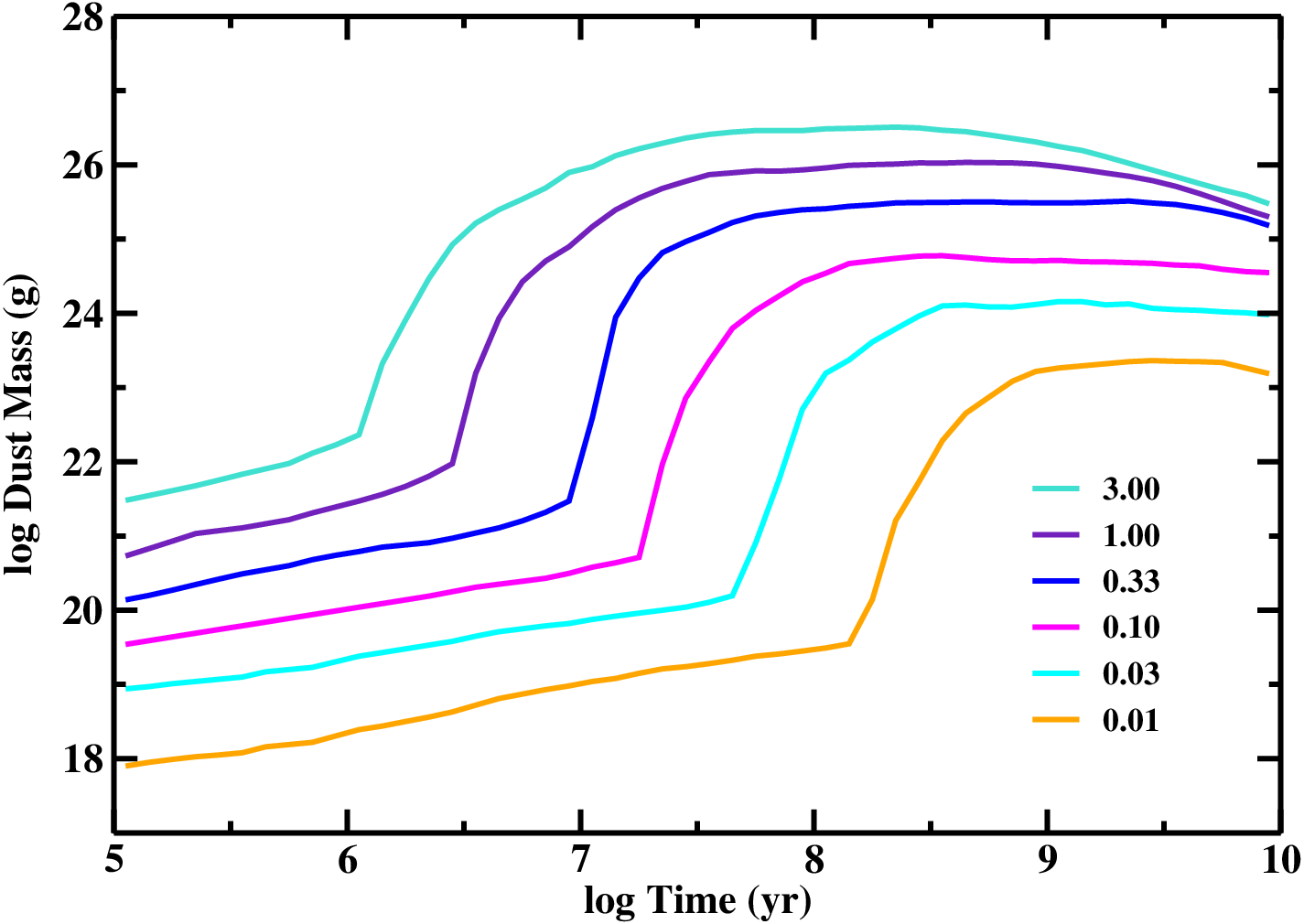}
\vskip 3ex
\caption{
As in Figure \ref{fig: mdust1-1msun} for disks with initial 
$\Sigma_d = 30 ~ x_m ~ a^{-3/2}$ g cm$^{-2}$.
\label{fig: mdust2-1msun}
}
\end{figure}

Figure \ref{fig: mdust2-1msun} shows the evolution of the total mass in small 
grains for disks with
$\Sigma \propto a^{-3/2}$. As in Figure \ref{fig: mdust1-1msun}, the dust mass
grows slowly during the early stages of the runaway, starts to rise rapidly
during the late stages of the runaway, and then reaches a fairly constant plateau 
throughout oligarchic growth. In both sets of calculations, the maximum mass in
small grains is a few lunar masses. This maximum mass scales with disk mass. 
Together with the results in KB08, we derive a typical maximum mass in small
grains
\begin{equation}
M_{max,small}~(M_{\oplus}) = \left\{ \begin{array}{rll}
    0.17 ~ x_m, &  &\Sigma = 30 ~ x_m ~ a^{-1}, ~ f_i = f_S \\
\\
    0.013 ~ x_m, & \hspace{5mm} &\Sigma = 30 ~ x_m ~ a^{-3/2}, ~ f_i = f_S  \\
\\
    0.019 ~ x_m, &  &\Sigma = 30 ~ x_m ~ a^{-3/2}, ~ f_i = f_W 
   \end{array} \right.
\label{eq: msmall}
\end{equation}
For large grains, the maximum mass is 30--40 times larger,
\begin{equation}
M_{max,large}~(M_{\oplus}) = \left\{ \begin{array}{rll}
    5.5 ~ x_m, &  &\Sigma = 30 ~ x_m ~ a^{-1}, ~ f_i = f_S \\
\\
    0.50 ~ x_m, & \hspace{5mm} &\Sigma = 30 ~ x_m ~ a^{-3/2}, ~ f_i = f_S  \\
\\
    0.62 ~ x_m, &  &\Sigma = 30 ~ x_m ~ a^{-3/2}, ~ f_i = f_W 
   \end{array} \right.
\label{eq: mlarge}
\end{equation}
In both sets of equations, the inter-quartile range for the maximum mass
is 10\%. Thus, the maximum mass is more sensitive to the fragmentation
parameters than stochastic variations in the evolution. Weak planetesimals 
produce more dust. The amount of dust is 
less sensitive to the surface density law. For disks with similar total
masses, the dust mass produced in disks with $\Sigma \propto a^{-1}$ 
relative to that in disks with $\Sigma \propto a^{-3/2}$ is 1.4:1 for 
small grains and 1.2:1 for large grains. 

\begin{figure} 
\includegraphics[width=6.5in]{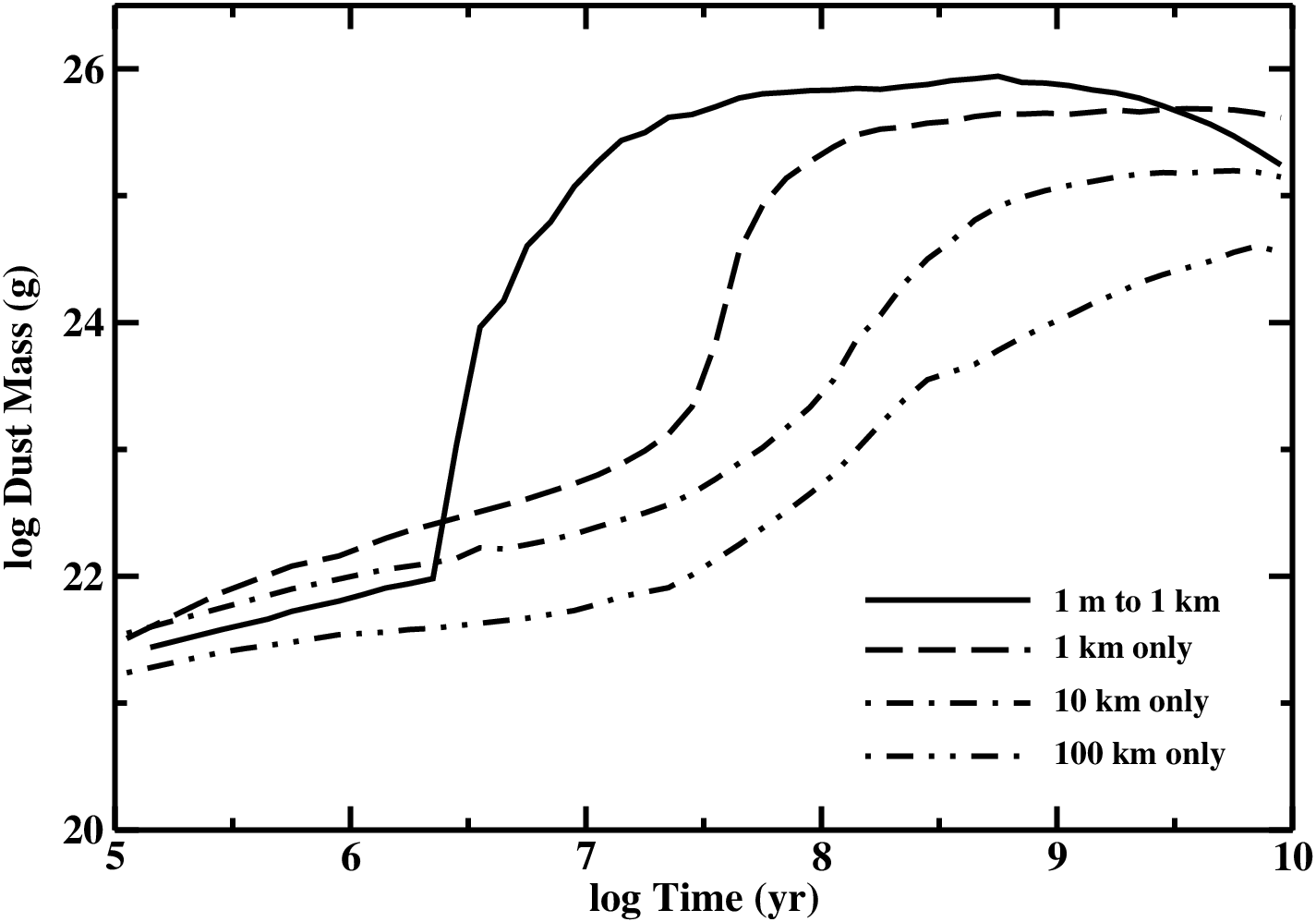}
\vskip 3ex
\caption{
As in Figure \ref{fig: mdust2-1msun} for disks with different
initial planetesimal sizes.  Disks with smaller planetesimals 
produce more dust faster than disks with larger planetesimals.
\label{fig: mdust3-1msun}
}
\end{figure}

The mass in small and large grains is also very sensitive to the initial 
sizes of planetesimals in the disk (Figure \ref{fig: mdust3-1msun}). For 
disks with $\Sigma = 30~a^{-3/2}$ g cm$^{-2}$ and strong planetesimals, 
an ensemble of 1~m to 1~km planetesimals produces a maximum dust mass 
of roughly a lunar mass in small grains at $t$ = 50~Myr to 1~Gyr. Disks 
with larger planetesimals produce less dust at later times. For an ensemble 
of 1~km planetesimals, the maximum dust mass is a factor of $\sim$ 
2 smaller and remains at this level from $t$ = 100~Myr to $t$ = 10~Gyr. 
However, ensembles of 10~km (100~km) planetesimals produce a factor of 3 
(15) less dust for much shorter periods of time. 

For the set of calculations starting with a single size planetesimal,
we derive dust masses of
\begin{equation}
M_{max,small}~(M_{\oplus}) = \left\{ \begin{array}{rll}
    0.009 ~ x_m, &  &r_0 = {1 ~ km}, ~ f_i = f_S \\
\\
    0.0026 ~ x_m, & \hspace{5mm} &r_0 = {10 ~ km}, ~ f_i = f_S \\
\\
    5 \times 10^{-4} ~ x_m, &  &r_0 = {100 ~ km}, ~ f_i = f_S \\
   \end{array} \right.
\label{eq: msmall-rall}
\end{equation}
for small grains and
\begin{equation}
M_{max,large} (M_{\oplus}) = \left\{ \begin{array}{rll}
   0.56 ~ x_m ~ &  &r_0 = {1 ~ km}, ~ f_i = f_S \\
\\
   0.20 ~ x_m ~ &  &r_0 = {10 ~ km}, ~ f_i = f_S \\
\\
   0.033 ~ x_m ~ &  &r_0 = {100 ~ km}, ~ f_i = f_S \\
   \end{array} \right.
\label{eq: mlarge-all}
\end{equation}
for large grains. In both sets of equations, the inter-quartile range in the
maximum mass is 10\%.

Calculations starting with small planetesimals produce dust masses consistent
with those observed in debris disks. The mass in dust grains detected in the
most luminous debris disks around solar-type main sequence stars is roughly a 
lunar mass throughout the stellar lifetime. For our adopted surface density 
laws, disks with 1~km (or smaller) planetesimals and either $x_m >$ 0.1 
($\Sigma \propto a^{-1}$) or $x_m \gtrsim$ 1 ($\Sigma \propto a^{-3/2}$) yield 
a lunar mass in small grains for a large fraction of the stellar lifetime. 
Disks with 10~km or larger planetesimals produce less than 0.2 lunar masses
of small grains late in the main sequence lifetime of the central star. 

\subsection{Icy Planet Formation Around 1.5--3 \msun\ Stars}

Stellar mass is an important aspect of icy planet formation.  With 
$t_{gro} \propto P / \Sigma_d$ (\S3.1.1), the formation timescale at fixed $a$ 
is $t_{gro} \propto \Sigma_d^{-1} M_\star^{-1/2}$.  In disks with identical 
surface density distributions, planets grow more rapidly around more massive 
stars (KB08).  In our models, the surface density of solids scales with the 
stellar mass as $\Sigma_d \propto M_\star$.  The growth time for icy planets 
at fixed $a$ is then $t_{gro} \propto M_\star^{-3/2}$.  Thus, icy planets form 
$\sim$ 3 times more rapidly around 2~\msun\ stars than around 1~\msun\ stars.

Despite this rapid growth, the collisional cascade is more efficient around 
lower mass stars (see also KB08).  The steep relation between the main sequence 
lifetime and the stellar mass, $t_{ms} \propto M_\star^{-n}$ with $n$ = 3--3.5 
\citep{ibe67, dem04}, allows collisions relatively more time to destroy
leftover planetesimals around low mass stars than around more massive stars.
Thus, disks around solar-type stars lose a larger fraction of their initial 
mass and have fewer leftover planetesimals than their equal mass counterparts 
around more massive stars.

\subsubsection{Growth of Large Objects}

\begin{figure} 
\includegraphics[width=6.5in]{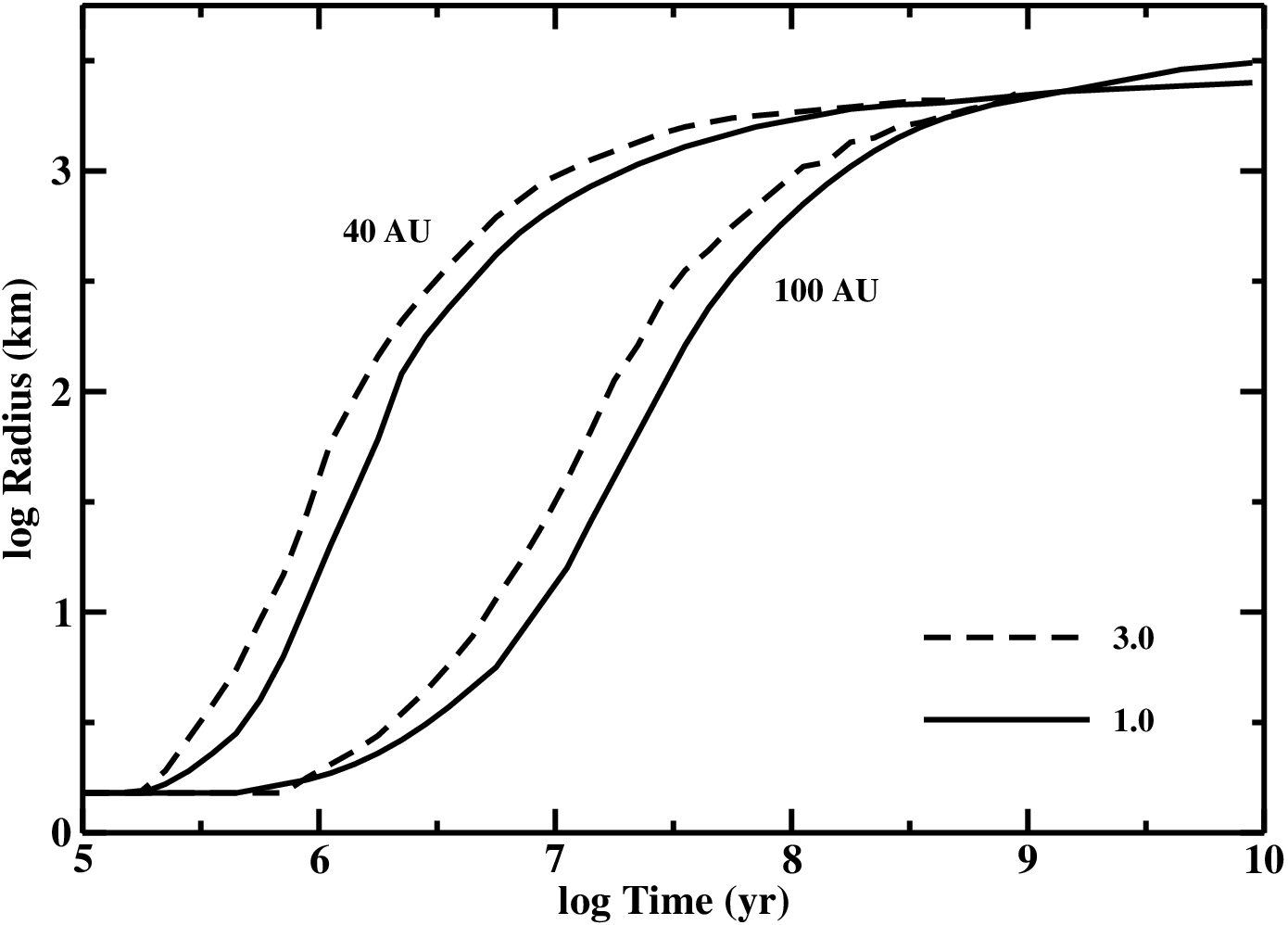}
\vskip 3ex
\caption{
Time evolution of the median radius of the largest object at 40~AU 
and at 100~AU for disks with initial dust surface density 
$\Sigma_d = 10 ~ a^{-1}$ g cm$^{-2}$ surrounding 1 \msun\ and 
3 \msun\ stars.  The legend indicates the stellar mass for 
each curve.  In identical disks, objects grow faster around 
more massive stars.
\label{fig: rad1-allm}
}
\end{figure}

To illustrate these points, we begin with the growth of large objects at 40~AU 
and at 100~AU around 1~\msun\ and 3~\msun\ stars.  For all stellar masses, the 
largest objects in an ensemble of 1~m to 1~km planetesimals at 30--150~AU grow 
slowly to sizes of 5--10~km, experience a short phase of runaway growth to sizes 
of 300--500~km, and then enter a long phase of oligarchic growth. During oligarchic 
growth, icy planets approach sizes of 2000--3000~km.  Figure \ref{fig: rad1-allm}
shows the evolution of the size of the largest object in disks with identical
surface density distributions, $\Sigma  = 10 ~ a^{-1}$ g cm$^{-2}$, around 
1~\msun\ and 3~\msun\ stars. In these disks, slow growth lasts $\sim$ 1~Myr at
40~AU and $\sim$ 10~Myr at 100~AU. Thus, the growth rate derived in \S3.1.1,
$t_{gro} \propto a^{-2.5}$ is preserved in disks around more massive stars.
Although the maximum radius of icy planets is roughly independent of stellar
mass in Figure \ref{fig: rad1-allm}, the plot also demonstrates that planets 
grow roughly 1.7 times faster around 3~\msun\ stars than around 1~\msun\ stars. 
This result confirms the expected scaling of growth time with stellar mass.

\begin{figure} 
\includegraphics[width=6.5in]{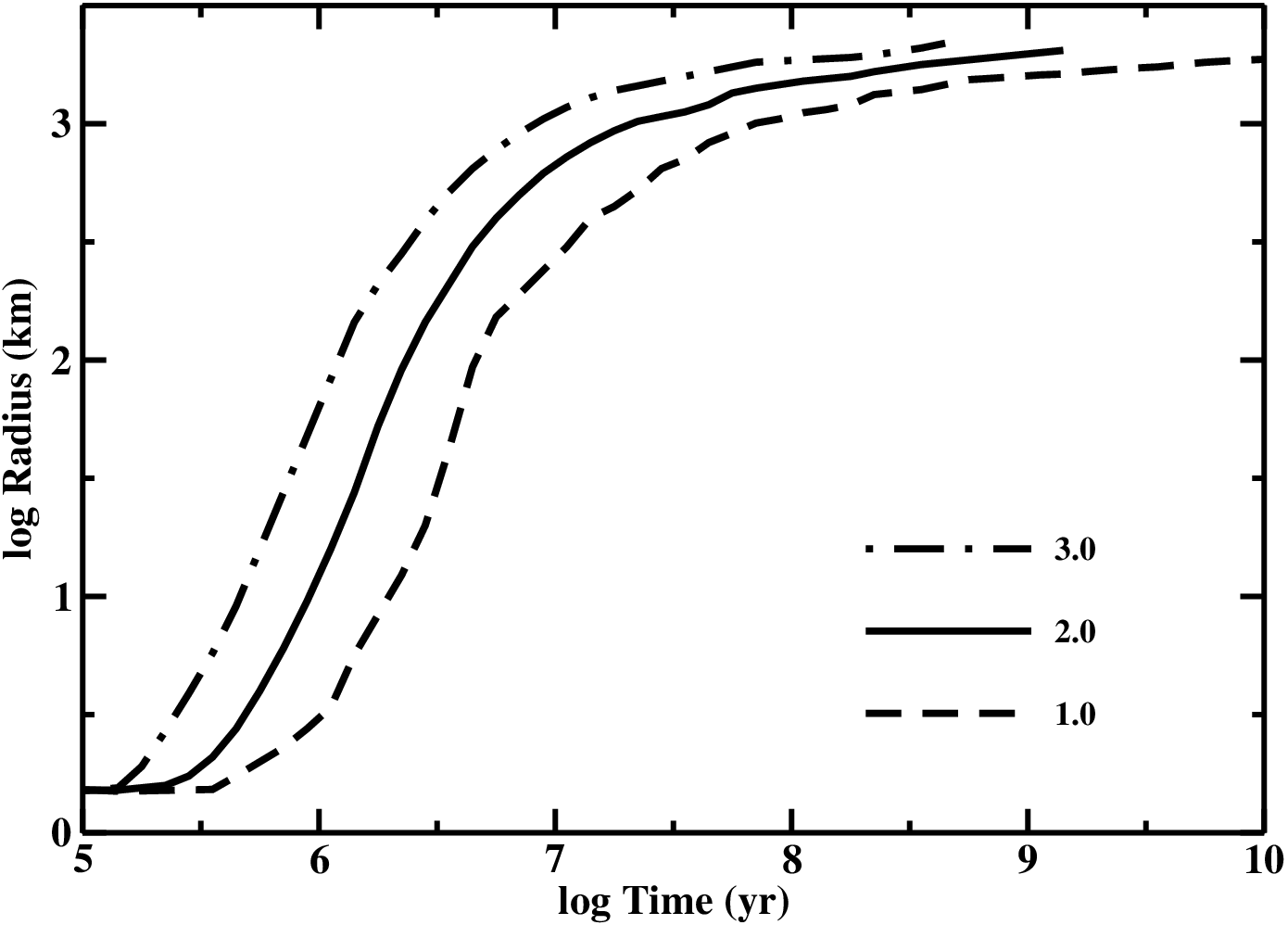}
\vskip 3ex
\caption{
As in Figure \ref{fig: rad1-allm} for planets at 40~AU in disks 
with $\Sigma_d = 3 ~ ( M_\star / M_\odot ) ~ a^{-3/2}$ g cm$^{-2}$. 
Planets grow more rapidly around more massive stars.
\label{fig: rad2-allm}
}
\end{figure}

Figure \ref{fig: rad2-allm} extends these conclusions for disks with scaled surface 
density distributions. In disks with $x_m \gtrsim$ 0.1, slow and runaway growth 
produce 300~km objects in 3--10~Myr around 1--3~\msun\ stars. During oligarchic 
growth, the largest objects slowly approach sizes of $\sim$ 2000~km. The growth
time clearly decreases around more massive stars. Compared to the timescale for
1~\msun\ stars, planets form roughly 3 times more rapidly around 2~\msun\ stars 
and roughly 5 times more rapidly around 3~\msun\ stars. These rates confirm 
our expectation of $t_{gro} \propto M_\star^{-3/2}$.

To quantify these general conclusions, we measure $t_{1000}$ and $r_{max}$ in 
every annulus as a function of input parameters. To derive the best-fitting 
exponents of equation (\ref{eq: tgrow}) for the growth time, we use least-squares 
fits. Although our results for the maximum radius of icy planets depend on the
initial disk mass ($x_m$), $r_{max}$ is independent of semimajor axis and stellar 
mass. Thus, we derive a simple scaling law between $r_{max}$ and $x_m$.  For 
calculations with an ensemble of 1~m to 1~km planetesimals, the third and fourth 
rows of Table \ref{tab:eqplanet1} list our results.

Our expressions for $t_{1000}$ and $r_{max}$ in disks around 1.5--3~\msun\ stars
agree with our results for 1~\msun\ stars. When planetesimals are weak, the
collisional cascade starts to destroy leftover planetesimals when oligarchs 
are relatively small. With less material to accrete, oligarchs take longer to
reach sizes of 1000~km and fail to reach sizes of 2000~km. When planetesimals
are strong, the cascade begins to remove planetesimals when oligarchs are larger.
These oligarchs have more material to accrete, reach sizes of 1000~km much faster,
and can then grow to sizes exceeding 2000~km. 

For any adopted planetesimal strength, the maximum radius scales with the initial 
surface density and is independent of stellar mass and semimajor axis.  In disks 
with identical masses, calculations with $\Sigma \propto a^{-1}$ or 
$\Sigma \propto a^{-3/2}$ yield the same $r_{max}$. Our results suggest that the 
most massive disks produce icy planets with $r_{max} \approx$ 3500~km.  Thus, 
the most massive icy planets have masses of 0.04 \mearth, $\sim$ 15 times 
the mass of Eris, the most massive Kuiper belt object in the Solar System
\citep{bro07}.

The relations for $t_{1000}$ and $r_{max}$ depend on the initial size distribution 
of planetesimals. The first and second rows of Table \ref{tab:eqplanet2} list our 
results for calculations with 1~km (column 1), 10~km (column 2) and 100~km (column 3) 
planetesimals and the $f_S$ fragmentation parameters. Although the growth of planets 
from planetesimals with a single size is much slower than with a range of initial 
planetesimal sizes, planets still grow faster around more massive stars. Because the 
growth time is longer than the gas dissipation time, gas drag is less important. Thus, 
the growth time is less sensitive to the initial surface density and more sensitive 
to the initial semimajor axis.

Independent of stellar mass, calculations with larger planetesimals produce larger
planets. For 1, 10, and 100~km planetesimals, the maximum radii of icy planets 
increase from $\sim$ 5500~km to $\sim$ 7500~km to $\sim$ 11,500~km. The masses 
of these planets range from $\sim$ 0.17 \mearth\ to $\sim$ 1.5 \mearth. 

Although planets form faster in disks around more massive stars, disks around lower 
mass stars generally produce planets more efficiently. To demonstrate this point, we 
consider the formation of 1000~km objects -- which we call `Plutos' -- as a function 
of semimajor axis and stellar mass.  For each set of calculations in our study, 
Tables \ref{tab:rad1000.1}--\ref{tab:rad1000.6} list the median number of Plutos $n_P$ 
as a function of $x_m$, \mstar, and $a$. For simplicity, we report $n_P$ in bins of 
semimajor axis; the width of each bin is $\delta a \approx 0.2 a$.

Throughout the disk, Pluto production correlates with initial disk mass.  For disks
with $n_P(x_m,a) \gtrsim$ 1--10, $n_P$ correlates with $x_m$. In the most massive disks, 
hundreds of Plutos form throughout the disk. In lower mass disks, Plutos are concentrated 
in the inner disk. 

Pluto production is also a strong function of the fragmentation parameters. For disks
with similar masses, calculations with strong planetesimals produce roughly 3 times as
many Plutos as calculations with weak planetesimals. In calculations with weak planetesimals,
Pluto formation is restricted to the inner regions of the most massive disks.

Plutos also form more efficiently in disks with small planetesimals and in disks with a 
range of planetesimal sizes.  The inner regions of very massive disks composed only of 100~km
planetesimals produce the most Plutos. However, the efficiency drops to zero in the outer 
disk and for all semimajor axes of lower mass disks. Thus, most disks with only 10--100~km 
planetesimals produce no Plutos. Disks with 1~km or smaller planetesimals almost always 
produce at least one Pluto and often produce Plutos throughout the disk.

These results demonstrate that icy planet formation at 30--150~AU is inefficient. In the most 
massive disks composed of 100~km planetesimals, objects larger than 1000~km contain 
10\% to 25\% of the initial solid mass at 30--35~AU.  This fraction drops to less than 
2\% at 50~AU and to less than 0.1\% at 60--70~AU.  For planetesimal sizes of 1--100~km, 
the percentage of initial mass in 1000~km objects falls by roughly a factor of two for 
every factor of 10 reduction in the initial size of planetesimals.  These fractions are 
fairly independent of the stellar mass. The ensemble of icy planets around 1~\msun\ stars 
is $\sim$ 25\% more massive than ensembles around 2--3~\msun\ stars. However, the mass 
fraction in 1000~km objects is very sensitive to the initial surface density. At 50--100~AU, 
our results suggest factors of 2--3 reduction in the mass fraction of 1000~km objects for 
every order of magnitude reduction in initial surface density.

\subsubsection{Evolution of Dust}

The evolution of dusty debris disks around 1.5--3~\msun\ stars generally follows the 
evolution for 1~\msun\ stars. During the late stages of runaway growth and the early
stages of oligarchic growth, stirring leads to a collisional cascade that grinds 
leftover planetesimals into dust grains. Runaway and oligarchic growth produce large
planets around more massive stars faster than around lower mass stars. Thus, debris
disks form first around more massive stars. Our scaling of the initial disk mass with
the stellar mass leads to larger dust production rates and larger dust masses around 
more massive stars.  Lower mass stars live longer than more massive stars.  Over the 
main sequence lifetime of the central star, lower mass stars produce more dust.  We 
will consider whether larger dust masses produce brighter debris disks in \S4. In 
this section, we show how dust production rates and total dust masses depend on 
stellar mass and the properties of planetesimals.

\begin{figure} 
\includegraphics[width=6.5in]{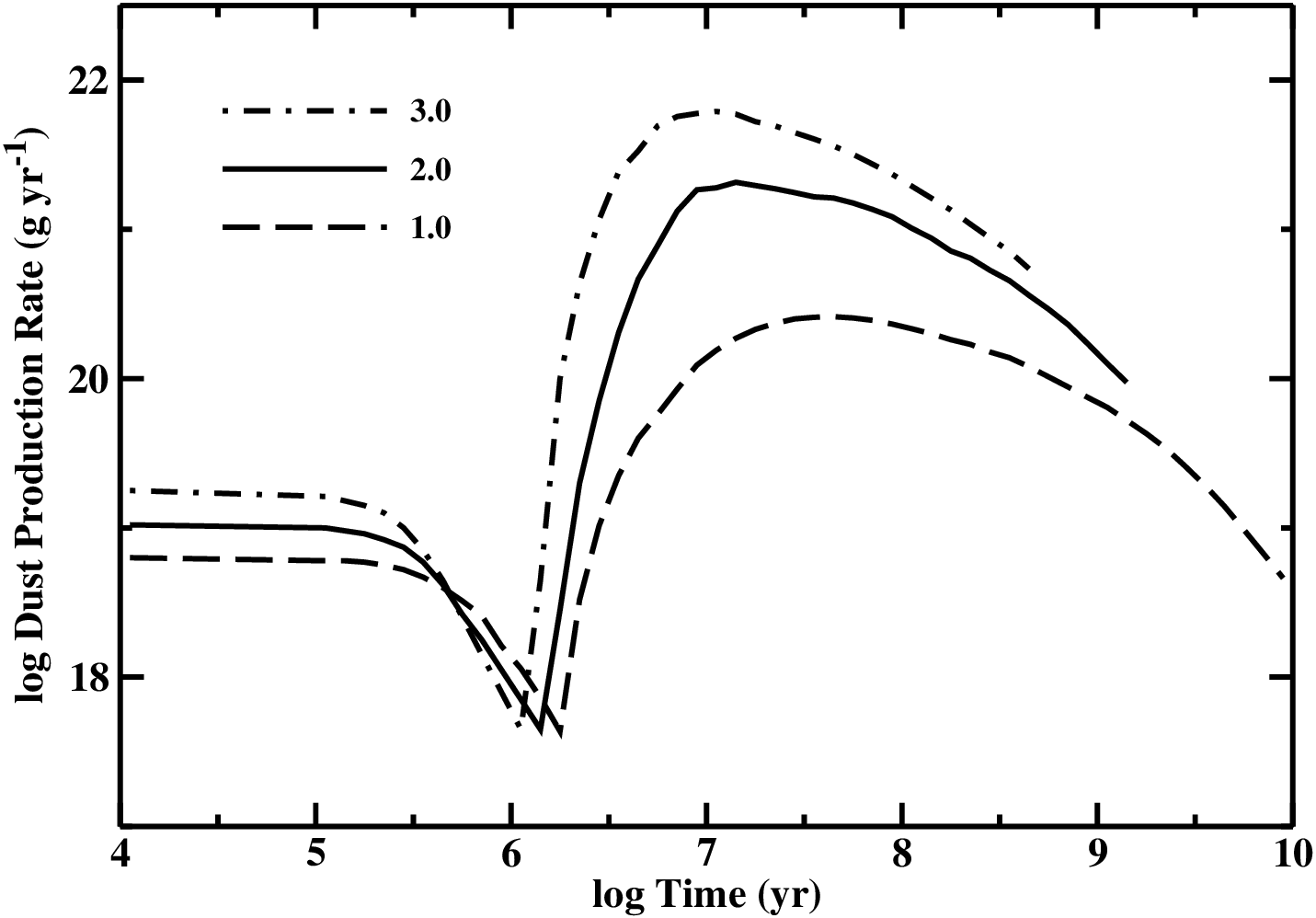}
\vskip 3ex
\caption{
Time evolution of the production rate of very small grains for disks with initial dust surface 
density $\Sigma_d = 3 ~ (M_{\star} / M_\odot) ~ a^{-1}$ g cm$^{-2}$ ($x_m$ = 0.1) 
surrounding 1--3 \msun\ stars.  The legend indicates the stellar mass in
\msun\ for each curve. Disks around more massive stars produce more dust 
faster than disks around lower mass stars.
\label{fig: mdot1-allm}
}
\end{figure}

Figure \ref{fig: mdot1-allm} compares productions rates of very small grains for disks with 
$\Sigma_d = 3 ~ (M_{\star} / M_\odot) ~ a^{-1}$ g cm$^{-2}$ around 1--3~\msun\ stars.
Initially, the higher mass disks around more massive stars have larger dust production 
rates. During runaway growth, dust production declines. Although the dust production 
rates decline fastest around more massive stars, disks with roughly similar masses
reach similar minimum dust production rates. During the transition from runaway growth
to oligarchic growth, the dust production rates increase. This transition occurs earlier
in disks around more massive stars. The larger initial disk masses around more massive
stars also lead to larger peak dust production rates at earlier times. Following the 
peak, dust production declines by more than an order of magnitude by the end of the
main sequence lifetime of the central star.

\begin{figure} 
\includegraphics[width=6.5in]{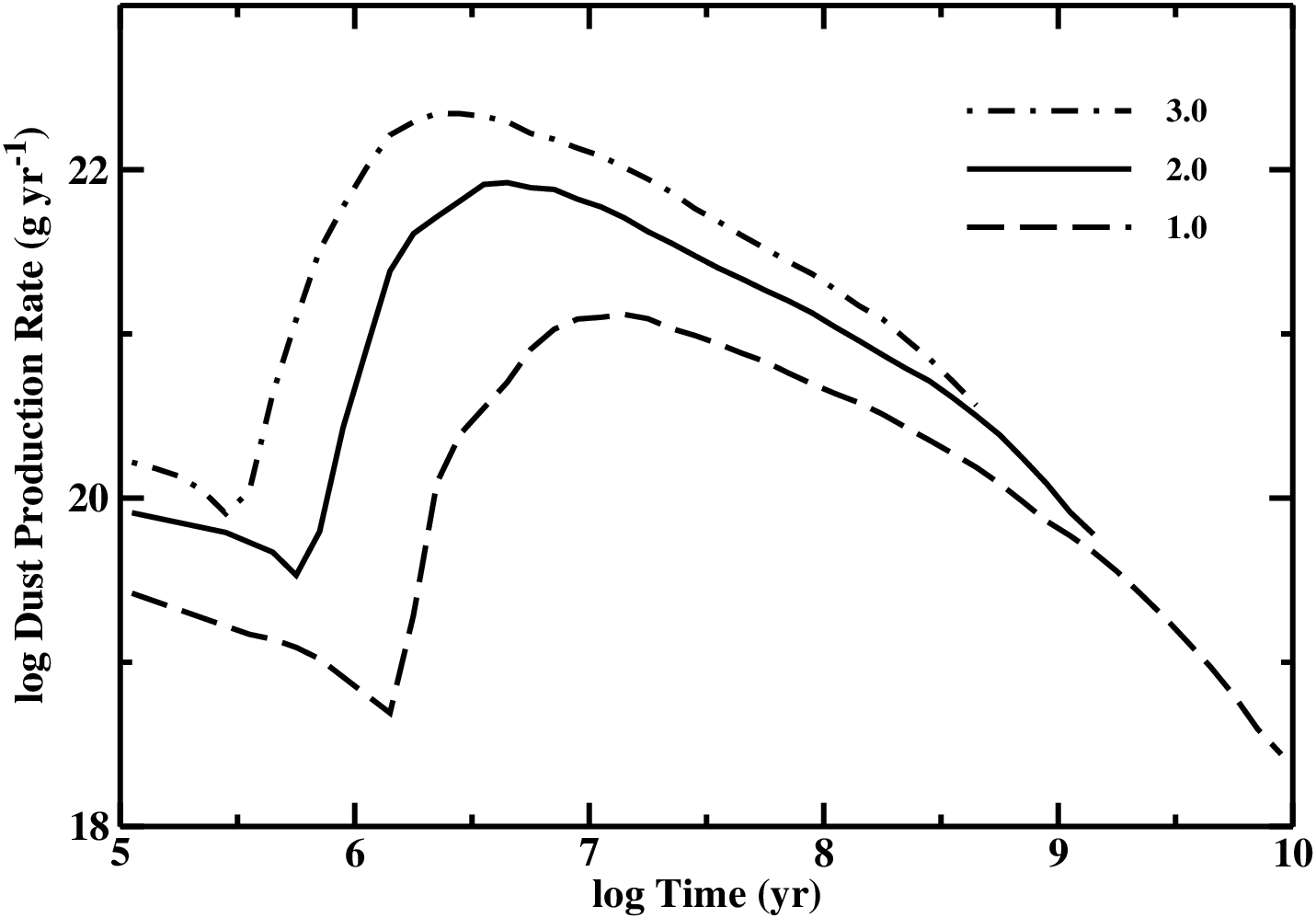}
\vskip 3ex
\caption{
As in Figure \ref{fig: mdot1-allm} for disks with 
$\Sigma_d = 30 ~ (M_{\star} / M_\odot) ~ a^{-3/2}$ g cm$^{-2}$
($x_m$ = 1).
\label{fig: mdot2-allm}
}
\end{figure}

Figure \ref{fig: mdot2-allm} repeats Figure \ref{fig: mdot1-allm} for disks with 
weak planetesimals and $\Sigma_d = 30 ~ (M_{\star} / M_\odot) ~ a^{-3/2}$ g cm$^{-2}$. 
During runaway growth,
the dust production rate declines less than the rates for disks with shallower surface
density distributions (compare with Figure \ref{fig: mdot1-allm}). Once oligarchic
growth begins, however, the dust production rate evolves as in Figure \ref{fig: mdot1-allm}.
More massive disks reach larger dust production rates sooner than lower mass disks. 
After reaching peak dust production, the rates decline by 1--2 orders of magnitude 
before the central star evolves off the main sequence.

The variation of dust production rate with the fragmentation parameters and the
initial sizes of planetesimals follows the results for 1~\msun\ stars.  For our 
two sets of fragmentation parameters, weak planetesimals produce twice as much 
dust roughly 10\% earlier than strong planetesimals. However, dust production is
much more sensitive to the initial planetesimal size. For the strong fragmentation
parameters, disks with an ensemble of 1~m to 1~km planetesimals produce dust at 
four times the rate of disks with only 1~km planetesimals. Calculations with the
large range of planetesimal sizes also reach peak dust production nearly 10 times
sooner than those with only 1~km planetesimals. These differences grow with increasing 
planetesimal size. Compared to the calculations with 1~m to 1~km planetesimals,
dust production rates for ensembles of 10~km (100~km) planetesimals are roughly
20 times (200 times) smaller and occur 50 times (500 times) later. 

The fifth and sixth rows of Table \ref{tab:eqplanet1} and the third and fourth rows
of Table \ref{tab:eqplanet2} summarize our scaling relations for the time and
magnitude of peak dust production. The time of peak dust production depends on the 
growth time, which scales inversely with disk mass and as $M_\star^{-3/2}$ (\S3.1.1).
The dust production rate depends on the number of collisions, which scales with
the square of the disk mass and $M_\star^{5/2}$ (KB08). In these relations, the
inter-quartile range is 5\% to 10\%.

\begin{figure} 
\includegraphics[width=6.5in]{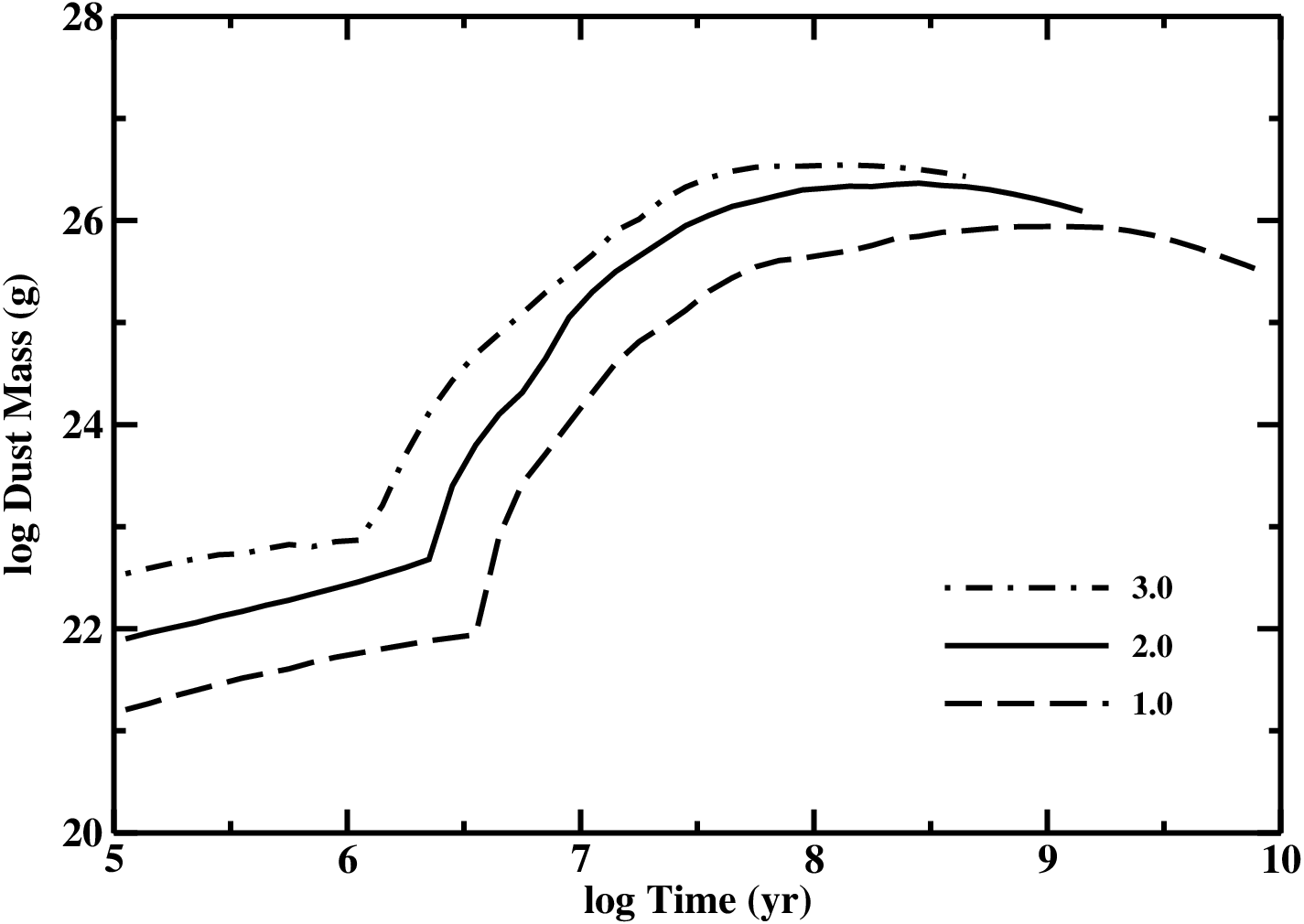}
\vskip 3ex
\caption{
Time evolution of the mass in 1 $\mu$m to 1 mm particles for disks with 
initial dust surface density $\Sigma_d = 3 ~ (M_{\star} / M_\odot) ~ a^{-1}$
g cm$^{-2}$ ($x_m$ = 0.1) surrounding 1--3 \msun\ stars. The legend indicates 
the stellar mass in \msun\ for each curve. 
\label{fig: mdust1-allm}
}
\end{figure}

Figure \ref{fig: mdust1-allm} illustrates the time variation of the mass in small dust 
grains for disks with initial $\Sigma_d = 3 ~ (M_{\star} / M_\odot) ~ a^{-1}$ g cm$^{-2}$ 
surrounding 1--3~\msun\ stars. The dust mass grows slowly throughout runaway growth,
increases dramatically during the transition from runaway to oligarchic growth, slowly 
reaches a plateau, and then declines with time. During the slow rise in dust mass,
collisions fragment small grains into very small grains. Radiation pressure rapidly ejects
these very small grains. As the dust mass reaches maximum, Poynting-Robertson drag starts
to remove small grains more and more rapidly. From this point on, Poynting-Robertson drag
removes more material from the disk than radiation pressure. Disks around more massive stars 
pass through this sequence more rapidly than disks around lower mass stars. Disks around more 
massive stars also have larger dust masses.

Figure \ref{fig: mdust1-allm} also shows the impact of stellar evolution on the dust
masses. Although more massive stars reach peak dust masses earlier, the central star
evolves off the main sequence before the dust mass declines significantly. Thus, older
disks around more massive stars have much more dust than older disks around less massive
stars. 

\begin{figure} 
\includegraphics[width=6.5in]{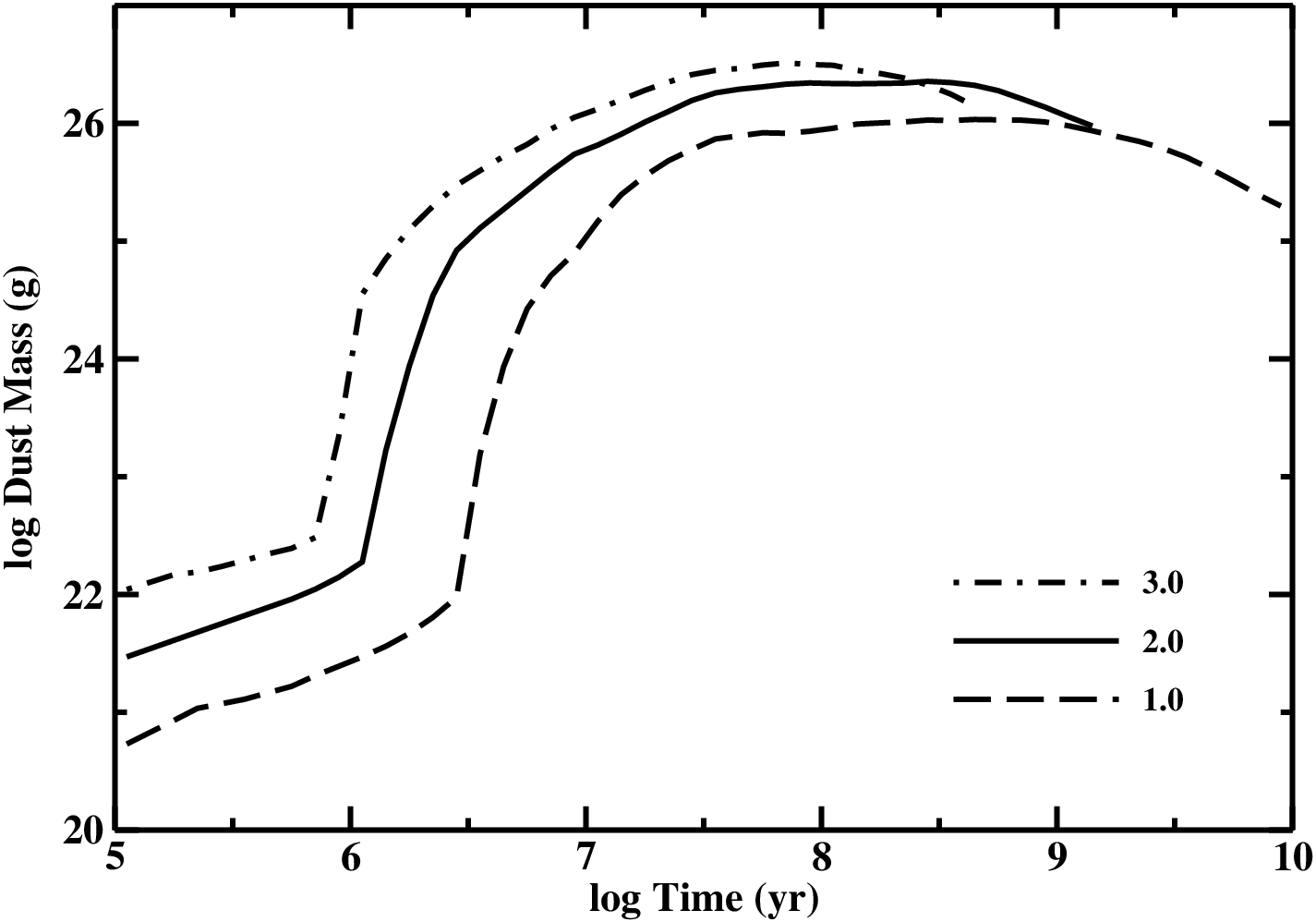}
\vskip 3ex
\caption{
As in Figure \ref{fig: mdust1-allm} for disks with 
$\Sigma_d = 30 ~ (M_{\star} / M_\odot) ~ a^{-3/2}$ g cm$^{-2}$
($x_m$ = 1).
\label{fig: mdust2-allm}
}
\end{figure}

Figure \ref{fig: mdust2-allm} repeats Figure \ref{fig: mdust1-allm} for disks with weak planetesimals 
and initial surface density $\Sigma_d = 30 ~ (M_{\star} / M_\odot) ~ a^{-3/2}$ g cm$^{-2}$. 
Throughout the evolution, disks with weak planetesimals have larger dust masses than disks
with strong planetesimals. For our set of fragmentation parameters, the difference in
peak dust mass is 50\% for small grains and 25\% for large grains. 

The dust masses in our debris disks are also sensitive to the initial mass distribution 
of planetesimals. For the $f_S$ parameters, the dust mass is roughly inversely proportional 
to the maximum initial size of planetesimals. For fixed initial disk mass, calculations with 
an ensemble of 1~m to 1~km planetesimals produce the largest dust masses; calculations with
100~km planetesimals yield the smallest dust masses. 

The seventh and eighth rows of Table \ref{tab:eqplanet1} and the fifth and sixth rows
of Table \ref{tab:eqplanet2} summarize our scaling relations for the masses in small
and large grains. All dust masses scale with the initial disk mass. For massive disks
with $x_m \approx$ 1 and small planetesimals, the mass in small grains is roughly a 
lunar mass.  This mass drops to 0.03--0.3 lunar masses for calculations with larger
(10~km to 100~km planetesimals). 

\subsection{Highlights of Icy Planet Formation at 30--150~AU}

\begin{figure} 
\includegraphics[width=6.5in]{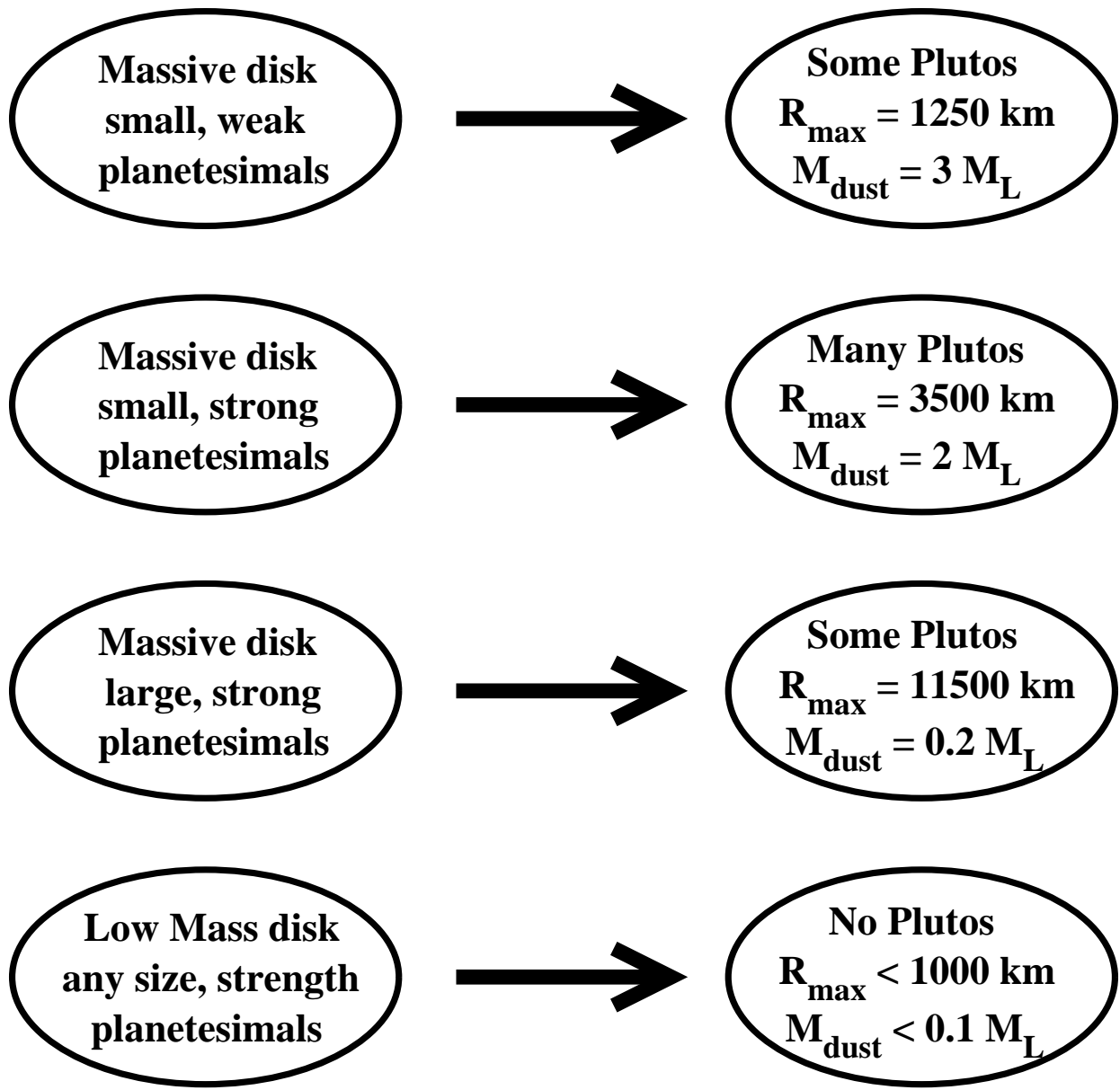}
\vskip 8ex
\caption{
Highlights of icy planet formation at 30--150~AU. Dust masses are in lunar masses.
\label{fig: schema1}
}
\end{figure}

In KB08, we highlighted the six main features of icy planet formation in disks composed
of 1~km planetesimals at 30--150~AU around 1--3~\msun\ stars.  To include conclusions 
based on our new calculations, we expand on these features here. Figure \ref{fig: schema1}
is a schematic summary of these highlights.

\begin{enumerate}

\item Following a short period of slow growth, runaway growth produces an ensemble of
oligarchs with radii of 500--1000~km. The timescale to produce oligarchs scales inversely
with the initial surface density and with the initial radii of planetesimals. Throughout 
runaway growth, oligarchs stir up the orbits of leftover planetesimals. Stirring reduces 
gravitational focusing factors and ends runaway growth.

\item Icy planet formation at 30--150~AU is self-limiting.  In calculations with small
planetesimals, the collisional cascade removes leftovers faster than oligarchs can accrete
them. In disks with larger planetesimals, the collisional cascade is not effective; however, 
the central star often evolves off the main sequence before oligarchs reach very large radii.
In both cases, the largest icy planets slowly reach a characteristic maximum radius which is 
independent of stellar mass.  This maximum radius depends on the initial sizes of the planetesimals.
For ensembles of 1~m to 1~km planetesimals, the maximum radius 
is $\sim$ 3500~km.  This maximum radius grows with the maximum initial radius of planetesimals. 
The largest planets -- $\sim 10^4$~km -- form out of 100~km planetesimals.  The maximum radius 
of an icy planet and the timescale to reach this radius also depend on the initial mass of 
solids in the disk. More massive disks make more massive planets more rapidly.  Although the maximum 
radius of a planet is less sensitive to the fragmentation parameters, weaker planetesimals produce 
smaller planets.  Tables \ref{tab:eqplanet1}--\ref{tab:eqplanet2} list the scaling relations 
for maximum radius and formation time as a function of initial disk mass and the mass of the 
central star.

\item As planets grow slowly, a collisional cascade grinds leftovers to dust.  Early on, 
radiation pressure ejects very small grains in an outflowing wind. Later, Poynting-Robertson
drag also removes larger grains from the disk. In our calculations, radiation pressure 
removes at least twice as much mass from the disk as Poynting-Robertson drag. When an annulus
is massive, radiation pressure dominates mass removal. As the mass in an annulus declines,
Poynting-Robertson drag removes a larger and larger fraction of the mass. The timescale 
for the collisional cascade to remove leftover planetesimals is close to the main sequence
lifetime of the central star. Thus, the cascade removes more material from the inner disk 
than from the outer disk.  Efficient collisional cascades require small planetesimals. Thus,
ensembles of large planetesimals lose less mass than ensembles of small planetesimals.

\item Icy planet formation is inefficient. Independent of the fragmentation parameters, 
only 1\% to 10\% of the initial mass in solids ends up in objects with radii of 1000~km 
or larger. Objects with radii larger than 100~km contain from 2\% to 100\% of the initial 
mass in solids. The efficiency of planet formation correlates with the initial radii
of planetesimals.  In ensembles of 1, 10, or 100~km planetesimals, objects with radii
exceeding 1000~km contain less than 1\%, 3\%, or 10\% of the initial mass in solids.
Only the inner regions of the most massive disks reach these levels.  In the outer 
regions of all disks and the inner regions of low mass disks, the largest objects 
have radii smaller than 1000~km. 

\item In nearly all disks, the collisional cascade produces an observable amount of dust.  
The outflowing winds of small particles are observable in the most massive disks.  The 
amount of debris scales inversely with the initial sizes of planetesimals.  At 30--150~AU, 
the mass in 1~$\mu$m to 1~mm particles is 0.01--3 lunar masses for disks with $x_m$ = 
0.01--3 (Figures \ref{fig: mdust1-1msun}--\ref{fig: mdust3-1msun} and 
\ref{fig: mdust1-allm}--\ref{fig: mdust2-allm}). The dust masses in luminous debris disks 
around A-type and G-type stars typically exceed 0.1 lunar masses. Our results suggest that 
the amount of dust rises from 1--10~Myr, maintains a roughly constant level for 10--100~Myr, 
and then slowly declines until the central star evolves off the main sequence.

\item Even with large planetesimals, dusty debris is a signpost of the formation of
a planetary system. This debris is present throughout the lifetime of the central star.

\end{enumerate}

\section{DEBRIS DISK EVOLUTION}

As in KB08, we convert model size distributions of dust grains, planetesimals, 
and planets into observable quantities. For each evolution time $t$, we calculate 
the stellar luminosity \lstar\ and effective temperature \tstar\ from the Y$^2$ 
stellar evolution models \citep{dem04}. We adopt standard values for the smallest 
stable grain size ($r_2$ = 1 $\mu$m) and the slope ($q$ = 1) of the emissivity law 
for small grains \citep[see also][]{bur79,art88,kim02,naj05,wil06}.  Using a simple
radiative transfer code, we then compute the radial optical depth $\tau(a)$ in 
each annulus of the model grid. The optical depth allows us to 
compute the fraction of \lstar\ absorbed by grains in each annulus. For each grain 
size in each annulus, we calculate the equilibrium grain temperature $T(r,a)$ and 
an emitted spectrum. Summing the spectra over $r$ and $a$ yields the predicted 
spectral energy distribution (SED) and the total dust luminosity $L_d$ as a function 
of time.  KB08 and \citet{kb04b} describe this calculation in more detail.

To describe the time evolution of observable quantities, we focus on $L_d$ 
and the excesses at IR and submm wavelengths.  The fractional dust luminosity 
\ldlstar\ measures the relative luminosity of the debris disk. For excesses at 
specific wavelengths, we quote the total emission of the disk and the central
star relative to the emission from the stellar photosphere,
$F_{\lambda} / F_{\lambda,0}$. With this definition, disks that produce
no excess have $F_{\lambda} / F_{\lambda,0}$ = 1; disks where the excess
emission is comparable to the emission from the central star have
$F_{\lambda} / F_{\lambda,0}$ = 2.

We begin this section with a discussion of excess emission for 1~\msun\ stars.  
After discussing results for 1.5--3~\msun\ stars, we conclude this section with 
a brief summary. To facilitate comparisons of our results with observations, 
Tables \ref{tab:mod-1p0-a1}--\ref{tab:mod-3p0-a15-s} list results for the fractional 
dust luminosity and excesses at 24--850~$\mu$m. The paper version lists the 
first five lines of results for $x_m$ = 0.01, 0.1, and 1. The electronic 
version includes all results for these $x_m$.

\subsection{Evolution for 1 \msun\ Stars}

\begin{figure} 
\includegraphics[width=6.5in]{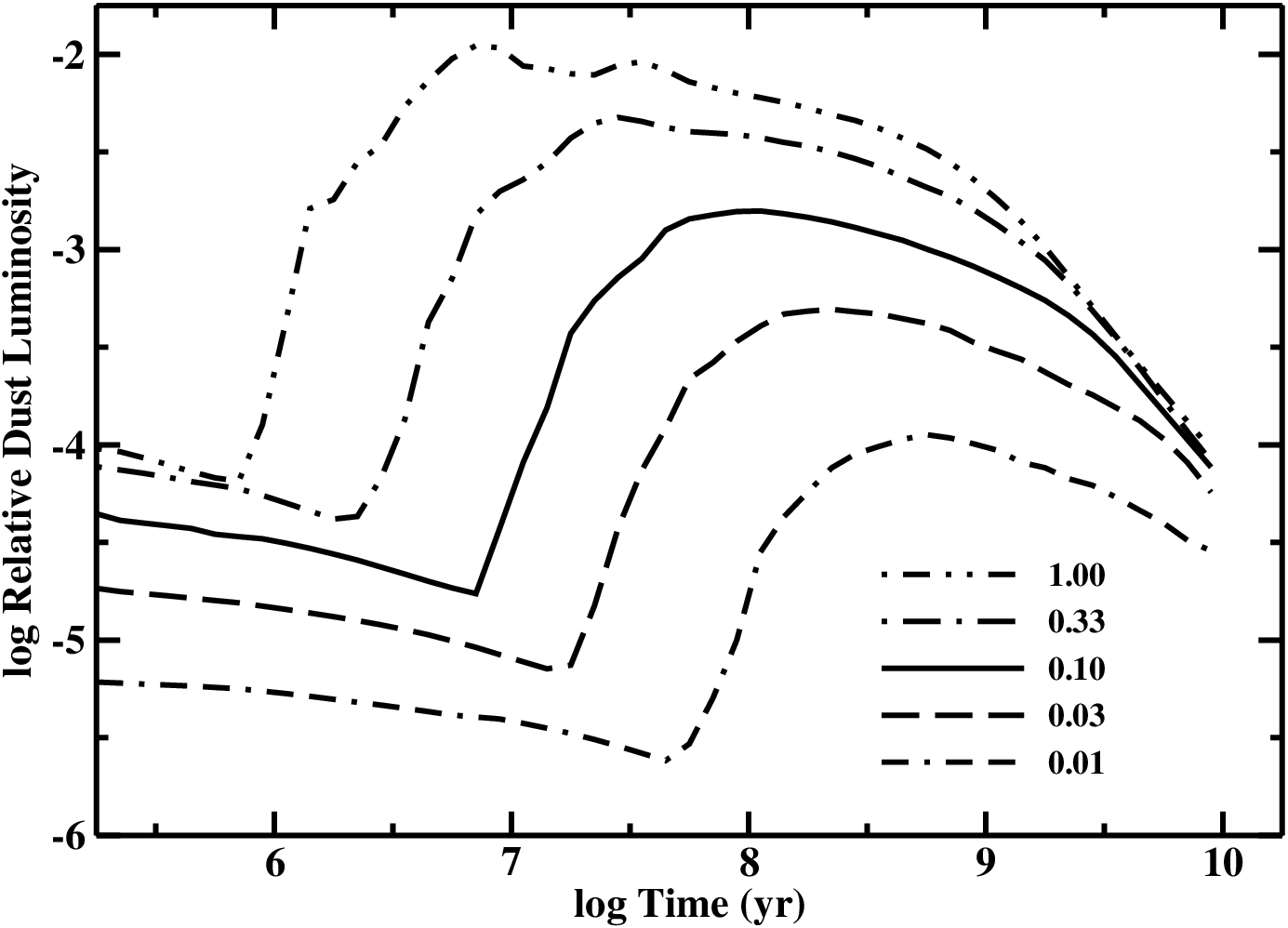}
\vskip 3ex
\caption{
Time evolution of the median dust luminosity relative 
to the central star ($L_d / L_\star$) for disks with initial 
dust surface density $\Sigma_d = 30 ~ x_m ~ a^{-1}$ g cm$^{-2}$
surrounding a 1 \msun\ star. The legend indicates the
scaling factor $x_m$ for each curve. More massive disks
achieve larger peak dust luminosity earlier than less
massive disks. The largest dust luminosities are
comparable to the dust luminosity of the most luminous
debris disks associated with solar-type stars.
\label{fig: ldust1-1msun}
}
\end{figure}

Figure \ref{fig: ldust1-1msun} shows the evolution of \ldlstar\ for disks with an
initial surface density
$\Sigma \propto a^{-1}$ and the $f_S$ fragmentation parameters. Early in the evolution, 
growth produces larger objects and little debris. Larger objects have a smaller surface
area per unit mass. The opacity and \ldlstar\ decline with time. Less massive disks have 
smaller dust masses and smaller \ldlstar. Planetesimals in less massive disks also grow 
more slowly (\S3). The decline of \ldlstar\ with time lasts longest in the lowest mass disks.

As oligarchic growth begins, \ldlstar\ increases by 1.5--2 orders of magnitude in 
10--500~Myr. Disks reach peak luminosities early in the oligarchic growth phase.
More massive disks reach larger peak dust luminosities earlier than less massive disks. 
Throughout oligarchic growth, debris production slowly declines with time.  All disks 
converge to roughly the same dust luminosity, \ldlstar\ $\sim 3-5 \times 10^{-5}$. 

\begin{figure} 
\includegraphics[width=6.5in]{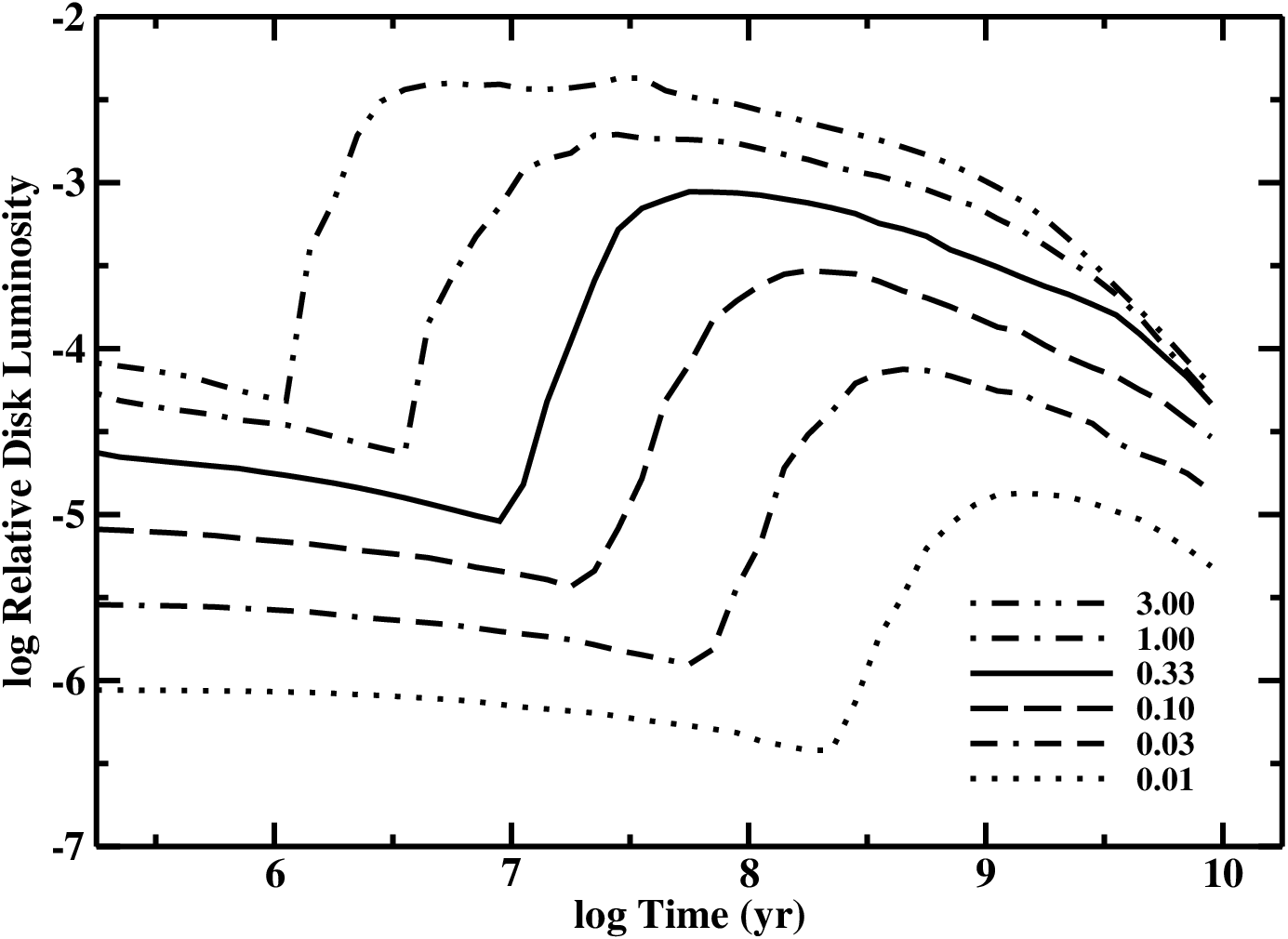}
\vskip 3ex
\caption{
As in Figure \ref{fig: ldust1-1msun} for disks with
$\Sigma_d = 30 ~ x_m ~ a^{-3/2}$ g cm$^{-2}$ and the weak fragmentation
parameters.
\label{fig: ldust2-1msun}
}
\end{figure}

Figure  \ref{fig: ldust2-1msun} repeats Figure \ref{fig: ldust1-1msun} for disks with 
initial $\Sigma \propto a^{-3/2}$ and the $f_W$ fragmentation parameters. These disks
follow the same evolution as in Figure \ref{fig: ldust1-1msun}. Throughout runaway 
growth, the dust luminosity slowly declines. During the transition from runaway
to oligarchic growth, the dust luminosity grows rapidly. After the disk reaches
peak luminosity, collisions produce less debris; the luminosity slowly fades with time. 
At late times, the most massive disks converge on the same \ldlstar.  The lowest mass 
disks evolve more slowly and have a broad range in \ldlstar\ at $t$ = 10 Gyr.

Combined with results for $\Sigma \propto a^{-3/2}$ and the $f_S$ fragmentation parameters
from KB08, Figures \ref{fig: ldust1-1msun} and \ref{fig: ldust2-1msun} show that debris 
disk evolution is remarkably independent of initial conditions. For the three sets of
calculations, we derive simple relations between the maximum dust luminosity and the
initial disk mass:
\begin{equation}
L_{d,max} / L_\star = \left\{ \begin{array}{rll}
     10^{-2} ~ x_m, &  &\Sigma = 30 ~ x_m ~ a^{-1}, ~ f_i = f_S \\
\\
     2 \times 10^{-3} ~ x_m , & \hspace{5mm} &\Sigma = 30 ~ x_m ~ a^{-3/2}, ~ f_i = f_S  \\
\\
     2 \times 10^{-3} ~ x_m , &  &\Sigma = 30 ~ x_m ~ a^{-3/2}, ~ f_i = f_W 
   \end{array} \right.
\label{eq: ldisk-1msun}
\end{equation}
The maximum dust luminosity is fairly independent of the surface density law. In our calculations, 
material in the inner disk at 30~AU produces the largest dust luminosity.  In disks with identical 
$\Sigma$ at 30~AU, disks with $\Sigma \propto a^{-3/2}$ are only 10\% more luminous than disks with
$\Sigma \propto a^{-1}$. 

Throughout the cascade, \ldlstar\ depends on the evolution of the vertical scale height $H$ and the
radial optical depth $\tau$ of the dust. When $\tau \lesssim$ 0.25, the emission depends only on the 
total dust mass. Low mass disks are optically thin; cascades with weaker planetesimals produce more 
dust. Thus, low mass, optically thin debris disks composed of weak planetesimals are more luminous 
than low mass, optically thin debris disks composed of strong planetesimals. When $\tau \gtrsim$ 0.25, 
dust emission depends on $H$ and $\tau$, \ldlstar\ $ \approx f H /a$, where $f = (1 - e^{-\tau})$. The 
vertical scale height $H$ is proportional to the escape velocity of the largest objects in an annulus.  
Disks with larger $H$ intercept more flux from the central star.  Disks with strong planetesimals 
produce more massive oligarchs; thus, $H_{strong} / H_{weak} \gtrsim$ 1.  Although 
$\tau_{strong} / \tau_{weak} \lesssim$ 1 and $f_{strong} / f_{weak} \lesssim$ 1, 
$(fH)_{strong} > (fH)_{weak}$. Thus, massive, optically thick disks composed of strong planetesimals 
are brighter than massive, optically thick disks composed of weak planetesimals.

The properties of the planetesimals are important at several phases of the evolution. During runaway 
growth, all disks are optically thin; disks with weaker planetesimals are brighter. Close to peak 
luminosity, $\tau \approx x_m$, with $\tau_{strong} / \tau_{weak} \approx$ 0.5--0.75; low mass (massive)
disks with weaker (stronger) planetesimals are brighter. Because the optical depth is roughly constant
with time after peak luminosity, these differences persist for much of the evolution.

The time $t_{d,max}$ of maximum dust luminosity is much more sensitive to the input parameters. 
For our results, we derive
\begin{equation}
t_{d,max} ~ ({\rm Myr}) = \left\{ \begin{array}{rll}
     7.5 ~ x_m^{-1}, &  &\Sigma = 30 ~ x_m ~ a^{-1}, ~ f_i = f_S \\
\\
     50 ~ x_m^{-1} , & \hspace{5mm} &\Sigma = 30 ~ x_m ~ a^{-3/2}, ~ f_i = f_S  \\
\\
     40 ~ x_m^{-1} , &  &\Sigma = 30 ~ x_m ~ a^{-3/2}, ~ f_i = f_W 
   \end{array} \right.
\label{eq: tldmax-1msun}
\end{equation}
Because weaker planetesimals fragment earlier in oligarchic growth, the dust luminosity 
peaks earlier. The difference in the time scale, 50~Myr vs. 40~Myr, is comparable to the
difference in the time scale of maximum dust production (equation (\ref{eq: tmdot})). In 
both cases, the maximum in
\ldlstar\ occurs a factor of $\sim$ 3.25--3.5 later than the maximum in the dust production
rate.  Although disks with $\Sigma \propto a^{-3/2}$ are 10\% more luminous than equal mass
disks with $\Sigma \propto a^{-1}$ (equation (\ref{eq: ldisk-1msun})), they reach maximum
luminosity 15\%--20\% earlier. In term of total energy emitted by dust, these two features 
of the evolution approximately cancel. Thus, the total emitted energy is roughly independent 
of the gradient of the surface density distribution.

\begin{figure} 
\includegraphics[width=6.5in]{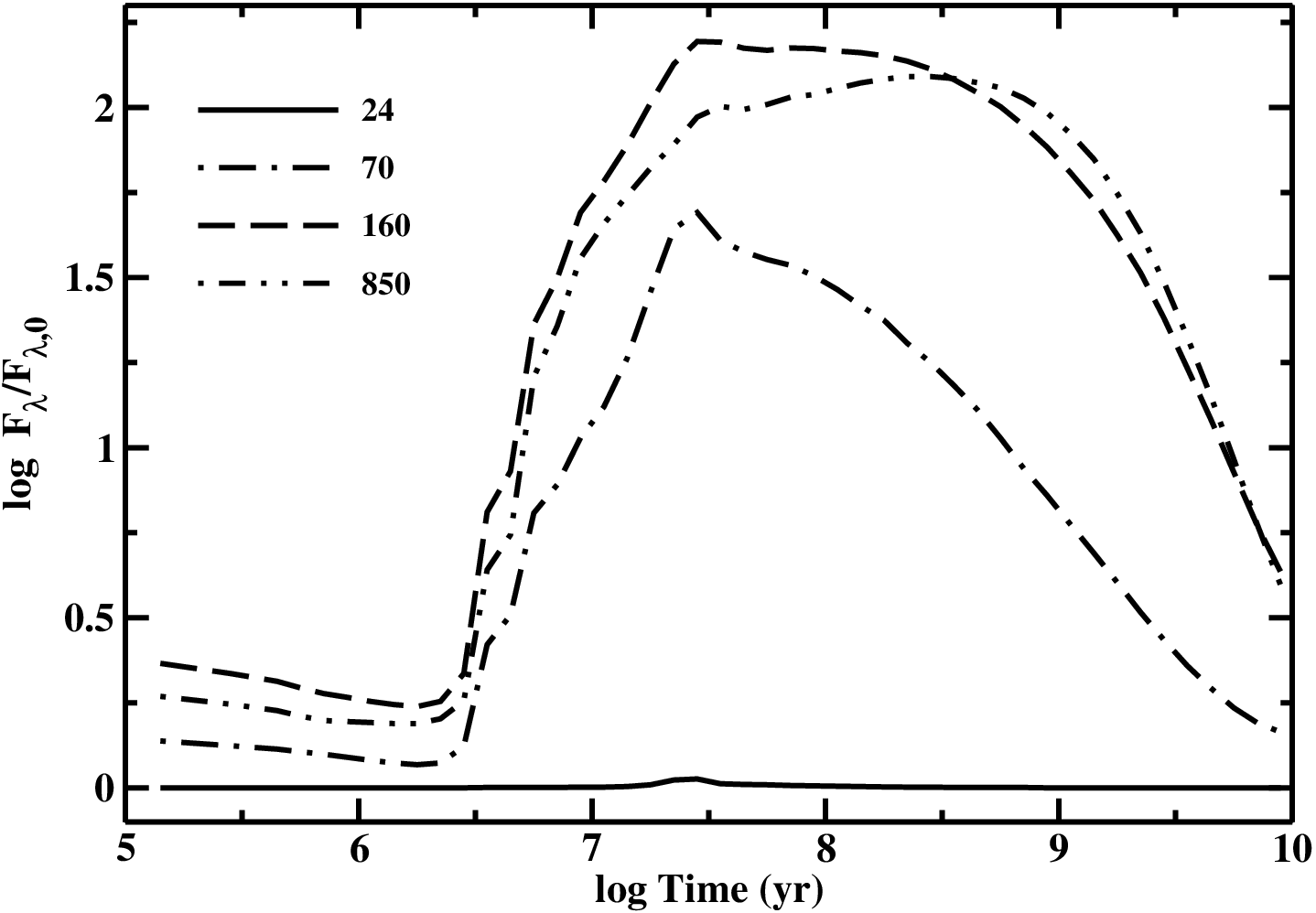}
\vskip 3ex
\caption{
Time evolution of the infrared excess at 24 $\mu$m, 70 $\mu$m,
160 $\mu$m, and 850 $\mu$m for a disk with 
$\Sigma_d = 10 ~ a^{-1}$ g cm$^{-2}$ surrounding a 1 \msun\ star.
For a 1 \msun\ central star,
dust at 30--150~AU is too cold to produce a 24 $\mu$m excess.
At longer wavelengths, dust produces large excesses; the
excess peaks at later times for longer wavelengths.
\label{fig: irall-1msun}
}
\end{figure}

\begin{figure} 
\includegraphics[width=6.5in]{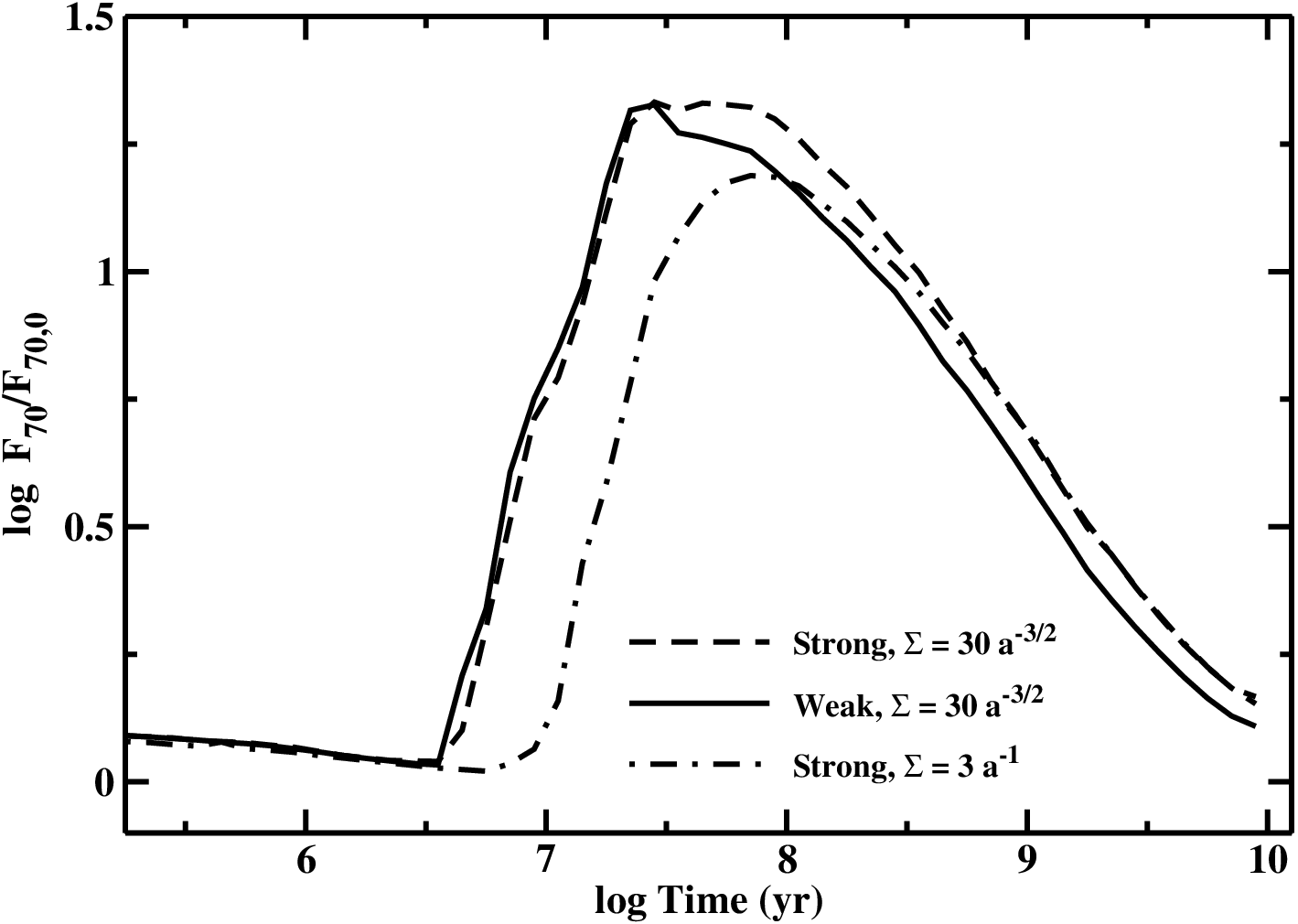}
\vskip 3ex
\caption{
Time evolution of the 70 $\mu$m excess for disks with similar
total dust masses around a solar-type star. In disks with
$\Sigma \propto a^{-3/2}$, weak planetesimals produce slightly
larger IR excesses at early times and much weaker IR excesses 
at late times. Disks with $\Sigma \propto a^{-1}$ produce 
relatively more dust emission at late times than disks with 
$\Sigma \propto a^{-3/2}$.
\label{fig: f70sigma-1msun}
}
\end{figure}

Figure \ref{fig: irall-1msun} shows the time evolution of the median 24--850 $\mu$m 
excesses for disks with initial $\Sigma \propto a^{-1}$ and $x_m$ = 1/3.  For 1~\msun\ stars, 
disks at 30--150~AU are rarely hot enough to produce observable 24 $\mu$m emission.
Aside from small, 1--3\% excesses at 20--50~Myr for the most massive disks, the observed
$F_{24} / F_{24,0}$ is always 1. At longer wavelengths, the excesses approximately track 
the evolution of \ldlstar\ in Figure \ref{fig: ldust1-1msun}. Hotter disk material close 
to the central star produces the rapid rise at 70--160 $\mu$m in Figure 
\ref{fig: f70sigma-1msun}. Because most of the 70 $\mu$m emission is produced by grains in 
the inner disk (30--60~AU), the 70 $\mu$m excess drops rapidly after the disk reaches peak
dust luminosity. At 50--100~AU, cooler material in the middle part of the disk emits
most of the 160 $\mu$m excess; this excess declines more slowly. Dust in the outer disk
at 100--150~AU produces most of the 850 $\mu$m excess; this emission rises slowly and
peaks at 300~Myr to 1~Gyr.  Once the collisional cascade reaches the outer disk at
$t \sim$ 1~Gyr, the excess emission declines rapidly at all wavelengths.  

Disks with initial $\Sigma \propto a^{-3/2}$ follow the same trends shown in 
Figure~\ref{fig: irall-1msun}. At 850 $\mu$m, the magnitude of the excess scales with
the total disk mass. Disks with similar total masses yield similar excesses at 850 $\mu$m 
throughout the main sequence lifetime of the central star. For 
$\Sigma \propto x_m ~ a^{-n}$, disks with ($n$, $x_m$) = (1, $x$) and (3/2, 9$x$) have 
similar masses ($x$ = 0.01--1; Table \ref{tab:massgrid}) and similar evolution at 850 $\mu$m.  
At shorter wavelengths, the peak excess and time of maximum excess depend on $\Sigma (a)$.
Disks with steeper surface density laws have more mass closer to the star. For disks with 
similar masses, calculations with $n$ = 3/2 reach oligarchic growth faster and produce larger 
70--160 $\mu$m excesses than the calculations with $n$ = 1 shown in Figure~\ref{fig: irall-1msun}. 

Figure \ref{fig: f70sigma-1msun} illustrates how IR excesses depend on the initial surface 
density law and the fragmentation parameters for disks with similar total masses. For disks 
with $\Sigma \propto a^{-3/2}$, the onset of oligarchic growth is fairly independent of 
the fragmentation parameters.  Calculations with weaker planetesimals reach the collisional
cascade somewhat sooner; the 70 $\mu$m excess rises a little faster for calculations with 
weaker planetesimals (solid line) than for calculations with stronger planetesimals (dashed line). 
In disks with $\Sigma \propto a^{-1}$, the smaller mass at 30--60~AU leads to a slower rise at
70 $\mu$m compared to disks with $\Sigma \propto a^{-3/2}$ (dot-dashed line). At late times,
the entire disk produces 70 $\mu$m emission. This emission depends on the disk mass and is
independent of the radial surface density profile. 

As the cascade proceeds, the IR flux depends on the evolution of $\tau$.  In low mass disks, 
the 70 $\mu$m flux tracks the disk mass. Thus, low mass disks with weak planetesimals are
brighter at 70 $\mu$m than low mass disks with strong planetesimals. In more massive disks, 
the dust is optically thick with $\tau \approx$ 1. Oligarchs are roughly a factor of four more 
massive in disks with strong planetesimals; thus, $H_{strong} / H_{weak} \approx$ 2.  Cascades 
with weaker planetesimals produce roughly 50\% more dust; $\tau_{strong} / \tau_{weak} \approx$
1/3. Combining the two results for disks with strong and weak planetesimals, the peak IR flux is 
roughly $(1 - e^{-\tau_{strong} / \tau_{weak}}) H_{strong} / H_{weak}$ $\approx$ 33\% larger in 
calculations with strong planetesimals. In massive disks, this difference remains for long 
periods after peak 70 $\mu$m emission. Despite the larger dust production, the 70 $\mu$m excess 
at late times is smaller when planetesimals are weak (compare the solid and dashed lines for 
log $t >$ 7.5 in Figure \ref{fig: f70sigma-1msun}).

At late times, the IR excess is independent of the surface density law. Disks with similar
total masses and similar strength planetesimals produce oligarchs with the same maximum size
(\S3.1). These disks also produce similar masses of dust. Because $\tau$ and $H$ are similar,
these disks produce similar IR excesses at late times (compare the dashed and dot-dashed lines 
for log $t >$ 8.5 in Figure \ref{fig: f70sigma-1msun}).  

\begin{figure} 
\includegraphics[width=6.5in]{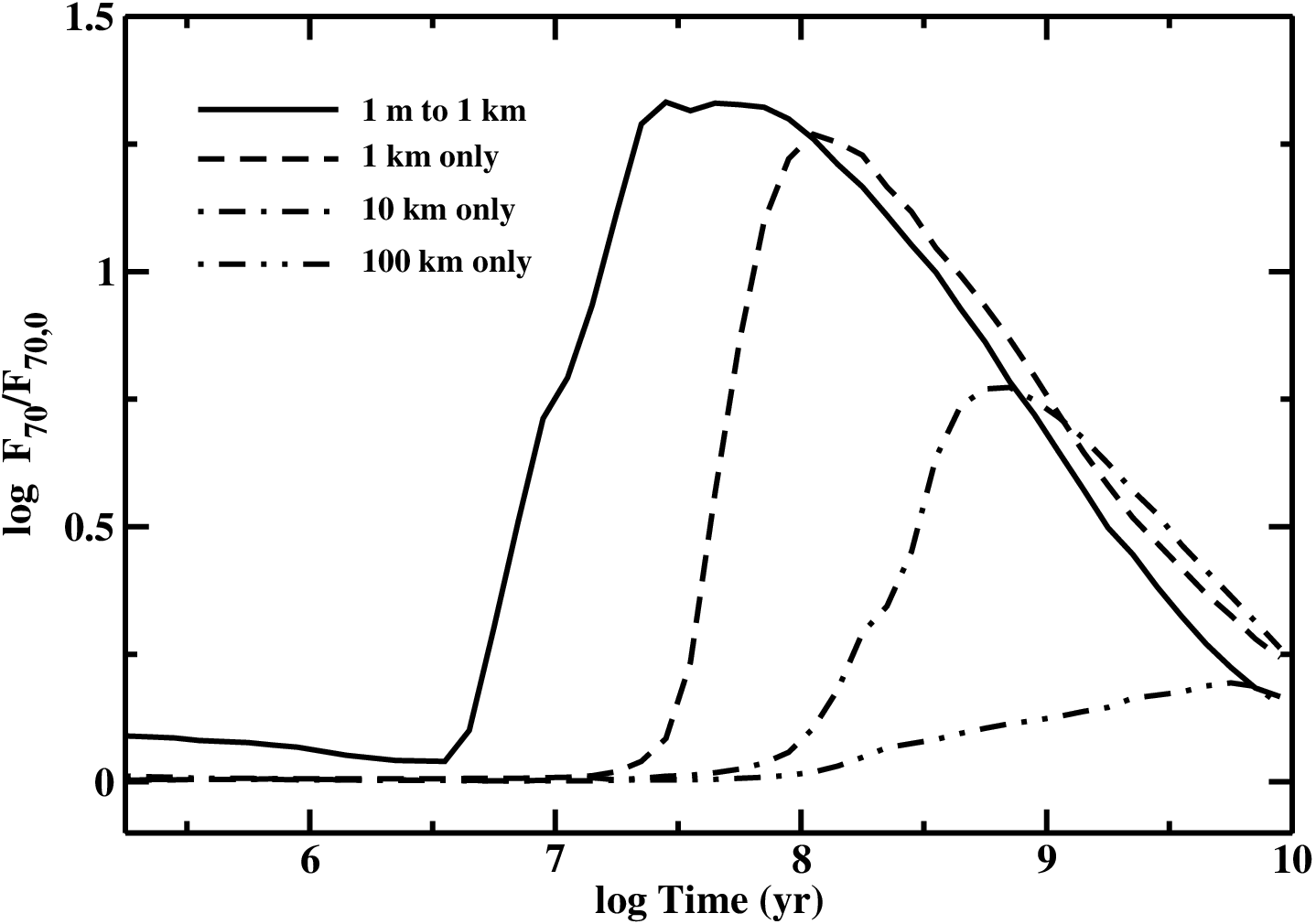}
\vskip 3ex
\caption[f22.eps]
{
Time evolution of the 70 $\mu$m excess for disks with a range
in the initial radius $r_0$ of the largest planetesimal. The 
legend indicates $r_0$. Disks with larger planetesimals produce
smaller peak IR excesses at later times than disks with smaller
planetesimals.
\label{fig: f70r0-1msun}
}
\end{figure}

Figure \ref{fig: f70r0-1msun} demonstrates that the initial sizes of planetesimals have a
larger impact on the IR excess than the fragmentation parameters. In an ensemble of 1~km
planetesimals, runaway and oligarchic growth are delayed by a factor of 10--20 relative to 
growth in an ensemble of 1~m to 1~km planetesimals (\S3.1). This delay in oligarchic growth 
produces a much later rise in the 70 $\mu$m excess, at $\sim$ 30~Myr instead of $\sim$ 3~Myr.
After the delay, the peak dust production rates differ by a factor of three.  Thus, 
calculations with 1~km planetesimals produce smaller IR excesses at later times than 
calculations with ensembles of 1~m to 1~km planetesimals.

In calculations with 10~km or 100~km planetesimals, this behavior is accentuated. Larger
planetesimals take longer to reach oligarchic growth. The rise in the IR excess occurs later 
and later for larger and larger planetesimals. Calculations with larger planetesimals also 
produce less debris at the peak of the collisional cascade. The peak IR excess declines as the 
initial planetesimal size grows. 

At late times, calculations with one initial planetesimal size produce larger objects. 
Because the vertical scale height is proportional to the escape velocity of the largest
oligarch, disks with larger planetesimals have larger $H$ at late times. For our derived
range in maximum oligarch mass (a factor of $\sim$ 8), we expect a factor of 2.8 range
in the maximum $H$.  In our disks, the range of dust masses at late times 
(Figure \ref{fig: mdust3-1msun}) leads to small reductions in the IR excess for calculations 
with 1~km planetesimals relative to our standard calculation with 1~m to 1~km planetesimals. 
The small dust masses of calculations with 10~km (100~km) planetesimals
leads to factor of three (ten) smaller excesses throughout the evolution.

\subsection{Evolution for 1.5--3 \msun\ Stars}

Three factors modify the evolution of the dust luminosity and the IR/submm excesses for 
1.5--3~\msun\ stars. The luminosities of 1--10~Myr old stars more massive than the Sun 
grow monotonically with time. Independent of the evolution of the collisional cascade,
this pre-main sequence evolution produces modest increases in the excess emission during 
slow and runaway growth. More massive stars are also more luminous than the Sun. 
Throughout the collisional cascade, dust grains around massive stars are warmer and 
emit more short wavelength radiation than grains around less massive stars.  More massive 
stars also evolve faster than the Sun. Because the collision timescale in the disk is
less sensitive to stellar mass than the evolutionary timescale of the central star,
massive stars have more dust at the end of their lifetimes than lower mass stars
(Figs. \ref{fig: mdust1-allm}--\ref{fig: mdust2-allm}). Thus, old, massive stars 
have relatively larger IR excesses than old, low mass stars.

\begin{figure}
\includegraphics[width=6.5in]{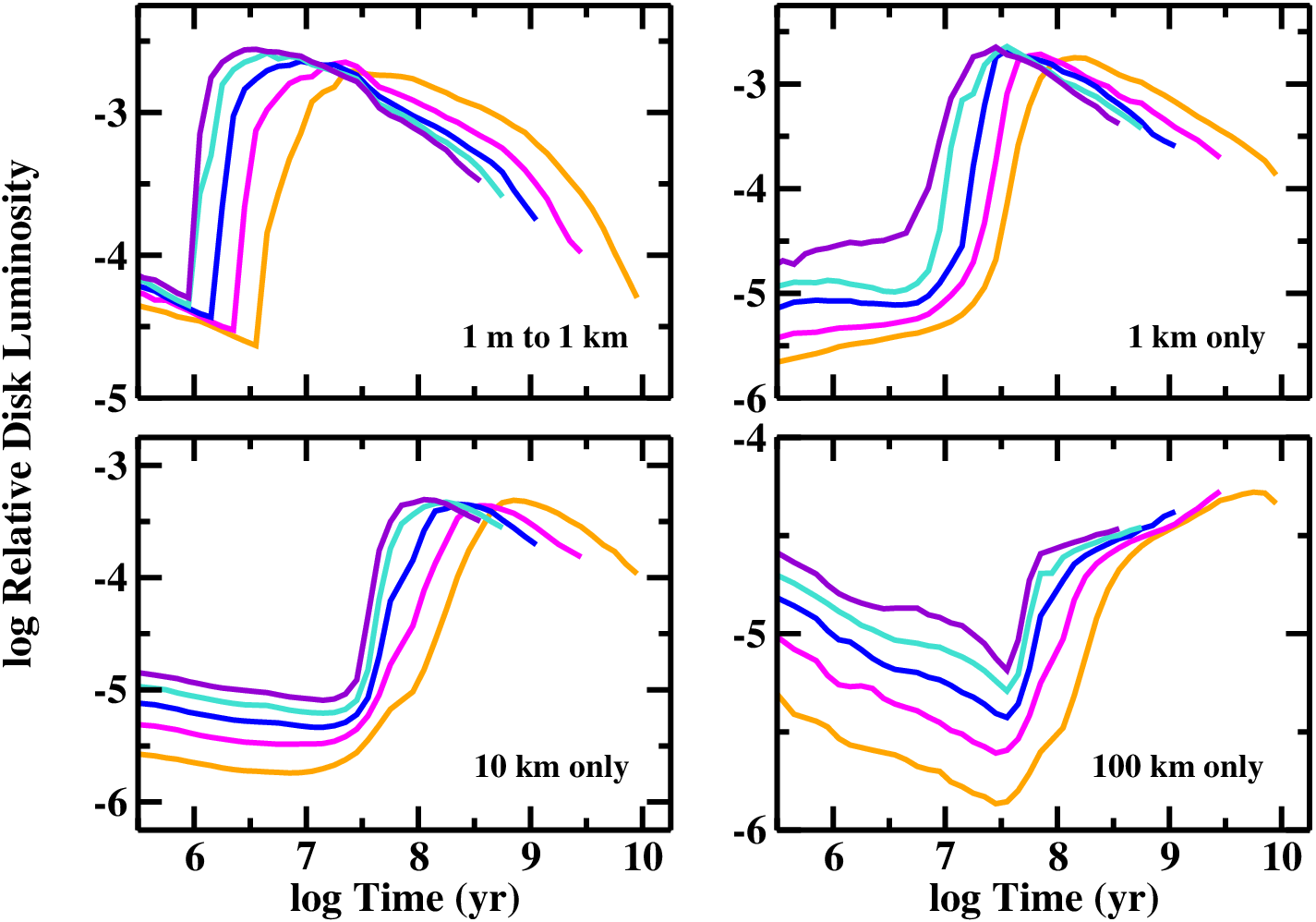}
\vskip 3ex
\caption[f23.eps]{
Time evolution of the disk luminosity for disks with 
$\Sigma_d = 30 ~ (M_\star / M_\odot) ~ a^{-3/2}$ g cm$^{-2}$
around 1 \msun\ (orange), 1.5 \msun\ (magenta), 2 \msun\ (blue),
2.5 \msun\ (turquoise), and 3 \msun\ (indigo) stars. The panels 
show the evolution for different initial size distributions of 
planetesimals.  The vertical scale changes in each panel.
{\it Upper left panel:} planetesimals have a range of initial
sizes between 1 m and 1~km;
{\it Upper right panel:} all planetesimals have $r_0$ = 1~km;
{\it Lower left panel:} all planetesimals have $r_0$ = 10~km;
{\it Lower right panel:} all planetesimals have $r_0$ = 100~km.
}
\label{fig:ldisk-all}
\end{figure}

To illustrate these points, we begin with the time evolution of \ldlstar\ for disks
with $\Sigma \propto a^{-3/2}$ around 1--3~\msun\ stars (Figure \ref{fig:ldisk-all}).
During runaway growth, the vertical scale height and the dust mass decline. The dust 
luminosity drops with time. When oligarchic growth and the collisional cascade begin, 
the dust luminosity increases dramatically. Oligarchs grow faster in more massive disks;  
the dust luminosity rises earlier in disks around more massive stars. Once the disks 
reach peak luminosity, the decline roughly follows a power law 
\ldlstar\ $\propto t^{-n}$ with $n$ = 0.6--0.8 (see also KB08).

The upper left panel of Figure \ref{fig:ldisk-all} shows that debris disks around more 
massive stars have somewhat larger peak luminosities than disks around lower mass stars.  
For disks with $\Sigma \propto a^{-3/2}$, the peak luminosity grows weakly with stellar mass, 
$L_{d,max}/L_\star \propto M_\star^{1/3}$. Once stars reach ages exceeding 100~Myr, debris
disks around less massive stars are more luminous than their coeval counterparts among more
massive stars. Among stars in a bound cluster, lower mass stars should have the most 
luminous debris disks. Once stars reach the end of their main sequence lifetimes, the dust
luminosity is more sensitive to stellar mass, with $L_d / L_\star (t = t_{ms}) \propto M_\star^{3/2}$.
Among stars with a range of ages in a fixed volume of the Galaxy, most main sequence stars
are near the main sequence turn-off. More masive stars should have the most luminous debris 
disks in this sample.

The remaining panels in the Figure demonstrate that disks with larger planetesimals produce 
less luminous debris disks.  Compared to disks composed of 1~m to 1~km planetesimals, disks 
with 1~km planetesimals have lower luminosities during runaway growth and comparable 
luminosities at the peak and throughout the decline (upper right panel).  However, the peak 
luminosity is a factor of 5--10 later in time.  As the initial size of the planetesimals 
increases, peak luminosities are smaller and occur much later in time. For 10~km planetesimals 
(lower left panel), the maximum values for \ldlstar\ are a factor of $\sim$ 4 smaller. Debris 
disks produced from ensembles of 100~km planetesimals (lower right panel) are a factor of 
10--50 fainter. Most of these disks cannot reach their peak luminosity before the central 
star evolves off the main sequence.

\begin{figure} 
\includegraphics[width=6.5in]{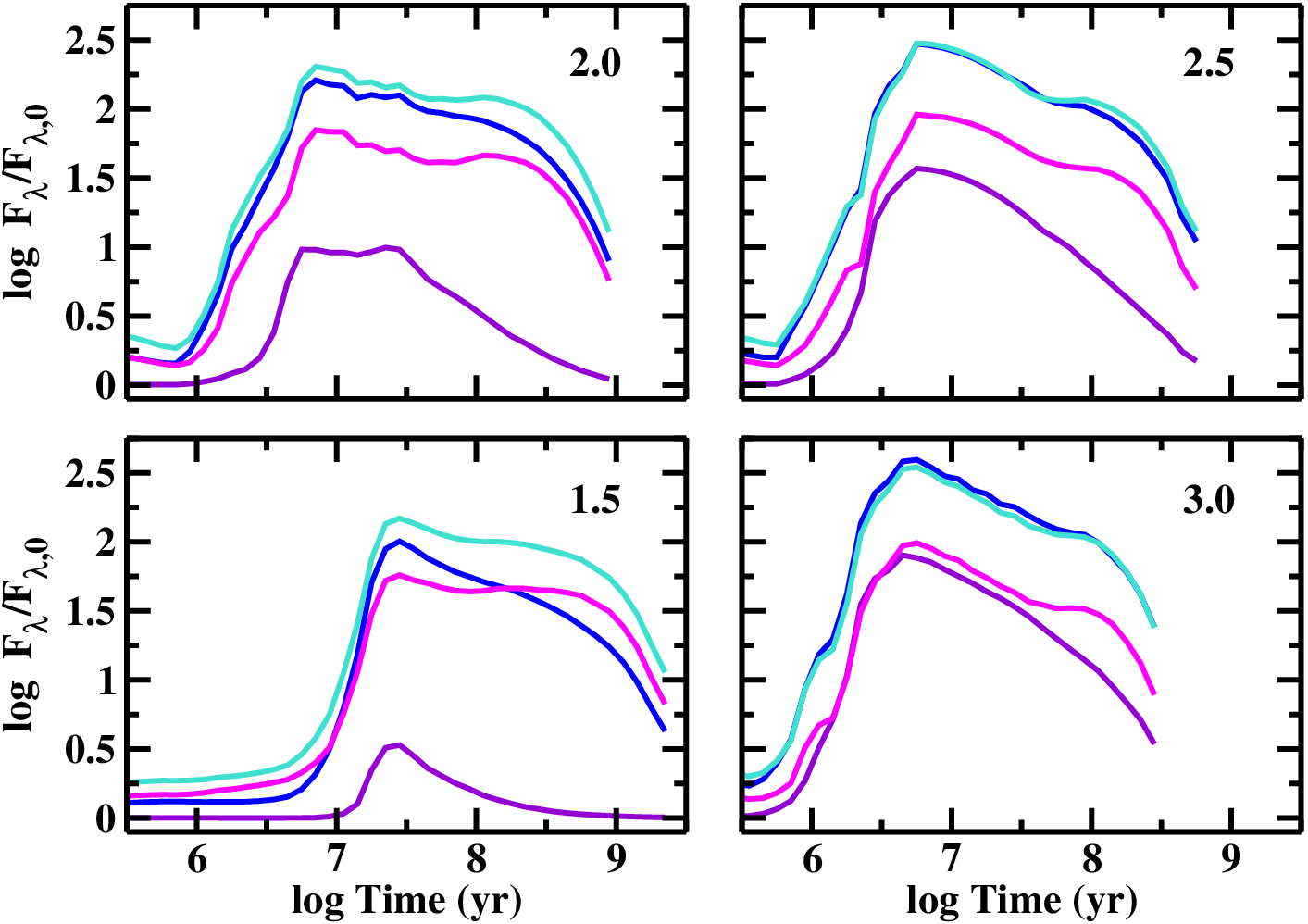}
\vskip 3ex
\caption[f24.eps]{
Time evolution of the infrared excess at 
24 $\mu$m (indigo curves), 70 $\mu$m (blue),
160 $\mu$m (turquoise), and 850 $\mu$m (magenta)
for disks with 1~m to 1~km planetesimals and 
$\Sigma_d = 10 ~ (M_\star / M_\odot) ~ a^{-1}$ g cm$^{-2}$ around 
1.5--3 \msun\ stars.  The number in the upper right corner 
of each panel indicates the stellar mass in \msun.  At 24 $\mu$m, 
more massive stars produce larger IR excesses at earlier times 
than lower mass stars. For all stars, the IR excess is largest 
at 70--160 $\mu$m.  
\label{fig: irall-allm}
}
\end{figure}

Figure \ref{fig: irall-allm} shows the evolution of the IR excesses at several wavelengths for 
disks with initial $\Sigma_d = 10 ~ (M_\star / M_\odot) ~ a^{-1}$ g cm$^{-2}$ around 1.5--3~\msun\ stars. 
The time evolution of IR emission follows the time evolution of planet formation.
Planets form more rapidly around more massive stars; IR emission peaks earlier in
disks around more massive stars.  For disks with comparable total masses, more massive, 
more luminous stars have warmer dust grains and larger 24 $\mu$m excesses (indigo curves). 
Nearly all of the 24 $\mu$m emission is produced in the inner disk; this emission 
declines more rapidly than emission at longer wavelengths. For 1.5~\msun\ stars, most of 
the 70 $\mu$m emission also comes from the inner disk and declines more rapidly than 
emission at 160--850 $\mu$m. For more massive stars, 70--850 $\mu$m emission is
produced throughout the disk. With typical disk temperatures of 20--50 K, emission
at 70--160 $\mu$m is brighter than emission at 850 $\mu$m. The 70--160 $\mu$m emission 
grows slowly with stellar mass; emission at 850 $\mu$m is independent of stellar mass.
Finally, more massive stars evolve off the main sequence sooner than lower mass stars. 
More massive stars have larger far-IR excesses at the ends of their main sequence lifetimes 
than lower mass stars.

\begin{figure} 
\includegraphics[width=6.5in]{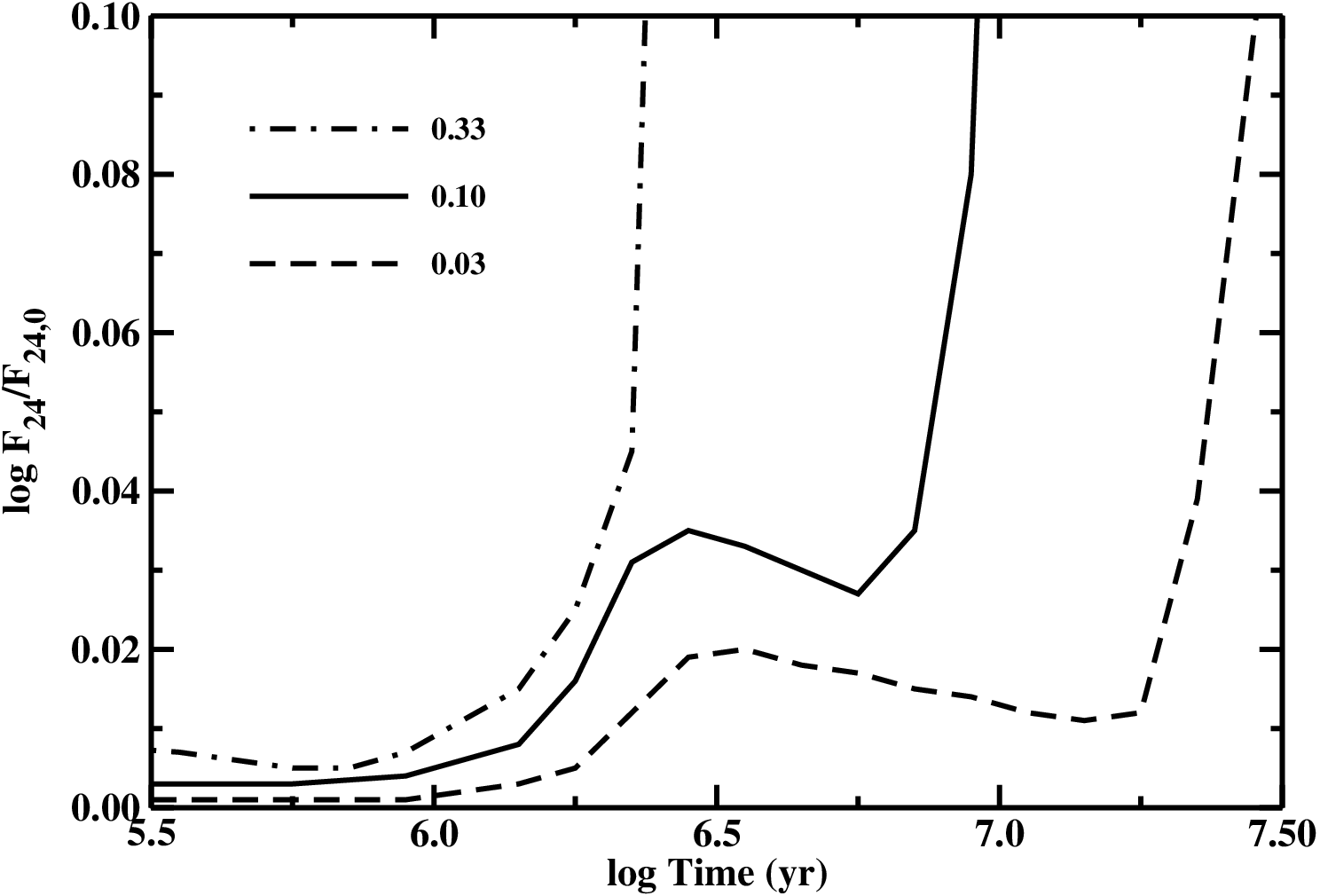}
\vskip 3ex
\caption[f25.eps]{
Time evolution of the 24 $\mu$m excess for disks with initial 
$\Sigma_d \propto a^{-3/2}$ around 3 \msun\ stars. An increasing
pre-main sequence stellar luminosity produces the small rises in 
excess at 3 Myr for disks with $x_m$ = 0.03 and 0.10. Debris 
production from the collisional cascade initiates the large rises
in excess for all disks.
\label{fig: 24m-pms}
}
\end{figure}

To explore the impact of stellar evolution in more detail, Figure \ref{fig: 24m-pms}
shows an expanded
view of the pre-main sequence evolution of the 24 $\mu$m flux for disks with 
$\Sigma \propto a^{-3/2}$ around 3~\msun\ stars. In the Y$^2$ models, the luminosity 
of a 3~\msun\ star increases by a factor of 3--4 during the first 3--4~Myr of the
evolutionary sequence.  Although \ldlstar\ declines by a factor of two as the star
approaches the main sequence, the change in luminosity increases the grain temperature
by 30\% to 40\% throughout the disk.  For low mass stars with $x_m$ = 0.03-0.1, this 
evolution produces modest, 1\%--4\%, increases in the IR excess at 2--4~Myr (solid
and dashed lines in the Figure). Once the star reaches the main sequence, the IR excess 
declines until the collisional cascade initiates a rapid and much larger rise in the 
amount of debris.  For more massive disks (dot-dashed line in the Figure), the large 
rise in emission from debris coincides (and sometimes precedes) the small rise in 
emission from stellar evolution. 

\begin{figure} 
\includegraphics[width=6.5in]{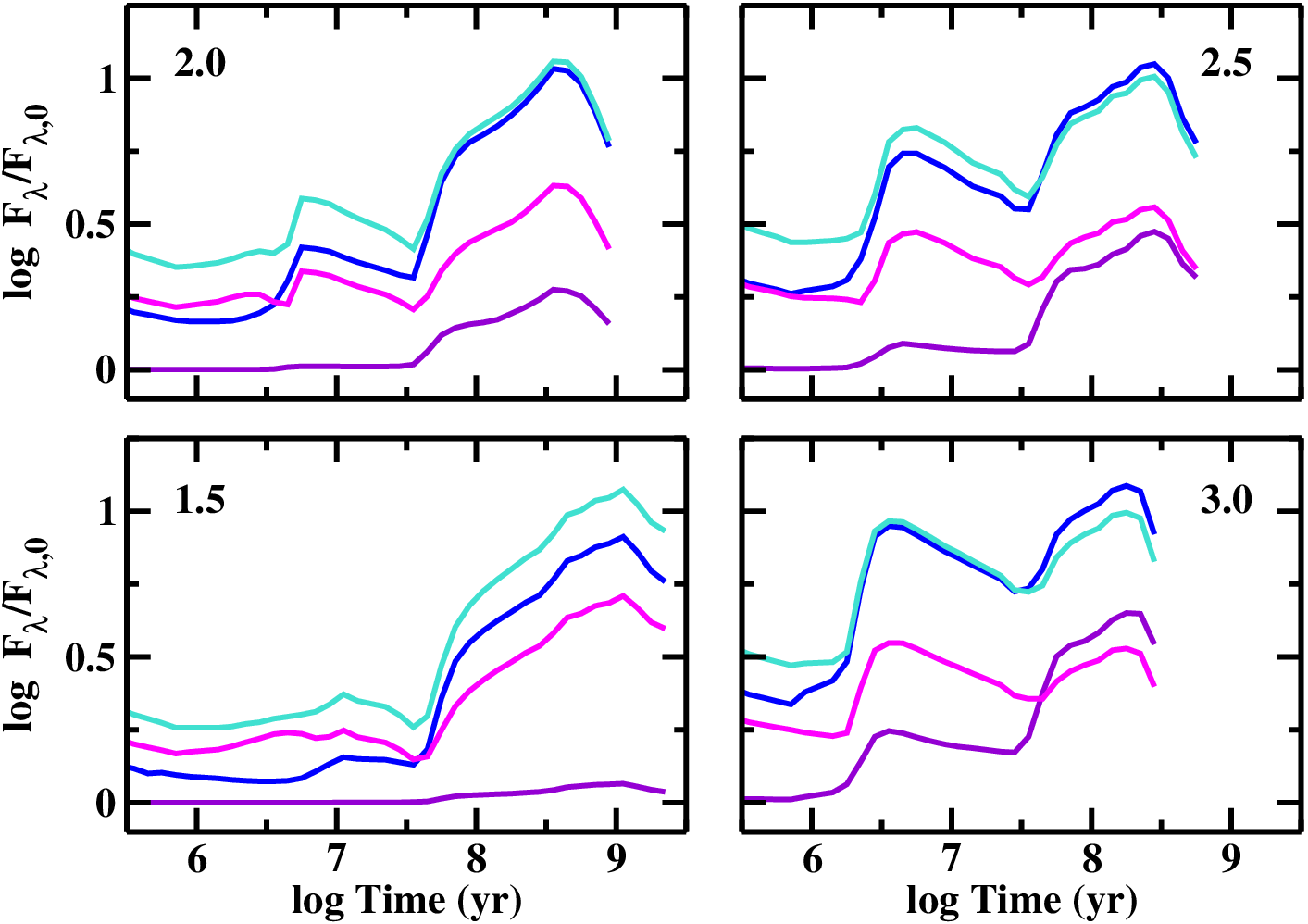}
\vskip 3ex
\caption[f26.eps]{
As in Figure \ref{fig: irall-allm} for disks with $r_0$ = 100~km.
The numbers in the upper left or right corner indicate the
stellar mass in \msun.
Disks with larger planetesimals produce smaller peak IR excesses 
at later times than disks with smaller planetesimals.
\label{fig: irallr0-allm}
}
\end{figure}

As the initial sizes of planetesimals increases, the collisional cascade produces smaller 
infrared excesses. For disks with 100~km planetesimals (Figure \ref{fig: irallr0-allm}),
the time evolution of the debris consists of four distinct phases. During the first 
1--2~Myr, viscous stirring excites planetesimals to large velocities. Collisions 
produce debris and modest IR excesses at all wavelengths. As the central star approaches 
the main sequence, temperatures rise throughout the disk. Infrared excesses rise.
Once the star is on the main sequence, runaway growth leads to a smaller dust 
production rate. The infrared excesses slowly decline. As systems make the transition 
from runaway growth to oligarchic growth, the collisional cascade begins to produce 
copious amounts of dust. The IR excesses then rise considerably. 

For disks with 100~km planetesimals, stellar evolution produces measurable infrared
excesses for 2--3~\msun\ stars. The excess is largest at 70--160 $\mu$m and smallest
at 24 $\mu$m.  The amplitude of the rise in the excess grows with stellar mass 
(Figure \ref{fig: irallr0-allm}). Disks with 10~km and smaller planetesimals produce little
dust at early times. Thus, the rise in the excess from stellar evolution is a few
percent or less in these systems.

In Tables \ref{tab:eqdebris1}--\ref{tab:eqdebris2}, we quantify the general conclusions
from Figures \ref{fig: ldust1-1msun}--\ref{fig: irallr0-allm}. Table \ref{tab:eqdebris1}
lists the scaling relations for debris disks produced from ensembles of 1~m to 1~km
planetesimals. Rows (1) and (2) list the fragmentation parameters and the surface
density law. Rows (3) and (4) list the maximum dust luminosity and the timescale
to reach this luminosity as a function of disk mass and stellar mass. Aside from a
weak dependence of dust luminosity on \mstar\ for weak planetesimals, the maximum
dust luminosity depends only on the initial mass in solids. The time to reach the
peak luminosity scales with disk mass and stellar mass.

The remaining rows of Table \ref{tab:eqdebris1} list the scaling relations for the
maximum excesses at 24--850 $\mu$m. All of the excesses depend on the initial disk
mass. The collisional cascade produces more debris in more massive disks. Thus, more 
massive disks have larger IR excesses. Because wavelengths at 24--70 $\mu$m are 
usually on the Wien
side of the disk spectral energy distribution, excesses at these wavelengths are more 
sensitive to the stellar luminosity. At longer wavelengths, the stellar luminosity
plays a more modest role in the magnitude of the IR excess.

Table \ref{tab:eqdebris2} lists scaling relations for debris disks produced from
ensembles of 1~km (column (2)), 10~km (column (3)), and 100~km (column(4)) 
planetesimals. For calculations with 1~km planetesimals, the properties of the
debris are similar to those derived from ensembles of 1~m to 1~km planetesimals.
However, the time to reach maximum debris disk fluxes is a factor of 5 longer.
As the maximum initial size of planetesimals grows, emission from the debris decreases
and occurs at later times. At 24--850 $\mu$m, the peak IR excess decreases by
factors of 10--50 as the initial size of planetesimals grows from 1~km to 100~km.

\begin{figure} 
\includegraphics[width=6.5in]{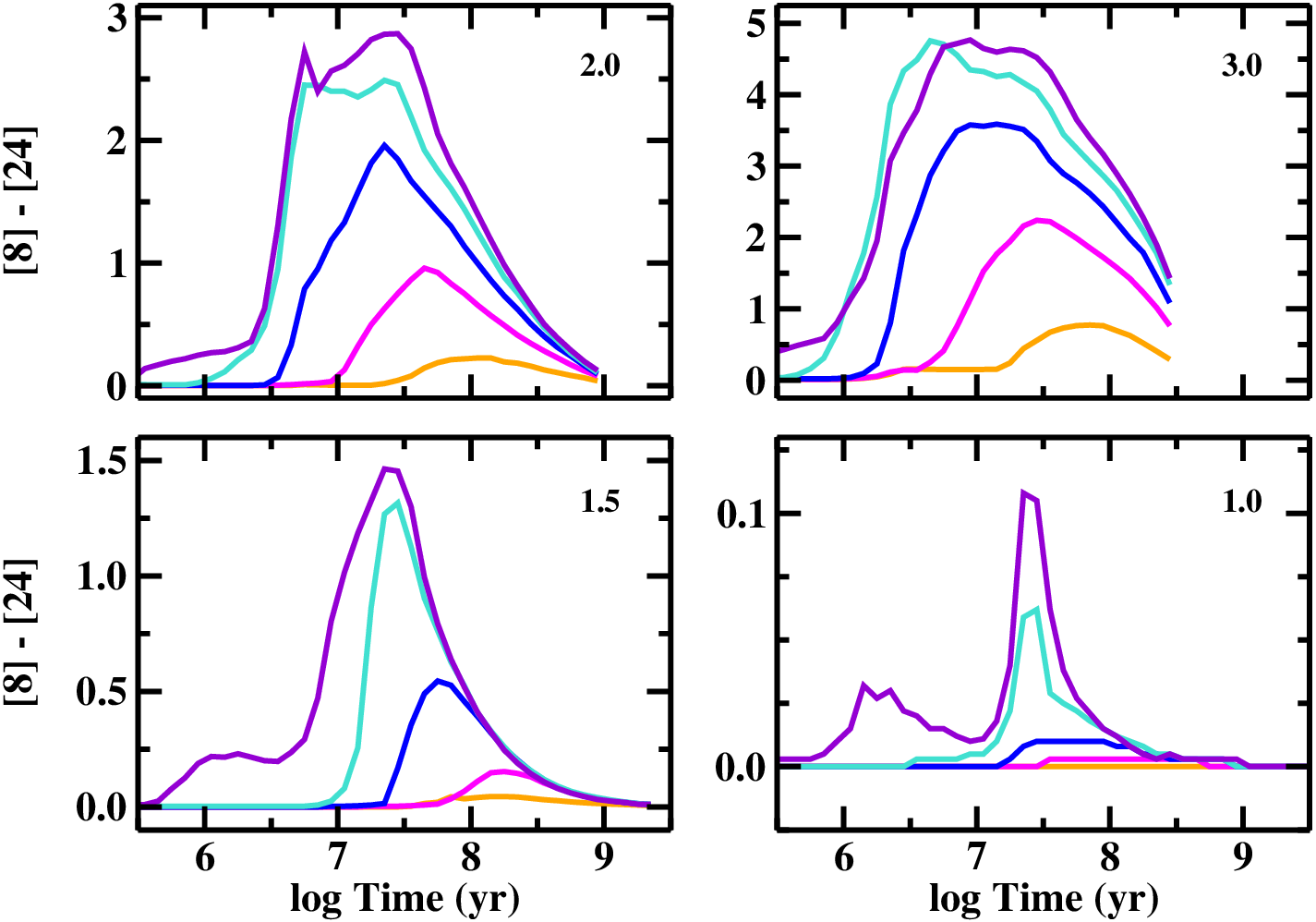}
\vskip 3ex
\caption[f27.eps]{
Time evolution of the median [8]--[24] color index for disks with initial
$\Sigma_d = 30~x_m~(M_\star / M_\odot) a^{-1}$ g cm$^{-2}$ around 1--3 \msun\ stars. 
The number in the upper right corner of each panel indicates the stellar
mass in \msun. Disks with $x_m$ = 0.01 (orange curves) and 0.03 (magenta)
produce small color excesses at late times. 
Disks with $x_m$ = 0.1 (blue), $x_m$ = 0.33 (turquoise), and $x_m$ = 1 
produce larger excesses at earlier times. Pre-main sequence stellar 
evolution is responsible for the modest increase in color at 1--10 Myr 
for 1--1.5~\msun\ stars.
\label{fig: c824-allm}
}
\end{figure}

To conclude this section, we examine several aspects of the color evolution of 
debris disks. Color indices provide a useful way to separate distinct phases of
planet growth without the need to measure the expected flux from the underlying
stellar photosphere. Figure \ref{fig: c824-allm} shows the evolution of the
[8]--[24] color as a function of stellar mass and initial disk mass. At early 
times (1--10~Myr), evolution to the main sequence dominates the color evolution
of low mass disks. As they approach the main sequence, low (high) mass stars 
become less (more) luminous. Disks become cooler (hotter); colors become redder
(bluer). When the collisional cascade begins, the entire disk produces dust. This 
dust is cooler than the dust from the inner disk which produces emission at the
start of the evolution. Thus, the colors become significantly redder. The maximum
color index scales with the initial disk mass. More massive disks produce more
dust at larger distances from the central star. More distant dust is cooler.
Thus, more massive disks have redder color indices.

\subsection{Highlights of Debris Disk Evolution for 1--3 \msun\ Stars}

\begin{figure} 
\includegraphics[width=6.5in]{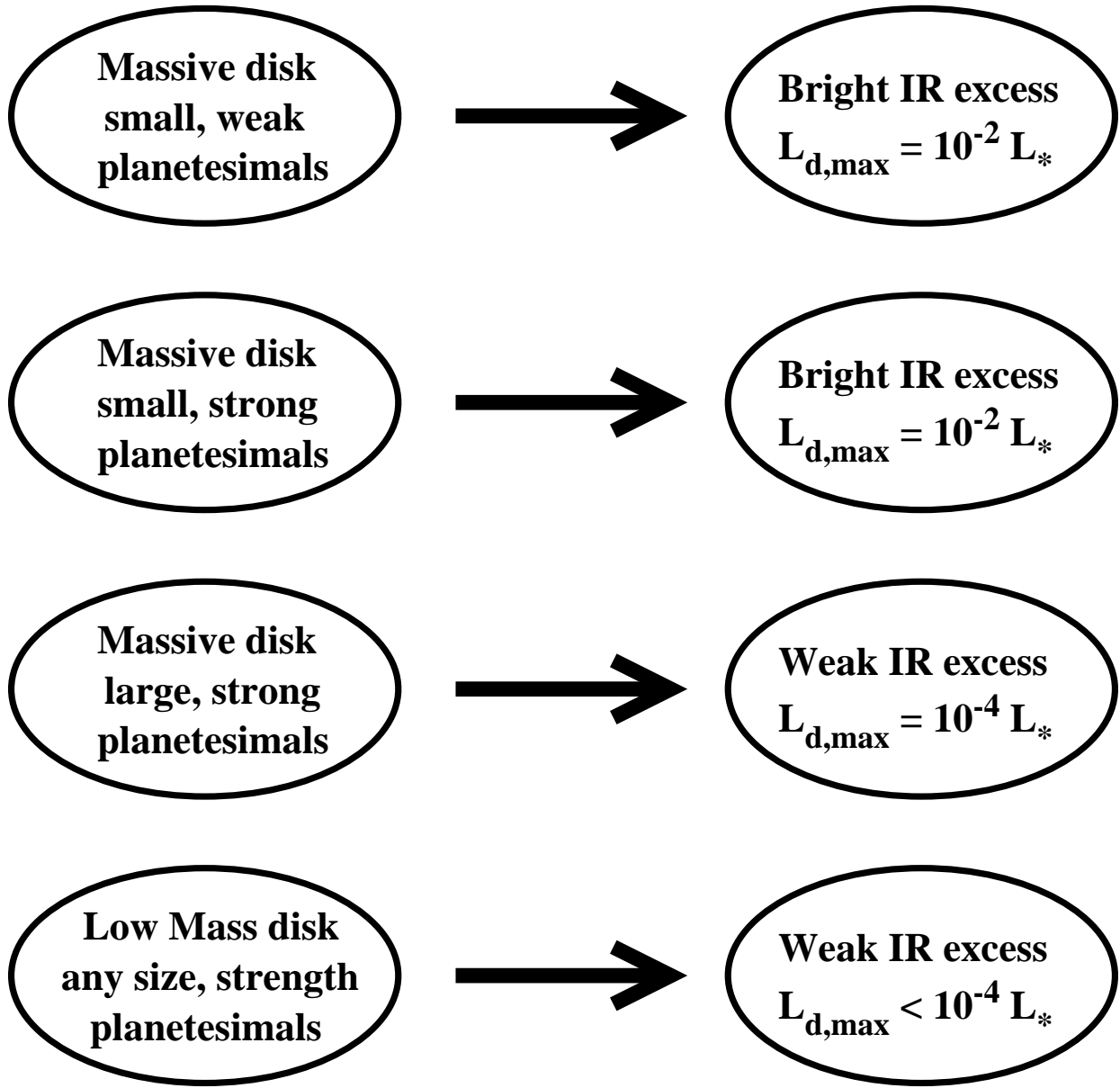}
\vskip 8ex
\caption{
Highlights of debris disk formation at 30--150~AU.
\label{fig: schema2}
}
\end{figure}

With current observational techniques, the properties of dusty debris disks are the 
only constraints on the formation of icy planets at 30--150~AU. 
Tables \ref{tab:eqdebris1}--\ref{tab:eqdebris2} list scaling relations which connect
the observable properties of the debris to the underlying physical properties of
the planetesimal disk.  Based on these relations and the results described in
Figures \ref{fig: ldust1-1msun}--\ref{fig: c824-allm}, we derive nine clear predictions 
for the evolution of debris disks throughout the formation and evolution of a planetary 
system. Figure \ref{fig: schema2} is a schematic summary of the most important predictions.

\begin{enumerate}

\item During runaway growth, the optical depth in large objects and the dust production 
rate decline with time. Thus, dust emission also declines with time.

\item As the central star evolves to the main sequence, an increase in the stellar 
effective temperature can produce small increases in dust emission at short wavelengths. 
Because the amount of dust in the disk decreases, the fractional dust luminosity 
\ldlstar\ decreases as the short wavelength emission rises.

\item Once oligarchic growth begins, dust emission rises dramatically. In more massive 
disks and in disks with smaller planetesimals, dust emission rises at earlier times to 
higher levels. Peak flux occurs roughly at the time when the inner disk produces several
Pluto mass objects. Massive disks with 1~km and smaller planetesimals produce the most 
luminous debris disks (\ldlstar\ $\approx 10^{-2}$) roughly at the time when the central 
star reaches the main sequence. Because the disks in our calculations are radially 
optically thick, these peak luminosities are roughly an order of magnitude smaller than 
results from analytic derivations \citep[e.g.,][]{wya07a,heng10}.
For 1--3~\msun\ stars, the maximum 24 $\mu$m excess grows by roughly a factor of 3 for 
every 0.5~\msun\ increase in stellar mass.  The maximum 70--160~$\mu$m excesses are roughly 
100--300 times the flux from the stellar photosphere. The largest 850 $\mu$m excesses are 
factors of 2--10 smaller. 

\item After the dust emission reaches peak flux, the emission declines slowly with time.
In our calculations, the decline is roughly a power law, \ldlstar\ $\propto t^{-n}$ with
$n \approx$ 0.6--0.8.  The time evolution of IR/submm excesses generally follow the trends in 
\ldlstar.  For typical disk temperatures $\lesssim$ 100 K, emission from wavelengths shorter 
than 20--30 $\mu$m are on the Wien side of the spectral energy distribution. This emission 
follows a power law decline in time with $n \approx$ 0.6--0.8.  Emission at longer wavelengths 
lies at the Planck peak or on the Rayleigh-Jeans tail of the energy distribution. Excesses at 
these wavelengths tend to decline more slowly with time ($n \approx$ 0.1--0.6).

\item The IR/submm excesses and \ldlstar\ are very sensitive to the initial sizes of 
planetesimals at 30--150~AU. Disks with small planetesimals ($r \lesssim$ 1~km)
transform into bright debris disks on relatively short timescales. Disks with
larger planetesimals never become bright debris disks. Our results suggest 
reductions of 4--40 in peak luminosity for disks with 10--100~km planetesimals.
Disks with large planetesimals produce low luminosity debris disks. With current
techniques, IR emission from these disks is not detectable.

\item IR/submm excesses are weakly sensitive to the fragmentation parameters.  In low mass,
optically thin disks, the emission depends only on the mass of dust. Low mass disks with weak
planetesimals produce more dust emission than low mass disks with strong planetesimals. In 
massive disks, the dust is opaque ($\tau \approx$ 1). The amount of dust emission then depends 
on the mass in dust and the scale height $H$ of the dust above the disk midplane. Disks with 
more massive planets have larger $H$. Thus, disks with stronger planetesimals produce more dust 
emission. 

\item Long main sequence lifetimes allow the collisional cascade to remove more material
from the disk. Thus, debris disks around older, lower mass stars are brighter than disks
around older, more massive stars.

\item Emission at short wavelengths depends on the mass of the central star. Stars with
$m_\star \lesssim$ 1~\msun\ are not luminous enough to produce significant emission from
debris at $a \gtrsim$ 30~AU. Thus, short wavelength emission from solar-type and lower
mass stars is produced from debris closer to the central star.

\item An ensemble of debris disks with a broad range of initial masses produces a broad
range of IR excess emission throughout the main sequence lifetime of the central star.

\end{enumerate}

\section{APPLICATIONS}

Our calculations make clear predictions for the evolution of debris disks as a function 
of stellar mass and initial disk mass.  To test these predictions, we now consider several 
applications to recent IR and millimeter data. 

We begin with a discussion of the time evolution of IR excesses at 24--70 $\mu$m. 
Here, the observed fluxes and the long-term trends depend mainly on the initial 
sizes of planetesimals and the initial mass in the disk.  Thus, our initial goals
are to see whether our predictions roughly match the observations and to infer
whether we can deduce approximate ranges for the initial properties of the disk.
In addition to these parameters, the outer disk radius and the typical sizes and 
emissivity law of the smallest grains also modify the predicted fluxes. Thus, our
second set of goals is to learn which changes in our assumptions are allowed by 
the data.

We conclude this section with a comparison of the observed and predicted detection 
frequencies of debris disks at IR and mm wavelengths. Here, the probability of
detecting a debris disk depends on the joint probability that a disk of a given
initial mass ($x_m$) produces a detectable excess and that a disk of that mass is 
common enough to appear in a typical survey with {\it Spitzer} or a ground-based
radio telescope. Thus, this comparison tests whether an adopted distribution of
initial disk masses can produce the observed distribution of known debris disks 
as a function of stellar age.

\subsection{Time evolution of debris disks}

In the past few years, several comprehensive {\it Spitzer} programs have searched for
debris disks around all classes of main sequence stars. Recent surveys of A-type stars
and solar-type stars span a broad range of stellar ages with sufficient statistics to
provide initial tests of our models. We begin with a discussion of A-type stars and then
consider comparisons between our models and observations of solar-type stars.

\subsubsection{A-type Stars}

Since the initial discovery of debris disks around Vega, Fomalhaut, and $\beta$ Pic,
IR observations have revealed debris around hundreds of A-type stars 
\citep[e.g.,][]{bac93,rie05,su06,cur08,reb08,gau08,car09b,mora09}.  For stars with ages of $\sim$ 
50~Myr to 1~Gyr, analyses of the 24--25 $\mu$m excesses of several large samples demonstrate 
a clear decline in the excess with time \citep{rie05,su06}.  Subsequent studies include
larger samples of younger stars. These results suggest a rise in the debris emission from 
$\sim$ 5~Myr to 10~Myr, a plateau in emission at 10--30~Myr, and a slow decline in emission 
for older stars \citep[][2007, 2009, Currie et al. 2008a, 2008b]{her06}.  
Although the significance of the rise in debris disk fluxes at 5--10~Myr is controversial 
\citep[e.g.,][]{car09b}, the long decline in the 24--25 $\mu$m flux with age is a robust 
feature of debris disk evolution around A-type stars \citep{wya07b,kenn10}.

To compare our predictions with observations, we consider data for A-type stars from
three compilations \citep{rie05,su06,curr09}. The combined sample from \citet{rie05} 
and \citet{su06} has 319 stars with 24--25 $\mu$m photometry, spectral types B7-A6, and 
ages 5--850~Myr. The \citet[][see also Currie et al. 2008a]{cur09} compilation adds 157 
stars with similar spectral types and ages 2.5--25~Myr. Based on comparisons of observed 
HR diagrams with stellar evolution models, the stars have masses of 1.7--2.5~\msun\ (KB08). 
Thus, we compare these data with our results for disks around 2~\msun\ and 2.5~\msun\ stars.

Figure \ref{fig: f24-astars1} shows several comparisons between our predicted fluxes and
observations of A-type stars. The data appear as blue points in each of the four panels. 
The solid curves show predicted fluxes for disks with strong planetesimals and initial 
$\Sigma \propto a^{-3/2}$ around 2.5~\msun\ stars. The legends in each panel indicate the 
initial planetesimal size; the legend in the upper left corner of the upper left panel 
indicates $x_m$ for all model curves.

All of the model curves show the same trend with time. Following a period with little
or no excess, the flux from the IR excess rises rapidly to a plateau value and then 
declines slowly with time. The flux level and the timing of the plateau correlates with 
initial disk mass and the initial planetesimal size. More massive disks and disks with
smaller planetesimals have larger excesses at 24~$\mu$m and reach peak excess earlier 
in time (see also Tables \ref{tab:eqdebris1}--\ref{tab:eqdebris2}).  

Two sets of model curves agree reasonably well with the data. In the left panels, the range
of predicted excesses for all initial disk masses encompasses most of the observed points.
Calculations with a range of initial planetesimal sizes provide a better match to observations 
of younger stars; low mass disks composed of 1~km planetesimals yield a better match to the
data for older stars. In both cases, the predicted fluxes for the ensemble of disks explain
97\% to 98\% of measured {\it Spitzer} fluxes for A-type stars.

Two sets of model curves do not match the observations. In the right panels, the range of
predicted fluxes for disks of 10--100~km planetesimals fail to explain the brightest sources 
with ages of 5--30~Myr. Disks of 100~km planetesimals also cannot explain the brightest
sources with ages of $\sim$ 100~Myr. Although both sets of models can match the range of
observations of older stars, sources with $F_{24} / F_{24,0} \gtrsim$ 2 require relatively
massive disks with $x_m \gtrsim$ 0.33 ($x_m \gtrsim$ 1) for models with 10~km (100~km)
planetesimals. Disks around young stars have typical $x_m \approx$ 0.03--0.10 \citep{and05}.
Thus, these calculations provide an unlikely match to the data.

Figure \ref{fig: f24-astars2} repeats the format of Figure \ref{fig: f24-astars1} for different
sets of models. In the left (right) panels, the curves plot predicted fluxes for disks with 
$\Sigma \propto a^{-1}$ ($\Sigma \propto a^{-3/2}$). The lower (upper) panels plot predicted
fluxes for 2~\msun\ (2.5~\msun) stars. The legends in the upper panels summarize the appropriate
$x_m$ for each model curve.

When disks are composed of small planetesimals, a range of initial disk masses can explain 
most of the observed range of debris disk fluxes. For systems with ages of 10--20~Myr, disks 
with $\Sigma \propto a^{-1}$ around 2.5~\msun\ stars match the observed fluxes for all but a 
few of the most luminous debris disks. Disks with $\Sigma \propto a^{-3/2}$ can also match 
the data.  Although disks around 2~\msun\ stars provide a poorer match to all observations of 
young stars, disks around lower mass stars match the data well for older stars. 

Variations in the initial surface density law yield a small range of debris disk fluxes. In 
calculations with small planetesimals, disks with different surface density laws but similar 
total masses within 150~AU produce somewhat different 24 $\mu$m fluxes. We show in the next 
sub-section that both of our adopted surface density laws can explain the observed frequency 
of debris disks around young A-type stars and G-type stars

Figure \ref{fig: c2470-astars} compares the predicted color evolution for model debris disks 
with observations from the \citet{su06} sample of A stars. The format follows the format of
the left panels in Figure \ref{fig: f24-astars2}. The upper panel shows results for models 
around 2.5~\msun\ stars; the lower panel shows results for models around 2~\msun\ stars. 
Disks with large planetesimals yield poor fits to the observed colors; disks with small
planetesimals and different surface density laws yield similar matches to the observed colors. 
Thus, we focus on models with $\Sigma \propto a^{-1}$ for 2~\msun\ and 2.5~\msun\ stars.

In each panel, the model colors show four clear trends. At 1--10~Myr, the colors redden as the
central star evolves to the main sequence. During this evolution, grains throughout the disk
become hotter and emit more flux.  This emission then dominates scattered light from the disk.
The [24]--[70] color becomes redder.  At the same time, the most massive disks start to produce 
copious amounts of debris.  More massive disks have more grains emitting at longer wavelengths.  
These disks have the reddest colors. As the collisional cascade develops, the flux at shorter
wavelengths initially rises faster than the flux at long wavelengths. The colors become bluer.
At later times, the flux at longer wavelengths rises more rapidly. The colors then become
redder. Finally, the emission starts to decline at all wavelengths. Radiation from the central 
star then contributes more and more to the IR excess. The colors become bluer.

The observed colors do not show any obvious trends with time. Although it is tempting to
suggest two peaks in the color distribution at $\sim$ 10~Myr and at $\sim$ 100~Myr, there 
are few stars with ages of 30--50~Myr.  Older stars appear to have more debris disks with
bluer colors than younger stars. However, this feature is probably due to a lack of young
stars with small excesses.

Despite the lack of trends in the data, both models match the observations in Figure 
\ref{fig: c2470-astars} reasonably well. At any stellar age, a broad range in disk masses 
explains most of the observed range in colors. Predicted colors for 2~\msun\ (2.5~\msun)
stars encompass 94\% (83\%) of the observed colors. For both stellar masses, none of the
models match the colors of the reddest sources. Most of these sources have ages of $\sim$ 
10~Myr and $\sim$ 50--100~Myr. Color measurements for stars with ages of 20--100~Myr would
provide a good test of the models for the reddest sources.

\subsubsection{G-type Stars}

Although {\it IRAS} and {\it ISO} data enabled the discovery of a few debris disks around
solar-type stars \citep{bac93,dec03,son05,rhe07a}, recent {\it Spitzer} observations have 
yielded a large sample of systems around FGK stars \citep[e.g.,][]{bei06,hil08,tri08,car09a}. 
As preparation for future searches for terrestrial planets, most surveys have targeted older,
nearby solar-type stars \citep[e.g.,][]{bei06,tri08}. One program targeted stars with a range 
of ages \citep[e.g.,][]{mey06}. Together these samples provide good initial tests of our
predictions.

To test our models with the data, we collected 70 $\mu$m data from \citet{bei06}, \citet{tri08}, 
and \citet[][see also Carpenter et al. 2009a]{hil08}. After eliminating stars with estimated masses 
smaller than 0.8~\msun, the combined 
sample has $\sim$ 100 stars with ages ranging from 10~Myr to 10~Gyr. We compare these data with
predictions for debris disks around 1~\msun\ stars. Predictions for disks around 1.5~\msun\ stars
yield similar conclusions.

Figure \ref{fig: f70-gstars} compares the observations with the predicted evolution for our models.
The left panel plots calculations for disks with $\Sigma \propto a^{-1}$, $x_m$ = 0.01--1, and
1~m to 1~km planetesimals.  The right panel includes calculations for disks with 
$\Sigma \propto a^{-3/2}$, $x_m$ = 0.01--3, and 1~km planetesimals. Predictions for disks with
larger planetesimals provide very poor matches to the data (e.g., Figure \ref{fig: f24-astars1}). 
Both sets of models show similar trends with time. During runaway growth, the IR excess remains
constant or declines slightly with time. Once oligarchic growth begins, the IR excess rises
dramatically. More massive disks rise earlier and reach larger peak fluxes than lower mass disks.
The peak flux and the time of peak flux also depend on the initial planetesimal size. Disks with
a range of small planetesimal sizes rise earlier than disks with a single planetesimal size.

Although the data clearly preclude large planetesimals, the observations do not discriminate
among possible models with small planetesimals \citep[see also][]{heng10}. Both sets of 
predictions provide reasonable
matches to the data. Roughly 75\% of the observed points lie within the area outlined by the 
models. For models with $\Sigma \propto a^{-1}$ ($\Sigma \propto a^{-3/2}$), disks with 
$x_m \approx$ 0.03 (0.33) have masses close to the median disk mass suggested from observations 
of pre-main sequence stars in Taurus-Auriga and Ophiuchus \citep[][2007b]{and05}. Thus, our 
calculations predict a large population of low luminosity debris disks in systems with low mass
disks.

In KB08, we noted that two changes to our assumptions yield a better match to observations.  Reducing 
the maximum stable grain size from $r_2$ = 1 $\mu$m to $r_2$ = 0.1 $\mu$m and changing the slope of 
the radiative emissivity law from $q$ = 1 to $q$ = 0.7 increase the predicted fluxes by
a factor of 3--10 at 70--160 $\mu$m. Observations currently provide little guidance on either
of these parameters for debris disks around solar-type. In the A-type stars Vega, $\gamma$ Oph, 
and HR 8799, {\it Spitzer} imaging and spectroscopic observations yield good constraints on the maximum 
stable grain size, with $r_2 \approx$ 5--10 $\mu$m \citep[][2008, 2009; M{\"u}ller et al. 2010]{su05}. 
Additional submm data can establish limits on $q$. Measurements of $r_2$ and $q$ for a sample of 
debris disks around solar-type stars would enable a better evaluation of our predictions.

In addition to $r_2$ and $q$, changes to $s$ -- the slope of the size distribution of the debris --
also yield a better match to observations. For weak planetesimals, the theoretical estimate of 
$s$ = 3.67 \citep{obr03,kob10} yields factor of 5--10 larger 70 $\mu$m fluxes throughout the
evolution. Thus, weak planetesimals with $s$ = 3.67 are favored over strong planetesimals with
$s$ = 3.5. Aside from matching the observed {\it Spitzer} fluxes, observations provide little 
guidance for $s$. However, larger samples of debris disks with ages of 10--100~Myr would yield
a good test. In the context of our calculations, detecting 10~Myr old systems with 
$F_{70} / F_{70,0} \approx$ 300 would provide some evidence for weak planetesimals with $s >$ 3.5.

\subsection{Frequency of Debris Disks}

Although the good agreement between predicted and observed fluxes is encouraging, Figures
\ref{fig: f24-astars1}--\ref{fig: f70-gstars} do not yield a quantitative measure of success
of our models. To make this measure, we consider the ability of our models to predict the
frequency of debris disks around stars with a range of masses. For this first attempt, we
construct the probability that an ensemble of disks with a range of initial masses produces
debris disks detectable with current technology. The sensitivity of current observations and
the predicted fluxes of debris disks depend on wavelength. Thus, we define this detection 
probability $p_{d,\lambda}$ as a function of wavelength.

To predict $p_{d,\lambda}$ for debris disks from our calculations, we adopt a simple model. 
We consider a sample of stars with stellar masses \mstar\ and disk masses $M_d$. 
For $\mu_d = {\rm log} ~ (M_d / M_\star)$, observed disk masses roughly follow a log normal 
probability distribution \citep[][2007b]{and05}.  Thus, we follow \citet{ale06} and adopt
\begin{equation}
p(\mu_d) = p_0 ~ e^{-((\mu_d - \mu_{d,0}) / \sigma_d)^2} ~ ,
\label{eq: mdisk}
\end{equation}
where $\mu_{d,0}$ = log ($M_{d,0} / M_\star$) defines the average disk mass, $\sigma_d$ is the 
dispersion, and $p_0$ is a normalization constant.  Following \citet{and07b}, $\mu_{d,0}$ = 
$-2.0$, $\sigma_d$ = 1.0, and $p_0$ = $1 / \sqrt{\pi}$.

To define $p_{d,\lambda}$, we use the results of our simulations to assign IR excesses -- 
log $F_{\lambda}/F_{\lambda,0}$ -- as a function of time $t$ to disks with initial masses $\mu_d$.
For any $t$, there is a range of disks with IR excesses larger than some detection
threshold, $\tau_{\lambda}$, 
\begin{equation}
F_{\lambda} / F_{\lambda,0} > t_\lambda ~ .
\end{equation}
Disks that satisfy this inequality have $\mu_d > \mu_{d,t}$. Thus the detection probability for
an adopted detection threshold $t_\lambda$ is
\begin{equation}
p_{d,\lambda} = \int_{\mu_{d,t}}^\infty ~ p_0 ~ e^{-((\mu_d - \mu_{d,0}) / \sigma_d)^2} ~ d \mu_d ~ .
\end{equation}

To set the detection threshold, we rely on the measurement uncertainties and detection limits
of existing {\it Spitzer} surveys. If $\sigma_\lambda$ is the uncertainty in the measured flux
at wavelength $\lambda$, stars 
with clear IR excesses have measured fluxes $F_{\lambda}$ that exceed the stellar photospheric 
flux $F_{\lambda,0}$ by several $\sigma_\lambda$. We define $s_\lambda$ as the signal-to-noise ratio
required to identify an excess \citep[e.g.,][]{rie05,bei06,bry06,su06,tri08}.  Thus, our criterion
for the detection threshold is 
\begin{equation}
F_{\lambda} > F_{\lambda,0} + s_\lambda \sigma_\lambda ~ .
\end{equation}
In our calculations, we derive the flux ratio $F_{\lambda} / F_{\lambda,0}$. Thus, we rearrange the
detection threshold as 
\begin{equation}
F_{\lambda} / F_{\lambda,0} > 1 + s_\lambda \sigma_\lambda / F_{\lambda,0} ~ .
\end{equation}
In this expression, the left-hand side is predicted by the models; the right-hand side is set by 
observations. Thus, our detection threshold is $t_\lambda = 1 + s_\lambda \sigma_\lambda / F_{\lambda,0}$.

There are two approaches to setting $t_{\lambda}$.  Both rely on the predicted mid-IR flux $F_{\lambda,p}$ 
from the stellar photosphere, derived from model atmosphere fits to optical and near-IR data.  When a set 
of observations is capable of detecting emission from the stellar photosphere, the average flux ratio 
$<F_{\lambda}/F_{\lambda,p}>$ for a subset of stars in the sample resembles a gaussian centered roughly 
on unity with a dispersion $\sigma_p$ \citep[e.g.,][]{rie05,bei06,su06}. In these stars, the observed flux is 
probably very close to the actual photospheric flux. Stars with 
$F_{\lambda}/F_{\lambda,p} >$ $1 + s_{\lambda} \sigma_p$ are then likely debris disks;
 $ t_\lambda = 1 + s_{\lambda} \sigma_p$ is the appropriate detection threshold for our models.  When 
observations cannot reliably detect photospheric emission in stars without debris disks, most analyses 
adopt the predicted photospheric flux as the actual photospheric flux. The appropriate threshold for
these studies is then $t_\lambda = 1 + s_{\lambda} \sigma_\lambda / F_{\lambda,p}$. 

\subsubsection{{\it Spitzer} Observations of A-type Stars}

To apply this model to observations, we first consider the \citet{su06} sample of A-type stars.  This sample 
has 160 stars with high quality {\it Spitzer} 24 $\mu$m and/or 70 $\mu$m photometry, stellar spectral types 
B7--A6, and ages 5--850~Myr.  Most stars in the sample have 24 $\mu$m and 70 $\mu$m fluxes consistent with 
the predicted photospheric fluxes. For these stars, \citet{su06} infer $<F_{\lambda}/F_{\lambda,p}>$ = 
0.98 $\pm$ 0.026 at 24 $\mu$m and $<F_{\lambda}/F_{\lambda,p}>$ = 1.11 $\pm$ 0.15 at 70 $\mu$m. Thus,
the detection thresholds are $t_{24} = 1 ~ + ~ 0.026 ~ s_{24}$ and  $t_{70} = 1.1 ~ + ~ 0.15 ~ s_{70}$.
For $s_\lambda$ = 3--5, roughly 33\% of the stars have IR excesses, at levels ranging from 10\% to 
several hundred times the flux from the stellar photosphere. 

Figure \ref{fig: dprob-a} compares our predictions for disks with $\Sigma \propto a^{-1}$ around 2~\msun\ stars 
with data from the \citet{su06} survey of A-type stars. To construct this plot, we adopt the conservative
estimate of $s_{24}$ = $s_{70}$ = 5 and compute predicted detection probabilities at 24~$\mu$m and at 70~$\mu$m.
For the data, we identify debris disks with excesses at the 5$\sigma$ level and calculate the observed
detection frequency. To improve statistics, we bin the observations in discrete age bins. For each point 
in the plot, the horizontal error bar indicates the range of stellar ages used for the bin; the vertical 
error bar indicates the 1$\sigma$ uncertainty from Poisson statistics. 

Despite the high quality of the \citet{su06} data, many sources have upper limits at 70 $\mu$m. To
provide an additional measure of the uncertainty in these data, we derive an upper limit to $p_{d,70}$
by assuming that every star with an upper limit has an IR excess. For the entire sample, this assumption
yields an upper limit of $p_{d,70}$ = 0.67 \citep[see also][]{su06}.

At both wavelengths, our predictions show similar trends.  The models predict a slow rise in the 
detection probability from 1~Myr to $\sim$ 50--100~Myr, a plateau at 100--300~Myr, and a sharp decline 
from 300~Myr to $\sim$ 1 Gyr.  Hotter dust in the inner disk evolves faster than cooler dust farther 
out in the disk. Thus, evolution at 24 $\mu$m is faster than at 70 $\mu$m. At both wavelengths, the 
slow rise in detection probability results from the detection of lower mass disks with smaller peak 
fluxes at later times.  For $t >$ 300~Myr, debris emission declines rapidly 
(Figures \ref{fig: f24-astars1}--\ref{fig: f24-astars2}).  This decline produces the steep drop in 
detection probability at later times.

Our predictions agree reasonably well with the observations.  At early times ($t \lesssim$ 10~Myr), the
model matches the 24 $\mu$m data and lies in between the observed and upper limit to the data at 70 $\mu$m.
At intermediate times ($t \approx$ 50--200~Myr), the data fall a factor of 2--3 below predictions.  At 
later times ($t >$ 500~Myr), the model again matches the 24 $\mu$m data but overpredicts the 70 $\mu$m data 
by a factor of 1.5--3.  For the full ensemble of observations, our models predict detection probabilities 
of $p_{d,24,p}$ = 0.55 $\pm$ 0.05 and $p_{d,70,p}$ = 0.81 $\pm$ 0.04. Both of these are somewhat larger 
than the observed detection rates of $p_{d,24} \approx p_{d,70} \approx$ 0.33 $\pm$ 0.05 \citep{su06}. 
Our prediction at 70 $\mu$m is close to the upper limit of $p_{d,70} \lesssim$ 0.67.

These conclusions do not depend on the initial surface density relation or the fragmentation parameters.
All our results suggest detection probabilities $p_{d,24,p} \approx$ 0.53--0.57 and $p_{d,70,p} \approx$ 
0.81--0.88.

Modest changes to our calculations can modify $p_d$ significantly.  To test the models in more detail,
we derived detection probabilities assuming larger values for the slope of the size distribution for 
small grains $s$ and for the maximum stable grain size $r_2$. Adopting $s$ = 3.67 increases predicted
fluxes at all wavelengths, increasing the predicted detection probabilities to more than 0.90. In the
context of our models, existing observations rule out this change. Adopting $r_2$ = 10 $\mu$m reduces 
the predicted fluxes at all wavelengths. This change yields $p_{d,24,p}$ = 0.40 $\pm$ 0.05 and 
$p_{d,70,p}$ = 0.62 $\pm$ 0.04, very close to the observed value at 24 $\mu$m and the upper limit 
at 70 $\mu$m. 

\subsubsection{{\it Spitzer} and Submillimeter Observations of G-type Stars}

To apply our model to G-type stars, we consider 70 $\mu$m data from {\it Spitzer} and 350--1200~$\mu$m 
data from several ground-based observatories. Most of these surveys did not detect the photospheres for 
a large fraction of program stars. Thus, we employ the second approach for the detection threshold in 
our analysis. 

We consider several samples of G-type stars at each wavelength. At 70 $\mu$m, the \citet{bei06} and 
\citet{tri08} surveys focus on nearby stars with typical ages of 1--15~Gyr. Data from the Formation and 
Evolution of Planetary Systems (FEPS) {\it Spitzer} Legacy program includes stars with a wide range of 
ages (1~Myr to 10~Gyr). At longer wavelengths, \citet{roc09} analyze 350 $\mu$m and 1.2 mm observations 
of older stars ($\gtrsim$ 10~Myr) from the FEPS survey.  Several deeper submm surveys describe data for 
nearby solar analogs \citep{gre09a} and for stars in the Pleiades \citep{gre09b}.

Figure \ref{fig: dprob-g} compares our predictions with the observed detection frequencies.  To construct 
this plot, we adopt a 5$\sigma$ detection threshold and compute observed and predicted detection rates.
To provide an estimate for the sensitivity of our results to the outer disk radius, we include predictions 
for large disks with radii of 150~AU and for small disks with outer radii of 75~AU. 

Our predictions agree well with observations at all wavelengths. At 70 $\mu$m, our predictions match 
surveys with large \citep{bei06}, intermediate \citep{tri08}, and small \citep{car08} detection rates.  
The close match between our models and these observations suggests that the broad range in observed 
detection rates is a result of small number statistics and differences in target selection. Although 
any set of model disks with outer radii $\gtrsim$ 75~AU agree with the 350~$\mu$m data, the small sample 
size precludes any clear test of the models. Data at longer wavelengths yield a more severe test. Small
disks match observations of nearby solar analogs \citep{gre09a} better than large disks. Predictions
for small and large disks agree with observations of FEPS and Pleiades stars.

\section{CONCLUSIONS}

Together with the discussion in KB08, the results described here provide a robust theoretical
picture for the formation of icy planets and debris disks from a disk of icy planetesimals 
surrounding 1--3~\msun\ stars.  These calculations set the context for the evolution of dusty 
debris in a dynamic system of planets and establish  a framework for interpreting existing 
observations of debris disks around intermediate mass stars. Our analysis also suggests new 
observational tests of this picture.

We describe a suite of numerical calculations of planets growing from ensembles of 
icy planetesimals at 30--150~AU in disks around 1--3~\msun\ stars. Using our hybrid 
multiannulus coagulation code, we solve for the evolution of sizes and orbits of 
objects with radii of $\sim$ 1~m to $\gtrsim$ 1000~km over the main sequence lifetime 
of the central star.  These results allow us to constrain the growth of planets as 
a function of disk mass, planetesimal size, surface density law, stellar mass, and
semimajor axis.

In disks with small planetesimals, debris disk formation is coincident with the 
formation of a planetary system.  All calculations of icy planet formation with
1--10~km planetesimals at 30--150~AU lead to a collisional cascade which produces 
copious amounts of dust on timescales of 5--30~Myr. This dust is observable 
throughout the lifetime of the central star. Because we consider a broad range of
input parameters, we derive the time evolution of (i) dust produced in the collisional 
cascade and (ii) the IR and submm emission from this dust as a function of the initial
parameters.

In disks with larger planetesimals, debris disks form near the end of the main sequence 
lifetime of the central star. The collisional cascade is weak and produces little dust.
This dust is barely observable with current technology. As long as the planetesimals are
initially large, this conclusion is independent of the initial mass or surface density 
of the disk, the semimajor axis, or the stellar mass.

We divide the rest of this section into
(i) theoretical considerations,
(ii) observable consequences, and
(iii) observational tests.
The theoretical considerations build on the highlights of icy planet
formation in \S3.3. Observable consequences and tests of the calculations
follow from the discussions in \S4.3 and \S5.

\subsection{Theoretical Considerations}

\begin{enumerate}

\item Icy planet formation at 30--150~AU is self-limiting. Starting with
a swarm of $\lesssim$ 10~km planetesimals, runaway growth produces a set
of 100--500~km protoplanets. As the protoplanets grow, they stir up
leftover planetesimals along their orbits. When the leftovers reach high
$e$, collisions produce debris instead of mergers. Protoplanets cannot accrete 
leftovers rapidly; a cascade of destructive collisions grinds the leftovers 
to dust. Poynting-Robertson drag and radiation pressure then remove the dust 
from the disk.

\item In disks composed of $\lesssim$ 10~km planetesimals, the maximum sizes 
of icy planets at 30--150~AU depend primarily on the initial disk mass.  The 
largest icy planets around 1--3~\msun\ stars have $r_{max} \sim$ 1500--3500~km 
and $m_{max} \sim$ 0.003--0.05~\mearth.  These objects contain $\lesssim$ 3\%--4\% 
of the initial disk mass.  Plausible ranges in the fragmentation parameters can 
reduce the maximum mass by a factor of $\sim$ 10. In addition to these robust 
limits on the maximum size of icy planets, the finite main sequence lifetimes of 
1--3~\msun\ stars limits the formation of many large planets in the outer disk. 
Thus, the inner disk produces many more Pluto-mass planets than the outer disk 
(Tables \ref{tab:rad1000.1}--\ref{tab:rad1000.6}).

\item In disks composed of $\gtrsim$ 10~km planetesimals, icy planets reach larger
sizes.  Our results suggest $r_{max} \sim$ 6000--11500~km and $m_{max} \sim$
0.2--1.5 \mearth. These planets form slowly. By the time the central star evolves
off the main sequence, large planets contain $\lesssim$ 10\% of the initial disk
mass. The weak collisional cascade produces little dust. Thus, Poynting-Robertson 
drag and radiation pressure remove little dust from the disk.

\item At 30--50~AU, collisional cascades remove a substantial fraction of small
planetesimals in the massive disks around 1--3~\msun\ stars.  In disks with 
$x_m \gtrsim$ 0.1, $\lesssim$ 10\% of the initial mass in small objects remains 
when the central star evolves off the main sequence. At larger disk radii
($\gtrsim$ 100~AU), more than 50\% of the initial disk mass is still in small
planetesimals at $t = t_{ms}$. Throughout lower mass disks with $x_m \lesssim$ 0.1, 
most of the mass remains in small planetesimals.

\item Collisional cascades produce copious amounts of dust. Dust starts to form 
during the transition from runaway to oligarchic growth, peaks when the largest
objects first reach sizes of $\sim$ 1000~km, and slowly declines as objects reach
their maximum sizes. This evolution is more sensitive to the initial sizes of the
planetesimals than on the initial surface density law or the fragmentation parameters.
In massive disks with small planetesimals, this evolution 
begins as the central star approaches the main sequence; in lower mass disks and
in disks with large planetesimals, peak dust formation occurs as the central star 
evolves off the main sequence. In ensembles of disks with $x_m$ = 0.01--1, the peak 
mass in 0.001--1 mm (0.001--1 m) particles ranges from 0.002 to 10 lunar masses
(0.002 to 10 \mearth). 

\item Radiative processes can remove large amounts of mass from the disk. During
the early stages of the collisional cascade, radiation pressure produces a wind 
of small particles. This wind contains 60\% to 100\% of the mass lost from the disk. 
During the end stages of the cascade, Poynting-Robertson drag pulls the rest of 
the lost mass into regions with $a <$ 30~AU. 

\item The impact of gas giant planets on debris disks is a major uncertainty in our
calculations. Gas giants at 5--20~AU \citep[e.g.,][]{kenn08} produce small-scale 
fluctuations in the orbits of planetesimals at 30--150~AU \citep[e.g.,][]{must09} and 
large-scale dynamical events throughout the disk \citep[e.g.,][]{gom05}. These 
perturbations modify the evolution of the collisional cascade and impose structure in 
the radial distribution of small objects \citep[e.g.,][]{wil02,mor05,qui06,boo09}.  
Current observational constraints on the frequency of gas giants around G stars 
\citep{cum08} and A-type stars \citep{john07} suggest gas giants probably form only in
the most massive disks. Thus, gas giants probably cannot change our predictions for
debris disks in low mass disks ($x_m \lesssim$ 0.1).  In massive disks around 
2--3~\msun\ stars, the rapid rise in dust production at 30--50~AU roughly coincides 
with the timescale for gas giant planet formation. Thus, gas giants probably enhance 
the dust production rate and may sometimes change the outcome of planet formation at 
30--150~AU.  In massive disks around 1--1.5~\msun\ stars, gas giants form well before 
the rise in the dust production.  Here, gas giants probably play a major role in the 
evolution of the debris disk. Quantifying this role requires simulations that include
gas giant planet formation in disks with dimensions of 3--150~AU. These simulations
are now possible on high performance computers.

\item Planetesimal formation is another uncertainty. Current simulations of the growth
of small, micron-sized dust grains in a turbulent disk require accurate treatments of
collisions, dust dynamics, and gas dynamics 
\citep[e.g.,][]{ric06,gar07,kre07,bra08,cuz08,lai08,you08,bir09}. Recent results are
inconclusive on the typical planetesimal size, with expected sizes ranging from meters 
to more than a thousand kilometers. Expected formation times are generally shorter than 
the typical evolution time of $\gtrsim$ 1~Myr. Thus, our assumption of initial ensembles 
of 1~m to 100~km planetesimals is probably reasonable. 

\item Our models provide some guidance for models of planetesimal formation.  Several 
recent numerical simulations of streaming instabilities in turbulent disks yield 
planetesimals with $r \gtrsim$ 100--1000~km \citep[e.g.,][2009]{joh07}.  \citet{mor09} 
demonstrate that an initial ensemble of large planetesimals may resolve several outstanding 
issues in the formation of the asteroid belt in the Solar System. In the context of our 
calculations, however, the observed evolution of debris disks around A-type and G-type 
stars strongly favors disks with a significant amount of solid material in small ($\lesssim$ 
1--10~km) planetesimals at 30--150~AU. Although forming small planetesimals in a turbulent
disk is challenge, the dead zones of protoplanetary disks are attractive locations for 
producing planetesimals with $r \sim$ 0.1--1~km on short timescales \citep[e.g.,][]{bra08}.  
Based on our results, identifying similar mechanisms to produce small planetesimals at 
30--150~AU may provide important insights into the formation of debris disks and planetary 
systems.

\end{enumerate}

\subsection{Observational Consequences}

Our calculations yield robust observational consequences of the collisional cascade.

\begin{enumerate}

\item The dusty debris from the collisional cascade is directly observable throughout 
the main sequence lifetime of the central star. For disks around 1--3~\msun\ stars, the 
maximum fractional dust luminosity of $L_d / L_{\star} \sim 10^{-2}$ is comparable to 
the maximum dust luminosities of known debris disks \citep{bac93,rie05,su06,rhe07a}. 
With typical radial optical depths of 1--10 at maximum, this limit is roughly an order 
of magnitude smaller than theoretical maximum \citep[e.g.,][]{wya07a,heng10}. The dust 
temperature at the inner edge of a 30--150~AU disk scales with the temperature of the 
central star; thus, the predicted 24~$\mu$m excess is very sensitive to the stellar
mass. At 70~$\mu$m, the predicted excesses scale roughly linearly with disk mass and 
stellar mass. The predicted 160--850~$\mu$m excesses depend on the disk mass but are 
nearly independent of the stellar mass.

\item The amount of dusty debris depends on the initial sizes of planetesimals. In
our calculations, bright debris disks at 5--30~Myr require planetesimals with initial
sizes $\lesssim$ 1~km. Calculations with 10--100~km planetesimals produce no debris
around young stars and little debris around older stars. Thus, our results suggest
planetesimals with initial sizes $\sim$ 1~km.  

\item The amount of debris depends weakly on the initial surface density law and the
fragmentation parameters. Disks with similar total masses produce similar amounts of
dust emission. Disks with stronger planetesimals produce more dust emission than 
disks with weaker planetesimals.

\item Stellar evolution limits the brightness evolution of debris disks. Among coeval 
stars with ages $\gtrsim$ 100~Myr (e.g., star clusters), lower mass stars have brighter 
debris disks. Among older stars near the main sequence turn-off (e.g., field stars), 
our results suggest that debris disks around massive stars are relatively brighter 
than debris disks around lower mass stars. 

\item The time evolution of \ldlstar, IR fluxes, and IR colors provide useful 
diagnostics of the source of the debris. When the collisional cascade begins, 
\ldlstar\ and IR excesses increase with time, reach a plateau, and then decline
with time. IR colors generally become redder when IR excesses rise and bluer when
IR excesses fall. This evolution contrasts with the expected evolution of a
pre-existing disk of small particles responding to changes in the properties of
the central star, where \ldlstar\ always declines with time.

\end{enumerate}

\subsection{Observational Tests}

Observations of debris disks around A-type and G-type stars provide good tests of our 
predictions.  More comprehensive comparisons with the data require calculations with
a broader range of grain parameters than discussed here. For example, \citet{kenn10} 
develop an elegant semi-analytic model that captures many aspects of our planet formation
calculations and apply this model to observations of A-type stars. By deriving the best
set of fragmentation parameters and dust properties to fit the data, they infer robust
constraints on the initial mass, radius, and surface density in the disk.  Here, we focus 
on comparisons that test the ability of our calculations to explain the observed evolution
and frequency of debris disks around A-type and G-type stars.

\begin{figure} 
\includegraphics[width=6.5in]{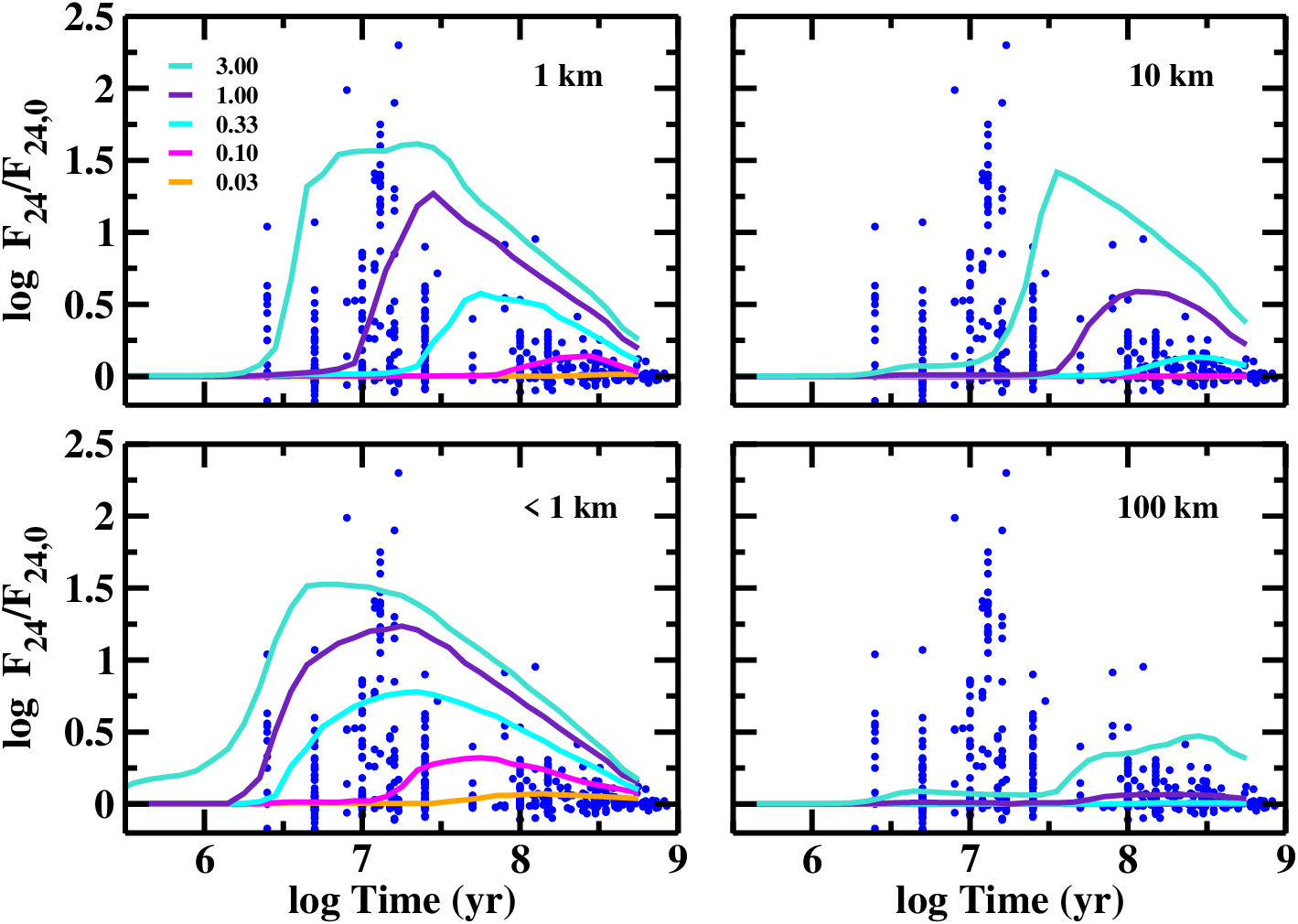}
\vskip 3ex
\caption[f29.eps]{
Observations of the 24 $\mu$m excess for A-type stars with known ages
\citep{rie05,su06,cur08}. The colored lines show the predicted evolution 
of the excess for debris disk models with $\Sigma \propto a^{-3/2}$ and
the strong fragmentation parameters around 2.5 \msun\ stars. In each panel,
the lines plot predictions for disks with 1 m to 1~km planetesimals (lower 
left), 1~km planetesimals (upper left), 10~km planetesimals (upper right), 
and 100~km planetesimals (lower right).  Values for $x_m$ are listed in the 
legend of the upper left panel.  Disks with a significant fraction of their 
initial mass in small planetesimals produce more debris at earlier times, as 
observed.
\label{fig: f24-astars1}
}
\end{figure}

\begin{figure} 
\includegraphics[width=6.5in]{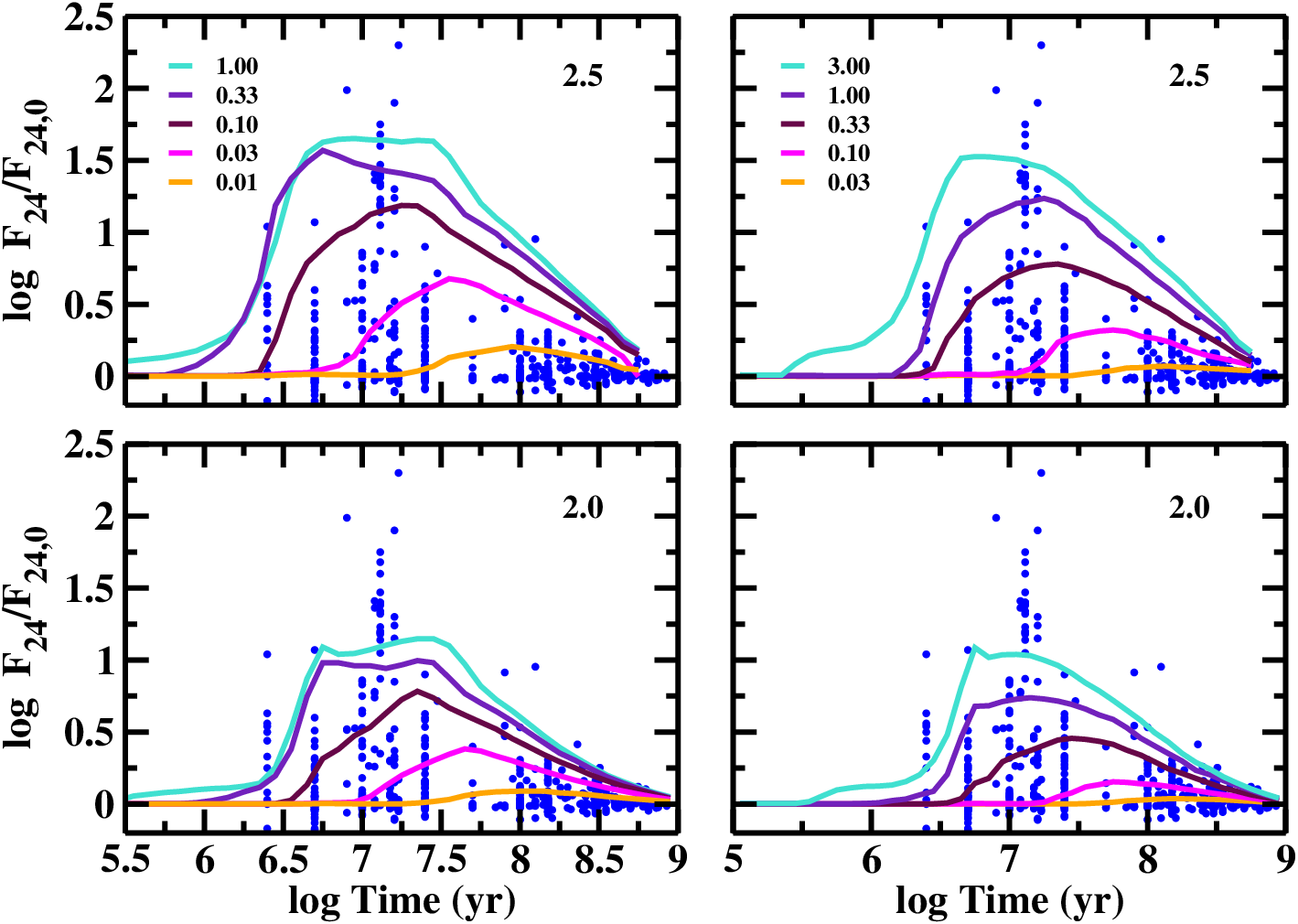}
\vskip 3ex
\caption[f30.eps]{
As in Figure \ref {fig: f24-astars1} for models with 
$\Sigma \propto a^{-1}$ (left panels) and 
$\Sigma \propto a^{-3/2}$ (right panels) around
2 \msun\ (lower panels) and 2.5 \msun\ (upper panels) stars.
Disks around more massive stars produce more luminous debris disks.
\label{fig: f24-astars2}
}
\end{figure}

\begin{figure} 
\includegraphics[width=5.0in]{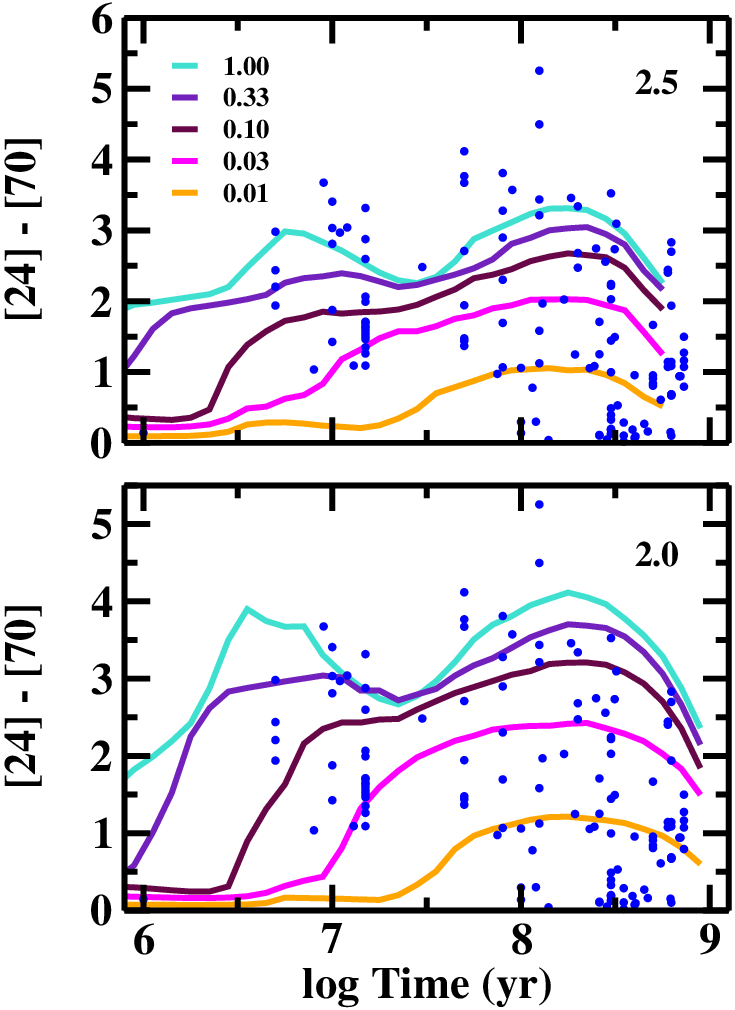}
\vskip 6ex
\caption[f31.eps]{
Observations of the [24]--[70] color excess for A-type stars with known ages
\citep{rie05,su06,cur08}. The colored lines show the predicted evolution 
of the excess for debris disk models around 2 \msun\ stars (lower panel) 
and 2.5 \msun\ stars (upper panel).  The lines plot predictions for disks with 
1 m to 1~km planetesimals, the $f_S$ fragmentation parameters, and 
$\Sigma \propto a^{-1}$ (equation (\ref{eq:sigma-dust})); values for 
$x_m$ are listed in the legend of the upper panel. 
\label{fig: c2470-astars}
}
\end{figure}

\begin{figure} 
\includegraphics[width=6.5in]{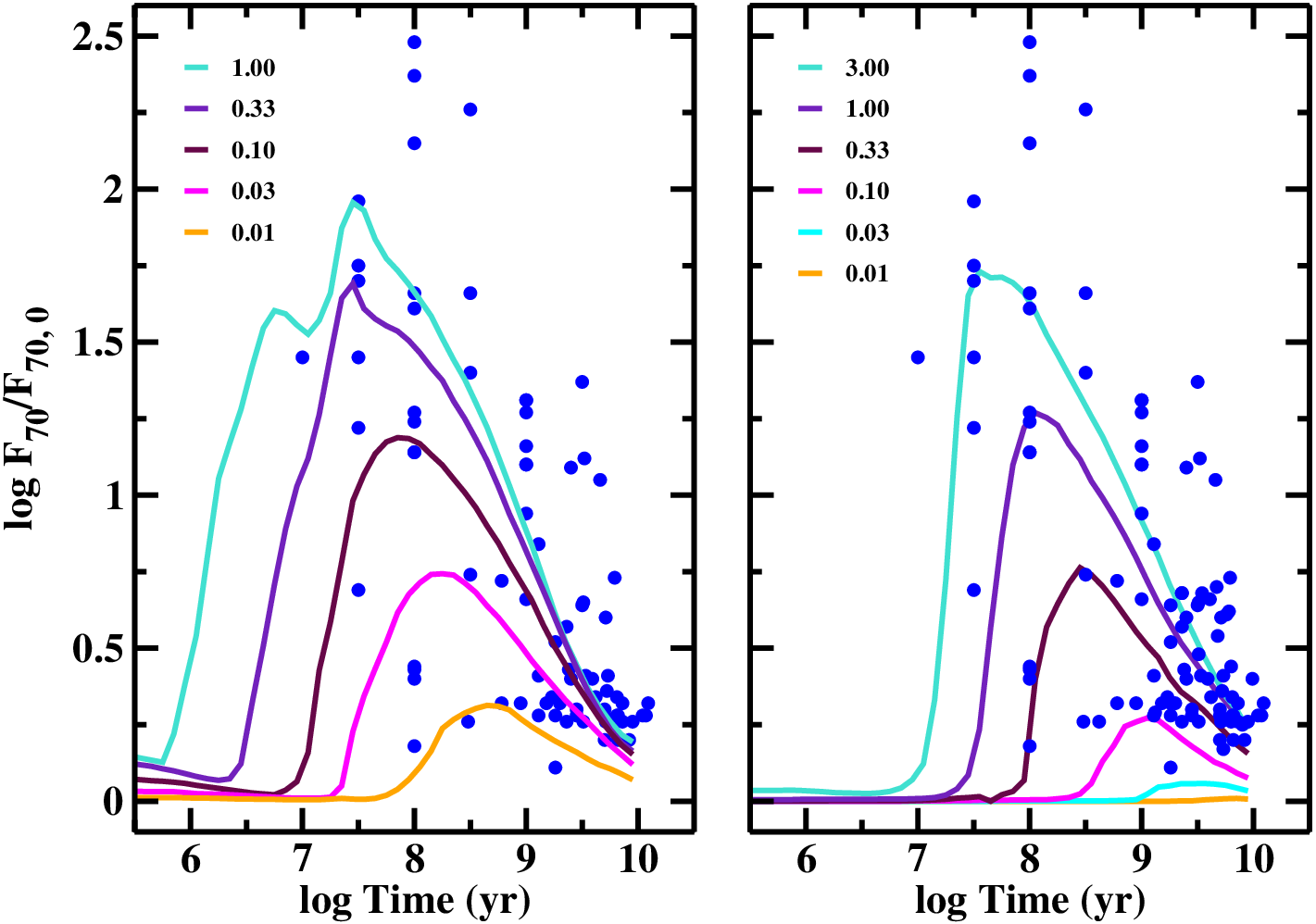}
\vskip 3ex
\caption[f32.eps]{
Observations of the 70 $\mu$m excess for G-type stars with known ages
\citep{bei06,hil08,tri08}. The colored lines show the predicted evolution of 
the excess for debris disk models around 1 \msun\ stars.
In the left panel, the lines plot predictions for disks with 1 m to 
1~km planetesimals, the $f_S$ fragmentation parameters, and 
$\Sigma \propto a^{-1}$ (equation (\ref{eq:sigma-dust})); values for $x_m$ are 
listed in the legend. In the right panel, the lines plot predictions for 
disks with $\Sigma \propto a^{-3/2}$, 1~km planetesimals, and the $f_S$ 
fragmentation parameters. 
\label{fig: f70-gstars}
}
\end{figure}

\begin{figure} 
\includegraphics[width=6.5in]{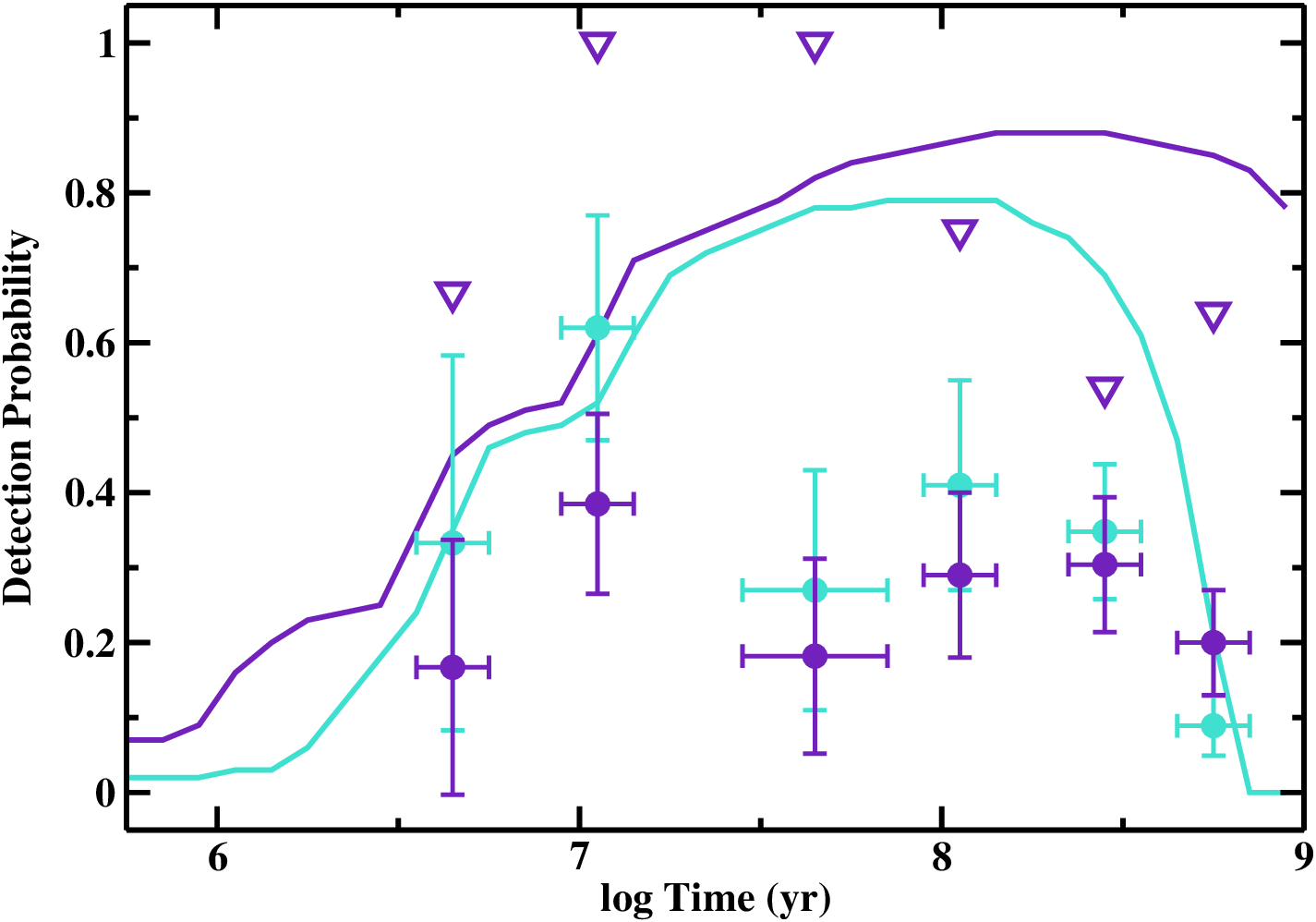}
\vskip 3ex
\caption[f33.eps]{
Detection probability for the excess emission at 24 $\mu$m (turquoise line 
and symbols) and at 70 $\mu$m (indigo line and symbols) from A-type stars 
in the \citet{su06} survey. The lines plot predictions for models with 
$\Sigma \propto a^{-1}$. The solid symbols plot results from \citet{su06};
triangles indicate upper limits at 70 $\mu$m.  The horizontal error bars 
indicate the range in ages; the vertical error bars indicate the 1$\sigma$ 
Poisson error in the observed detection rate.
\label{fig: dprob-a}
}
\end{figure}

\begin{figure} 
\includegraphics[width=6.5in]{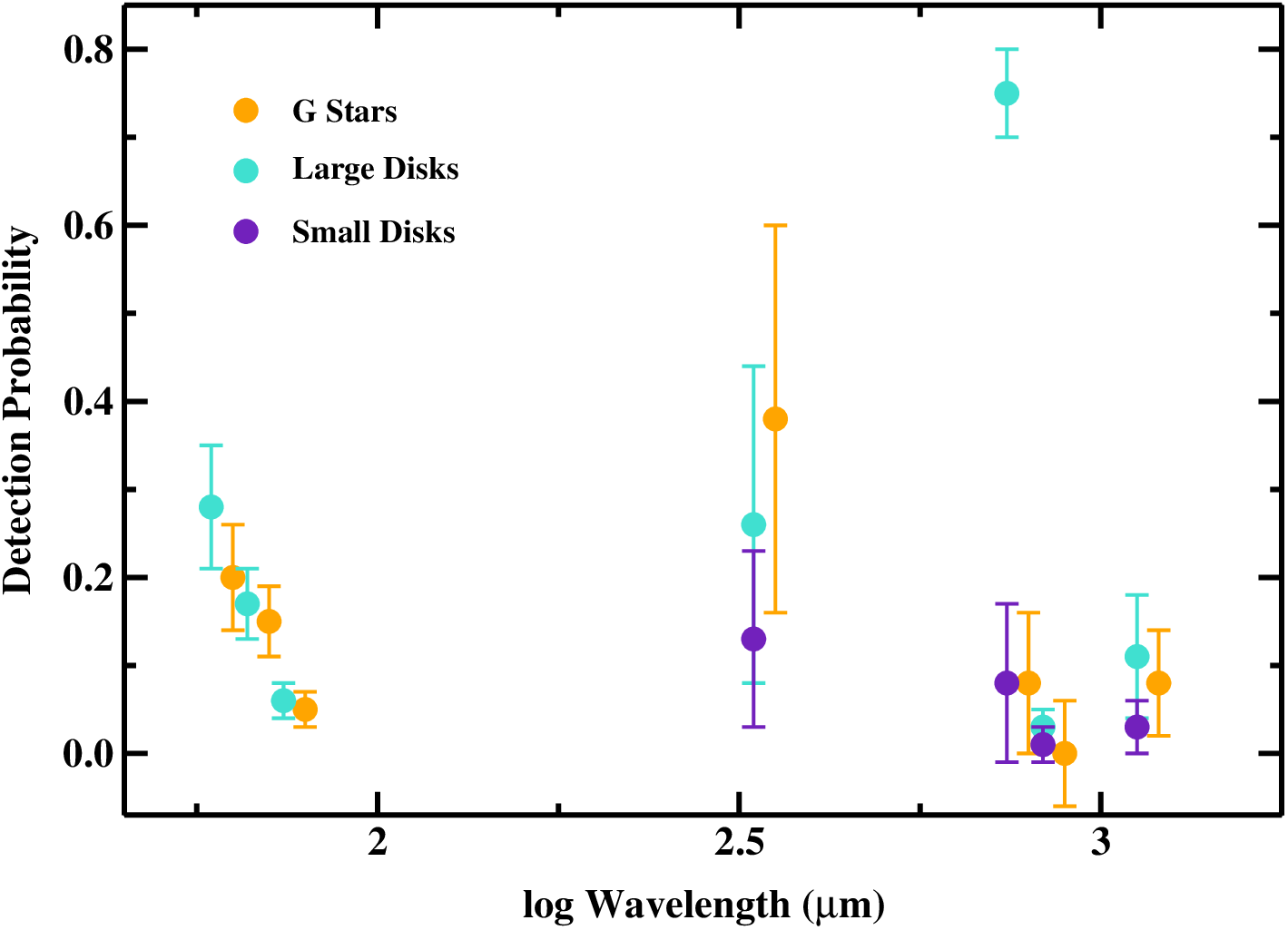}
\vskip 3ex
\caption[f34.eps]{
Comparison of observed and predicted detection probabilities for debris disks 
around G stars at 70--1200 $\mu$m. The filled orange circles with error bars 
show results from recent surveys. The turquoise and indigo symbols plot 
predictions for large disks with outer radii of 150~AU (turquoise) and small
disks with outer radii of 75~AU (indigo). The vertical error bars indicate 
1$\sigma$ Poisson errors in the detection rates. Some points have been 
displaced horizontally for clarity.
\label{fig: dprob-g}
}
\end{figure}

\begin{enumerate}

\item Our calculations explain the observed time evolution of debris disk fluxes 
for solar-type stars (Figure \ref{fig: f70-gstars}). For 
an initial ensemble of 1~km planetesimals, the predicted evolution of the 70 $\mu$m 
excess follows the apparent rise in source fluxes at 10--100~Myr, the peak at $\sim$ 
100--300~Myr, and the decline from $\sim$ 300~Myr to $\sim$ 10~Gyr.  Larger samples
of young G stars \citep[e.g.,][2007b]{cur07a} would provide a better discrimination 
among models with different initial conditions.  Although predicted fluxes for models 
with a smallest stable grain size $r_2 \lesssim$ 1 $\mu$m, a slope of the emissivity 
law for small grains $q \lesssim$ 1, and a slope for the size distribution of small 
grains $p \gtrsim$ 3.5 improve our match to the data, our standard models explain 75\% 
of the observed fluxes. These models also predict an undiscovered population of low 
luminosity debris disks. More sensitive surveys with {\it Herschel} can test this 
prediction and provide better guidance on our adopted values for $r_2$ and $q$.
Better constraints on $r_2$ and $q$ should also yield better guidance on $p$.

\item Our models predict very small 24 $\mu$m excesses from dust at 30--150~AU around 
G-type stars.  Matching the observed levels of 24 $\mu$m emission requires another 
source of dust emission.  Dust in the terrestrial zone is the best candidate. For young
stars, predicted fluxes from terrestrial debris disks typically exceed observed fluxes
\citep[e.g.,][2005]{kb04b}. Although the average flux in model debris disks drops rapidly, 
catastrophic collisions can produce observable levels of debris around stars with ages
exceeding $\sim$ 100 Myr \citep[e.g.,][]{gro01,wya02,kb05,wya07a,smi08}. Resolved observations 
of debris disk emission around G-type stars at 24 $\mu$m would test model predictions for
dust emission from the terrestrial zone and from material at 30--150~AU.

\item Our calculations also match the observed time evolution for debris disks around 
A-type stars (Figures \ref{fig: f24-astars1}--\ref{fig: f24-astars2}). For ensembles of 
planetesimals with initial sizes $\lesssim$ 1~km, our results match the apparent rise 
in fluxes at 5--10~Myr, the plateau at 10--30~Myr, and the decline from $\sim$ 30 Myr 
to $\sim$ 1 Gyr.  The predicted rate of the decline, \ldlstar\ $\propto \lambda^{-n}$ 
with $n \approx$ 0.6--0.8, agrees with the observed rate of decline 
\citep[$n \approx$ 0.5--1;][]{gre03,rie05,rhe07a,wya08}.
Models for disks around 2.5~\msun\ stars match the peak fluxes of A-type stars at 
10--30~Myr; models for 2~\msun\ stars provide a better match to fluxes during the 
decline. Accurate estimates of stellar mass would improve our evaluation of these 
matches. 

\item Predicted colors for debris disks around A-type stars also agree with the 
observations reasonably well (Figure \ref{fig: c2470-astars}). Colors for disks 
with initial surface densities $\Sigma \propto a^{-1}$ match the data somewhat better than 
colors for disks with steeper initial surface density laws. Models around 2~\msun\ stars 
also provide better matches to the data. These models cannot match the colors of the
reddest sources; colors for debris disks with ages of 30--100 Myr can test the models
in more detail. These models yield clear predictions for the fluxes at longer wavelengths 
as a function of the stellar mass and the initial disk surface density relation. Large 
surveys -- such as the proposed JCMT Legacy Survey \citep{mat07}, several Herschel key 
programs, and submm observations with ALMA and SOFIA -- can test these predictions. 

\item Our predictions provide reasonable matches to the observed detection probabilities
of debris disks around A-type and solar-type stars. For solar-type stars, our predictions
for $p_{d,70}$ and $p_{d,850}$ agree with {\it Spitzer} and submm observations. Disks with 
outer radii of 75--100~AU may match the data better than disks with outer radii of 150~AU.
Deeper surveys with {\it Herschel} will provide better constraints on the outer disk radius 
and will test our ability to predict detection probabilities for low luminosity debris disks. 
For young A-type stars, predictions for $p_{d,24}$ and $p_{d,70}$ agree well with observations; 
among older stars, we overpredict the detection rates by a factor of 2--3.  Models with
$r_2 \approx$ 5--10 $\mu$m provide a better match to the data. Calculations which include
the dynamical evolution of giant planets probably also change the detection probability at
late times.  More sensitive observations of A-type stars at 70--160 $\mu$m would enable better 
tests of models with our standard values for $r_2$ and $q$ and would discriminate among models 
with different choices for these two parameters and models with gas giant planets.

\item In their comprehensive analytic study of planetesimal disks, \cite{heng10} show that
emission from long-lived disks of m-sized planetesimals can mimic debris disks produced by
a collisional cascade. Imaging and multi-wavelength spectral energy distributions can
distinguish between these two possibilities. Planetesimals at a distance $a$ (in AU) from 
the central star emit as blackbodies with equilibrium grain temperatures, $T_g \approx$ 278 
$L_\star^{1/4} a^{-1/2}$. Grains in debris disks are hotter than blackbody grains and have 
an emissivity law with $q >$ 0 \citep[e.g.,][]{bac93,heng10}. Recent {\it Spitzer} data suggest
debris disks have small grains \citep[e.g.,][2008, 2009]{su05}. Data from {\it Herschel} will
yield direct measurements of $q$ and enable searches for long-lived planetesimal disks. 

\end{enumerate}

From the broad suite of calculations in KB08 and this paper, we conclude that debris disks 
are the inevitable outcome of icy planet formation in a disk of solid objects. The full set
of models explains many of the general properties of debris disks, including the time evolution
of IR excesses and colors and the frequency of debris disks as a function of stellar mass and 
time. Our results also lead to clear predictions which can be tested with new and existing 
observations.

\acknowledgements
We acknowledge a generous allotment, $\sim$ 10 cpu years, of computer
time on the 1024 cpu Dell Xeon cluster `cosmos' at the Jet Propulsion
Laboratory through funding from the NASA Offices of Mission to Planet
Earth, Aeronautics, and Space Science.  We thank M. Werner for his strong
support of this project.  We also acknowledge use of $\sim$ 2 cpu years
on the CfA cluster `hydra.' Advice and comments from J. Carpenter, C. Chen,
T. Currie, M. Geller, G. Kennedy, M. Meyer, G. Rieke, K. Su, and an anonymous
referee greatly
improved our presentation.  Portions of this project were supported by
the {\it NASA } {\it Astrophysics Theory Program,} through grant
NAG5-13278, the {\it NASA} {\it TPF Foundation Science Program,} through
grant NNG06GH25G, and the {\it Spitzer Guest Observer Program,} through
grant 20132.

\clearpage

\begin{deluxetable}{lcccccc}
\tablecolumns{6}
\tablewidth{0pc}
\tabletypesize{\footnotesize}
\tablecaption{Grid of Initial Disk Masses\tablenotemark{1} (\msun)}
\tablehead{
  \colhead{} &
  \multicolumn{5}{c}{Stellar Mass in $M_{\odot}$} 
\\
  \colhead{$x_m$} &
  \colhead{~~1.0~~} &
  \colhead{~~1.5~~} &
  \colhead{~~2.0~~} &
  \colhead{~~2.5~~} &
  \colhead{~~3.0~~}
}
\startdata
\cutinhead{$\Sigma_s \propto a^{-1}$}
0.01 & 0.003 & 0.004 & 0.005 & 0.006 & 0.008 \\
0.03 & 0.008 & 0.013 & 0.017 & 0.021 & 0.025 \\
0.10 & 0.025 & 0.038 & 0.050 & 0.063 & 0.075 \\
0.33 & 0.083 & 0.125 & 0.167 & 0.208 & 0.250 \\
1.00 & 0.250 & 0.375 & 0.500 & 0.625 & 0.750 \\
\cutinhead{$\Sigma_s \propto a^{-3/2}$}
0.01 & 0.0003 & 0.0004 & 0.0006 & 0.0007 & 0.0009 \\ 
0.03 & 0.001 & 0.0015 & 0.002 & 0.024 & 0.003 \\
0.10 & 0.003 & 0.004 & 0.006 & 0.007 & 0.009 \\
0.33 & 0.010 & 0.015 & 0.019 & 0.024 & 0.029 \\
0.50 & 0.014 & 0.022 & 0.029 & 0.036 & 0.043 \\
1.00 & 0.029 & 0.044 & 0.057 & 0.071 & 0.086 \\
2.00 & 0.057 & 0.088 & 0.114 & 0.142 & 0.172 \\
3.00 & 0.086 & 0.131 & 0.171 & 0.214 & 0.258 \\
\\
$t_{ms}$\tablenotemark{2} & 10.00 & 2.90 & 1.22 & 0.65 & 0.39 \\ 
\enddata
\tablenotetext{1}{Total disk mass at 30--150~AU assuming a gas to dust ratio of 100:1}
\tablenotetext{2}{Main sequence lifetime in Gyr \citep{dem04}}
\label{tab:massgrid}
\end{deluxetable}
\clearpage

\begin{deluxetable}{lccccc}
\tablecolumns{6}
\tablewidth{0pc}
\tabletypesize{\footnotesize}
\tablecaption{Grid of Debris Disk Calculations\tablenotemark{1}}
\tablehead{
  \colhead{} &
  \multicolumn{5}{c}{Stellar Mass in $M_{\odot}$} 
\\
  \colhead{$x_m$} &
  \colhead{~~1.0~~} &
  \colhead{~~1.5~~} &
  \colhead{~~2.0~~} &
  \colhead{~~2.5~~} &
  \colhead{~~3.0~~}
}
\startdata
\cutinhead{$\Sigma_s \propto a^{-1}$, $N(r) \propto r^{-1}$, $r_0$ = 1~km, $f_i = f_S$}
0.01 & 18 & 14 & 24 & 25 & 20 \\
0.03 & 16 & 17 & 15 & 26 & 15 \\
0.10 & 17 & 17 & 14 & 19 & 15 \\
0.33 & 18 & 17 & 17 & 16 & 15 \\
1.00 & 16 & 17 & 13 & 15 & 15 \\
\cutinhead{$\Sigma_s \propto a^{-3/2}$, $N(r) \propto r^{-1}$, $r_0$ = 1~km, $f_i = f_S$}
0.01 & 14 & 37 & 16 & 16 & 18 \\
0.03 & 16 & 44 & 15 & 15 & 21 \\
0.10 & 15 & 16 & 15 & 27 & 21 \\
0.33 & 19 & 15 & 20 & 18 & 18 \\
0.50 & 18 & 18 & 17 & 17 & 17 \\
1.00 & 15 & 22 & 17 & 15 & 15 \\
2.00 & 15 & 30 & 17 & 16 & 15 \\
3.00 & 18 & 12 & 22 & 15 & 21 \\
\cutinhead{$\Sigma_s \propto a^{-3/2}$, $N(r) \propto r^{-1}$, $r_0$ = 1~km, $f_i = f_W$}
0.01 & 7 & 8 & 7 & 7 & 8 \\
0.03 & 7 & 9 & 7 & 7 & 7 \\
0.10 & 7 & 9 & 7 & 10 & 7 \\
0.33 & 7 & 7 & 7 & 9 & 7 \\
1.00 & 14 & 8 & 7 & 9 & 12 \\
3.00 & 13 & 7 & 7 & 9 & 7 \\
\cutinhead{$\Sigma_s \propto a^{-3/2}$, $N(r) \propto r^{0.17}$, $r_0$ = 1~km, $f_i = f_W$}
0.01 & 7 & 7 & 7 & 7 & 8 \\
0.03 & 7 & 7 & 12 & 7 & 7 \\
0.10 & 7 & 7 & 7 & 7 & 7 \\
0.33 & 8 & 7 & 8 & 7 & 7 \\
1.00 & 8 & 7 & 7 & 9 & 7 \\
3.00 & 7 & 7 & 7 & 8 & 7 \\
\cutinhead{$\Sigma_s \propto a^{-3/2}$, $N(r) \propto r^{0.17}$, $r_0$ = 10~km, $f_i = f_W$}
0.01 & 14 & 9  & 7  & 7  & 7  \\
0.03 & 7 & 8  & 7  & 11  & 7  \\
0.10 & 12 & 10  & 7  & 11  & 8  \\
0.33 & 14 & 7  & 7  & 7  & 9  \\
1.00 & 15 & 8  & 9  & 8  & 8  \\
3.00 & 7 & 7  & 7  & 7  & 8  \\
\cutinhead{$\Sigma_s \propto a^{-3/2}$, $N(r) \propto r^{0.17}$, $r_0$ = 100~km, $f_i = f_W$}
0.01 & 7  & 8  & 7  & 8  & 7  \\
0.03 & 7  & 9  & 8  & 8  & 7  \\
0.10 & 7  & 7  & 12  & 7  & 7  \\
0.33 & 9 & 7  & 10  & 9  & 10  \\
1.00 & 7 & 7  & 10  & 10  & 7  \\
3.00 & 7 & 7  & 7  & 7  & 7  \\
\enddata
\tablenotetext{1}{Number of independent calculations for each 
combination of $M_{\star},x_m$}
\label{tab:modgrid}
\end{deluxetable}
\clearpage

\begin{deluxetable}{lcccccc}
\tablecolumns{6}
\tablewidth{0pc}
%\tabletypesize{\small}
\tablecaption{Scaling Relations for Planet Formation\tablenotemark{1,2}}
\tablehead{
  \colhead{Parameter} & 
  \colhead{(1)} &
  \colhead{(2)} &
  \colhead{(3)} &
  \colhead{Notes}
}
\startdata
$\Sigma$ (g cm$^{-2}$) & 30 $x_m ~  M_{\star,2} ~ a^{-1}$ & 30 $x_m ~  M_{\star,2} ~ a^{-3/2}$ & 30 $x_m ~ M_{\star,2} ~ a^{-3/2}$ & (3) \\
\\
$f_i$ & $f_S$ & $f_S$ & $f_W$ & \\
\\
$t_{1000}$ (Myr) & 10 $x_m^{-1.1} ~ M_{\star,2}^{-3/2} ~ a_{80}^{2.5}$ &
145 $x_m^{-1.1} ~ M_{\star,2}^{-3/2} ~ a_{80}^{3}$ &
400 $x_m^{-1.1} ~ M_{\star,2}^{-3/2} ~ a_{80}^{3}$ & (4) \\
\\
$r_{max}$ (km) & 3500 $x_m^{0.2}$ & 2000 $x_m^{0.25}$ & 1250 $x_m^{0.25}$ & \\
\\
$\dot{M}_{max}$ (g yr$^{-1}$) & $1.7 \times 10^{23} x_m^2 M_{\star,2}^{5/2}$ & $3.5 \times 10^{21} x_m^2 M_{\star,2}^{5/2}$ & $7.7 \times 10^{21} x_m^2 M_{\star,2}^{5/2}$ \\
\\
$t_{\dot{M}_{max}}$ (Myr) & 2 $x_m^{-1} M_{\star,2}^{-3/2}$ & 5 $x_m^{-1} M_{\star,2}^{-3/2}$ & 4.5 $x_m^{-1} M_{\star,2}^{-3/2}$ & \\
\\
$M_{max,small}$ (\mearth) & 0.33 $x_m M_{\star,2}$ & 0.026 $x_m M_{\star,2}$ & 0.038 $x_m M_{\star,2}$ & \\
\\
$M_{max,large}$ (\mearth) & 11 $x_m M_{\star,2}$ & 1.0 $x_m M_{\star,2}$ & 1.25 $x_m M_{\star,2}$ & \\
\enddata
\tablenotetext{1}{Results for calculations with a range of planetesimal sizes, 1 m to 1~km}
\tablenotetext{2}{Inter-quartile ranges for the coefficients are 15\%--20\% for $t_{1000}$,
10\% for $r_{max}$, 5\%--10\% for $\dot{M}_{max}$ and $t_{\dot{M}_{max}}$, and 10\% for 
$M_{max,small}$ and $M_{max,large}$.}
\tablenotetext{3}{$M_{\star,2} = M_{\star} / 2~M_\odot$}
\tablenotetext{4}{$a_{80} = a / {\rm 80~AU}$ }
\label{tab:eqplanet1}
\end{deluxetable}
\clearpage

\begin{deluxetable}{lcccccc}
\tablecolumns{6}
\tablewidth{0pc}
%\tabletypesize{\small}
\tablecaption{Scaling Relations for Planet Formation\tablenotemark{1, 2}}
\tablehead{
  \colhead{Parameter} & 
  \colhead{1~km only} &
  \colhead{10~km only} &
  \colhead{100~km only} &
  \colhead{Notes}
}
\startdata
$t_{1000}$ (Myr) & 325 $x_m^{-1.05} ~ M_{\star,2}^{-3/2} ~ a_{80}^{3.3}$ &
1200 $x_m^{-1.05} ~ M_{\star,2}^{-3/2} ~ a_{80}^{3.3}$ &
7500 $x_m^{-1} ~ M_{\star,2}^{-3/2} ~ a_{80}^{3.5}$ & (3) \\
\\
$r_{max}$ (km) & 3000 $x_m^{0.30}$ & 4000 $x_m^{0.25}$ & 6000 $x_m^{0.25}$ & \\
\\
$\dot{M}_{max}$ (g yr$^{-1}$) & $10^{21} x_m^2 M_{\star,2}^{5/2}$ & $2 \times 10^{20} x_m^2 M_{\star,2}^{5/2}$ & 
$2 \times 10^{19} x_m^2 M_{\star,2}^{5/2}$ & (4) \\
\\
$t_{\dot{M}_{max}}$ (Myr) & 45 $x_m^{-1} M_{\star,2}^{-3/2}$ & 275 $x_m^{-1} M_{\star,2}^{-3/2}$ & 2000 $x_m^{-1} M_{\star,2}^{-3/2}$ (3) \\
\\ 
$M_{max,small}$ (\mearth) & 0.017 $x_m M_{\star,2}$ & 0.0037 $x_m M_{\star,2}^{1/2}$ & $5 \times 10^{-4} x_m$ & \\
\\
$M_{max,large}$ (\mearth) & 0.56 $x_m M_{\star,2}$ & 0.20 $x_m M_{\star,2}$ & 0.033 $x_m M_{\star,2}^{1/2}$ & \\
\enddata
\tablenotetext{1}{Results for calculations with one initial planetesimal size,
the weak fragmentation parameters, and $\Sigma = 30 ~ x_m ~  M_{\star,2} ~ a^{-3/2}$}
\tablenotetext{2}{Inter-quartile ranges for the coefficients are 15\% for $t_{1000}$,
10\% for $r_{max}$, 5\%--10\% for $\dot{M}_{max}$ and $t_{\dot{M}_{max}}$, and 10\% for 
$M_{max,small}$ and $M_{max,large}$.}
\tablenotetext{3}{Relation for 100~km only calculations valid only for 1~\msun\ stars}
\tablenotetext{4}{Relation for 100~km only calculations valid only for $x_m$ = 1--3}
\label{tab:eqplanet2}
\end{deluxetable}
\clearpage

\begin{deluxetable}{lccccccccc}
\tablecolumns{10}
\tablewidth{0pc}
\tabletypesize{\scriptsize}
\tablecaption{Median number of Plutos at t = $t_{ms}$/3 for disks around 1--3 \msun\ stars\tablenotemark{a, b}}
\tablehead{
  \colhead{$M_\star$} &
  \colhead{$x_m$} &
  \colhead{30--37~AU} &
  \colhead{37--45~AU} &
  \colhead{45--55~AU} &
  \colhead{55--67~AU} &
  \colhead{67--82~AU} &
  \colhead{82--100~AU} &
  \colhead{100--123~AU} &
  \colhead{123--146~AU}
}
\startdata
1.0 & 0.01 &   5 &  5 &  5 &  1 &  0 &  0 &  0 &  0 \\
    & 0.03 &  25 & 30 & 22 & 26 & 23 & 13 &  2 &  0 \\
    & 0.10 &  69 & 72 & 59 & 66 & 50 & 55 & 31 & 14 \\
    & 0.33 & 137 &224 &174 &216 &201 &115 & 97 & 58 \\
    & 1.00 & 220 &363 &367 &555 &465 &490 &415 &165 \\
\\
1.5 & 0.01 &  3 &  4 &  7 &  0 &  5 &  0&   0 &  0 \\
    & 0.03 & 17 & 19 & 14 &  3 &  0 &  0&   0 &  0 \\
    & 0.10 & 44 & 39 & 27 & 21 & 26 &  8&   0 &  0 \\
    & 0.33 & 98 & 53 & 63 &136 & 57 & 26&  11 &  6 \\
    & 1.00 &390 &413 &539 &668 &957 &877&1253 &279 \\
\\
2.0 & 0.01 &  0 &  0 &  0&   0 &  0 &  0&   0 &  0 \\
    & 0.03 & 62 & 48 & 31&  23 &  5 &  0&   0 &  0 \\
    & 0.10 &155 &151 &134& 137 &103 & 68&   6 &  0 \\
    & 0.33 &247 &274 &301& 379 &335 &287& 220 & 59 \\
    & 1.00 &358 &257 &591&1264 &872 &822&1261 &476 \\
\\
2.5 & 0.01 &  3 &  1 &  0 &  0&   0&   0&   0 &  0 \\
    & 0.03 & 74 & 66 & 46 & 33&  14&   0&   0 &  0 \\
    & 0.10 &146 &194 &177 &243& 171& 130&  73 &  0 \\
    & 0.33 &224 &348 &378 &478& 543& 500& 288 &126 \\
    & 1.00 &443 &743 &827 &763&1142&1083&1218 &219 \\
\\
3.0 & 0.01 &   1 &  0&   0 &  0&   0&   0&   0 &  0 \\
    & 0.03 &  59 & 89&  59&  28&   5&   0&   0 &  0 \\
    & 0.10 & 142 &250& 194& 196& 240& 104&  38 &  0 \\
    & 0.33 & 207 &459& 431& 433& 605& 760& 523 &214 \\
    & 1.00 & 543 &685&1039&1270&1625&2046&1428 &862 \\
\enddata
\label{tab:rad1000.1}
\tablenotetext{a}{For calculations with $\Sigma \propto a^{-1}$,
planetesimals with initial radii of 1 m to 1~km, and the $f_S$ 
fragmentation parameters}
\tablenotetext{b}{In Tables 5--10, the typical inter-quartile 
range in the number of Plutos ($N$) is roughly $\sqrt{N}$.}
\end{deluxetable}

\begin{deluxetable}{lccccccccc}
\tablecolumns{10}
\tablewidth{0pc}
\tabletypesize{\scriptsize}
\tablecaption{Median number of Plutos at t = $t_{ms}$/3 for disks around 1--3 \msun\ stars\tablenotemark{a}}
\tablehead{
  \colhead{\mstar} &
  \colhead{$x_m$} &
  \colhead{30--37~AU} &
  \colhead{37--45~AU} &
  \colhead{45--55~AU} &
  \colhead{55--67~AU} &
  \colhead{67--82~AU} &
  \colhead{82--100~AU} &
  \colhead{100--123~AU} &
  \colhead{123--146~AU}
}
\startdata
1.0 & 0.01 &   0 &  0 &  0 &  0 &  0 &  0 &  0 &  0 \\
    & 0.03 &   0 &  0 &  0 &  0 &  0 &  0 &  0 &  0 \\
    & 0.10 &   1 &  0 &  1 &  0 &  0 &  0 &  0 &  0 \\
    & 0.33 &   8 & 11 & 15 &  6 &  5 &  5 &  1 &  0 \\
    & 1.00 &  48 & 55 & 32 & 47 & 42 & 20 & 19 &  5 \\
    & 3.00 & 100 &187 &115 &228 & 95 &103 & 92 & 23 \\
\\
1.5 & 0.01 &   0 &  0 &  0 &  0 &  0 &  0 &  0 &  0 \\
    & 0.03 &   0 &  0 &  0 &  0 &  0 &  0 &  0 &  0 \\
    & 0.10 &   0 &  0 &  0 &  0 &  0 &  0 &  0 &  0 \\
    & 0.33 &  16 & 11 & 15 & 15 &  5 &  0 &  0 &  0 \\
    & 1.00 &  57 & 82 & 71 & 55 & 43 & 30 & 17 &  2 \\
    & 3.00 & 137 &152 &202 &167 &264 &143 & 69 & 21 \\
\\
2.0 & 0.01 &   0 &  0 &  0 &  0 &  0 &  0 &  0 &  0 \\
    & 0.03 &   0 &  0 &  0 &  0 &  0 &  0 &  0 &  0 \\
    & 0.10 &   0 &  0 &  0 &  0 &  0 &  0 &  0 &  0 \\
    & 0.33 &  27 & 15 &  2 &  1 &  0 &  0 &  0 &  0 \\
    & 1.00 &  47 & 95 & 48 & 46 & 35 &  5 &  0 &  0 \\
    & 3.00 & 118 &212 &202 &219 &130 &152 & 51 &  6 \\
\\
2.5 & 0.01 &   0 &  0 &  0 &  0 &  0 &  0 &  0 &  0 \\
    & 0.03 &   0 &  0 &  0 &  0 &  0 &  0 &  0 &  0 \\
    & 0.10 &   0 &  0 &  0 &  0 &  0 &  0 &  0 &  0 \\
    & 0.33 &  21 & 20 &  7 &  5 &  1 &  0 &  0 &  0 \\
    & 1.00 &  90 & 99 & 87 & 93 & 61 & 17 &  3 &  0 \\
    & 3.00 & 153 &218 &225 &247 &179 &204 &159 & 28 \\
\\
3.0 & 0.01 &   0 &  0 &  0 &  0 &  0 &  0 &  0 &  0 \\
    & 0.03 &   0 &  0 &  0 &  0 &  0 &  0 &  0 &  0 \\
    & 0.10 &   0 &  0 &  0 &  0 &  0 &  0 &  0 &  0 \\
    & 0.33 &  33 & 23 &  5 &  1 &  0 &  0 &  0 &  0 \\
    & 1.00 & 105 &104 &100 & 79 & 59 & 15 &  0 &  0 \\
    & 3.00 & 233 &179 &362 &311 &266 &159 &158 &  0 \\
\enddata
\label{tab:rad1000.2}
\tablenotetext{a}{For calculations with $\Sigma \propto a^{-3/2}$,
planetesimals with initial radii of 1 m to 1~km, and the $f_W$ 
fragmentation parameters}
\end{deluxetable}

\begin{deluxetable}{lccccccccc}
\tablecolumns{10}
\tablewidth{0pc}
\tabletypesize{\scriptsize}
\tablecaption{Median number of Plutos at t = $t_{ms}$/3 for disks around 1--3 \msun\ stars\tablenotemark{a}}
\tablehead{
  \colhead{\mstar} &
  \colhead{$x_m$} &
  \colhead{30--37~AU} &
  \colhead{37--45~AU} &
  \colhead{45--55~AU} &
  \colhead{55--67~AU} &
  \colhead{67--82~AU} &
  \colhead{82--100~AU} &
  \colhead{100--123~AU} &
  \colhead{123--146~AU}
}
\startdata
1.0 & 0.01 &  0 &  0 &  0 &  0 &  0 &  0 &  0 &  0 \\
    & 0.03 &  8 &  0 &  0 &  0 &  0 &  0 &  0 &  0 \\
    & 0.10 & 11 & 10 &  8 &  2 &  0 &  0 &  0 &  0 \\
\\
1.5 & 0.01 &  0 &  0 &  0 &  0 &  0 &  0 &  0 &  0 \\
    & 0.03 &  0 &  0 &  0 &  0 &  0 &  0 &  0 &  0 \\
    & 0.10 & 15 & 14 &  7 &  1 &  0 &  0 &  0 &  0 \\
\\
2.0 & 0.01 & 0 &  0 &  0 &  0 &  0 &  0 &  0 &  0 \\
    & 0.03 & 0 &  0 &  0 &  0 &  0 &  0 &  0 &  0 \\
    & 0.10 & 7 &  3 &  0 &  0 &  0 &  0 &  0 &  0 \\
\\ 
2.5 & 0.01 &  0 &  0 &  0 &  0 &  0 &  0 &  0 &  0 \\
    & 0.03 &  0 &  0 &  0 &  0 &  0 &  0 &  0 &  0 \\
    & 0.10 & 16 & 11 &  2 &  0 &  0 &  0 &  0 &  0 \\
\\
3.0 & 0.01 & 0  & 0 &  0 &  0 &  0 &  0 &  0 &  0 \\
    & 0.03 & 0  & 0 &  0 &  0 &  0 &  0 &  0 &  0 \\
    & 0.10 & 7  & 5 &  3 &  0 &  0 &  0 &  0 &  0 \\
\enddata
\label{tab:rad1000.3}
\tablenotetext{a}{For calculations with $\Sigma \propto a^{-3/2}$,
planetesimals with initial radii of 1 m to 1~km, and the $f_S$ 
fragmentation parameters}
\end{deluxetable}

\begin{deluxetable}{lccccccccc}
\tablecolumns{10}
\tablewidth{0pc}
\tabletypesize{\scriptsize}
\tablecaption{Median number of Plutos at t = $t_{ms}$/3 for disks around 1--3 \msun\ stars\tablenotemark{a}}
\tablehead{
  \colhead{\mstar} &
  \colhead{$x_m$} &
  \colhead{30--37~AU} &
  \colhead{37--45~AU} &
  \colhead{45--55~AU} &
  \colhead{55--67~AU} &
  \colhead{67--82~AU} &
  \colhead{82--100~AU} &
  \colhead{100--123~AU} &
  \colhead{123--146~AU}
}
\startdata
1.0 & 0.01 & 0  & 0  & 0  & 0  & 0  & 0  & 0   &0 \\
    & 0.03 & 7  & 0  & 0  & 0  & 0  & 0  & 0   &0 \\
    & 0.10 & 11 & 10 &  3 &  0 &  0 &  0 &  0  & 0 \\
    & 0.33 & 13 &  8 & 13 &  5 &  6 &  0 &  0  & 0 \\
    & 1.00 & 32 & 13 & 17 &  5 &  8 &  8 &  1  & 0 \\
    & 3.00 & 77 & 51 & 42 & 26 & 17 &  6 &  8  & 0 \\
\\
1.5 & 0.01 &   0 &  0 &  0 &  0 &  0  & 0 &  0 &  0 \\
    & 0.03 &   1 &  0 &  0 &  0 &  0  & 0 &  0 &  0 \\
    & 0.10 &  10 &  5 &  2 &  0 &  0  & 0 &  0 &  0 \\
    & 0.33 &  23 & 26 & 10 & 10 &  0  & 0 &  0 &  0 \\
    & 1.00 &  66 & 38 & 25 & 19 & 19  & 5 &  0 &  0 \\
    & 3.00 & 134 &133 &118 & 69 & 46  &17 & 10 &  0 \\
\\
2.0 & 0.01 &   0 &  0 &  0 &  0 &  0 &  0 &  0 &  0 \\
    & 0.03 &   0 &  0 &  0 &  0 &  0 &  0 &  0 &  0 \\
    & 0.10 &  10 &  0 &  0 &  0 &  0 &  0 &  0 &  0 \\
    & 0.33 &  34 & 25 &  6 &  3 &  0 &  0 &  0 &  0 \\
    & 1.00 &  92 & 50 & 23 & 27 &  7 &  0 &  0 &  0 \\
    & 3.00 & 243 &219 &138 &112 &100 & 33 &  1 &  0 \\
\\
2.5 & 0.01 &   0 &  0 &  0 &  0  & 0 &  0 &  0 &  0 \\
    & 0.03 &   0 &  0 &  0 &  0  & 0 &  0 &  0 &  0 \\
    & 0.10 &  17 &  3 &  0 &  0  & 0 &  0 &  0 &  0 \\
    & 0.33 &  30 & 33 & 29 &  5  & 0 &  0 &  0 &  0 \\
    & 1.00 & 119 &127 & 77 & 27  & 6 &  2 &  0 &  0 \\
    & 3.00 & 378 &243 &378 &107  &71 & 57 & 11 &  0 \\
\\
3.0 & 0.01 &  0 &  0 &  0 &  0 &  0 &  0  & 0 &  0 \\
    & 0.03 &  0 &  0 &  0 &  0 &  0 &  0  & 0 &  0 \\
    & 0.10 & 15 &  2 &  0 &  0 &  0 &  0  & 0 &  0 \\
    & 0.33 & 35 & 51 & 20 &  0 &  0 &  0  & 0 &  0 \\
    & 1.00 &152 &177 &109 & 80 & 20 &  0  & 0 &  0 \\
    & 3.00 &233 &553 &232 &202 &186 & 48  & 5 &  0 \\
\enddata
\label{tab:rad1000.4}
\tablenotetext{a}{For calculations with $\Sigma \propto a^{-3/2}$,
planetesimals with initial radii of 1~km only, and the $f_W$ 
fragmentation parameters}
\end{deluxetable}

\begin{deluxetable}{lccccccccc}
\tablecolumns{10}
\tablewidth{0pc}
\tabletypesize{\scriptsize}
\tablecaption{Median number of Plutos at t = $t_{ms}$/3 for disks around 1--3 \msun\ stars\tablenotemark{a}}
\tablehead{
  \colhead{\mstar} &
  \colhead{$x_m$} &
  \colhead{30--37~AU} &
  \colhead{37--45~AU} &
  \colhead{45--55~AU} &
  \colhead{55--67~AU} &
  \colhead{67--82~AU} &
  \colhead{82--100~AU} &
  \colhead{100--123~AU} &
  \colhead{123--146~AU}
}
\startdata
1.0 & 0.01 &  0 &  0 &  0 &  0 &  0 &  0 &  0 &  0 \\
    & 0.03 &  0 &  0 &  0 &  0 &  0 &  0 &  0 &  0 \\
    & 0.10 &  8 &  1 &  0 &  0 &  0 &  0 &  0 &  0 \\
    & 0.33 & 16 & 13 &  6 &  0 &  0 &  0 &  0 &  0 \\
    & 1.00 & 35 & 29 & 42 & 17 &  1 &  0 &  0 &  0 \\
    & 3.00 & 54 & 16 & 26 & 29 &  3 &  0 &  0 &  0 \\
\\
1.5 & 0.01 & 0  & 0  & 0  & 0  & 0  & 0  & 0  & 0 \\
    & 0.03 & 0  & 0  & 0  & 0  & 0  & 0  & 0  & 0 \\
    & 0.10 & 3  & 0  & 0  & 0  & 0  & 0  & 0  & 0 \\
    & 0.33 & 23 & 13 &  8 &  0 &  0 &  0 &  0 &  0 \\
    & 1.00 & 59 & 49 & 32 & 14 &  0 &  0 &  0 &  0 \\
    & 3.00 & 97 &139 &158 & 66 &174 &  0 &  0 &  0 \\
\\
2.0 & 0.01 &   0 &  0 &  0  & 0  & 0 &  0 &  0 &  0 \\
    & 0.03 &   0 &  0 &  0  & 0  & 0 &  0 &  0 &  0 \\
    & 0.10 &   0 &  0 &  0  & 0  & 0 &  0 &  0 &  0 \\
    & 0.33 &  14 &  1 &  0  & 0  & 0 &  0 &  0 &  0 \\
    & 1.00 &  35 & 34 &  8  & 0  & 0 &  0 &  0 &  0 \\
    & 3.00 & 251 &277 &177  &66  & 0 &  0 &  0 &  0 \\
\\
2.5 & 0.01 &   0 &  0 &  0 &  0 &  0 &  0 &  0 &  0 \\
    & 0.03 &   0 &  0 &  0 &  0 &  0 &  0 &  0 &  0 \\
    & 0.10 &   0 &  0 &  0 &  0 &  0 &  0 &  0 &  0 \\
    & 0.33 &  19 &  8 &  0 &  0 &  0 &  0 &  0 &  0 \\
    & 1.00 &  68 & 93 & 30 &  1 &  0 &  0 &  0 &  0 \\
    & 3.00 & 177 & 66 &100 &317 & 31 &  0 &  0 &  0 \\
\\
3.0 & 0.01 &   0 &  0 &  0 &  0 &  0 &  0 &  0 &  0 \\
    & 0.03 &   0 &  0 &  0 &  0 &  0 &  0 &  0 &  0 \\
    & 0.10 &   0 &  0 &  0 &  0 &  0 &  0 &  0 &  0 \\
    & 0.33 &  27 &  6 &  0 &  0 &  0 &  0 &  0 &  0 \\
    & 1.00 &  54 & 88 & 36 &  1 &  0 &  0 &  0 &  0 \\
    & 3.00 & 305 &396 &252 &225 & 20 &  0 &  0 &  0 \\
\enddata
\label{tab:rad1000.5}
\tablenotetext{a}{For calculations with $\Sigma \propto a^{-3/2}$,
planetesimals with initial radii of 10~km only, and the $f_W$ 
fragmentation parameters}
\end{deluxetable}

\begin{deluxetable}{lccccccccc}
\tablecolumns{10}
\tablewidth{0pc}
\tabletypesize{\scriptsize}
\tablecaption{Median number of Plutos at t = $t_{ms}$/3 for disks around 1--3 \msun\ stars\tablenotemark{a}}
\tablehead{
  \colhead{\mstar} &
  \colhead{$x_m$} &
  \colhead{30--37~AU} &
  \colhead{37--45~AU} &
  \colhead{45--55~AU} &
  \colhead{55--67~AU} &
  \colhead{67--82~AU} &
  \colhead{82--100~AU} &
  \colhead{100--123~AU} &
  \colhead{123--146~AU}
}
\startdata
1.0 & 0.01 &   0 &  0 &  0 &  0 &  0 &  0 &  0&   0 \\
    & 0.03 &   0 &  0 &  0 &  0 &  0 &  0 &  0&   0 \\
    & 0.10 &   0 &  0 &  0 &  0 &  0 &  0 &  0&   0 \\
    & 0.33 &   8 &  0 &  0 &  0 &  0 &  0 &  0&   0 \\
    & 1.00 & 324 &177 & 23 &  0 &  0 &  0 &  0&   0 \\
    & 3.00 & 748 &796 &824 &613 &231 &  7 &  0&   0 \\
\\
1.5 & 0.01 & 0   &0   &0   &0   &0   &0   &0   &0 \\
    & 0.03 & 0   &0   &0  & 0   &0   &0   &0   &0 \\
    & 0.10 & 0   &0   &0  & 0   &0   &0   &0   &0 \\
    & 0.33 & 0   &0   &0  & 0   &0   &0   &0   &0 \\
    & 1.00 & 363 &125 &  7&   0 &  0 &  0 &  0 &  0 \\
    & 3.00 & 1083&1235& 993& 468&  14&   0&   0&   0 \\

\\
2.0 & 0.01 &    0&   0 &  0 &  0 &  0 &  0 &  0 &  0 \\
    & 0.03 &    0&   0 &  0 &  0 &  0 &  0 &  0 &  0 \\
    & 0.10 &    0&   0 &  0 &  0 &  0 &  0 &  0 &  0 \\
    & 0.33 &    0&   0 &  0 &  0 &  0 &  0 &  0 &  0 \\
    & 1.00 &   83&   0 &  0 &  0 &  0 &  0 &  0 &  0 \\
    & 3.00 & 1288&1076 &255 &  1 &  0 &  0 &  0 &  0 \\
\\
2.5 & 0.01 &   0&   0 &  0 &  0 &  0 &  0  & 0 &  0 \\
    & 0.03 &   0&   0 &  0 &  0 &  0 &  0  & 0 &  0 \\
    & 0.10 &   0&   0 &  0 &  0 &  0 &  0  & 0 &  0 \\
    & 0.33 &   0&   0 &  0 &  0 &  0 &  0  & 0 &  0 \\
    & 1.00 & 296&  26 &  0 &  0 &  0 &  0  & 0 &  0 \\
    & 3.00 &1729&1629 &870 & 51 &  0 &  0  & 0 &  0 \\

\\
3.0 & 0.01 &    0&   0 &  0 &  0 &  0 &  0 &  0 &  0 \\
    & 0.03 &    0&   0 &  0 &  0 &  0 &  0 &  0 &  0 \\
    & 0.10 &    0&   0 &  0 &  0 &  0 &  0 &  0 &  0 \\
    & 0.33 &    0&   0 &  0 &  0 &  0 &  0 &  0 &  0 \\
    & 1.00 &  381&  57 &  0 &  0 &  0 &  0 &  0 &  0 \\
    & 3.00 & 1981&1741 &790 & 19 &  0 &  0 &  0 &  0 \\
\enddata
\label{tab:rad1000.6}
\tablenotetext{a}{For calculations with $\Sigma \propto a^{-3/2}$,
planetesimals with initial radii of 100~km only, and the $f_W$ 
fragmentation parameters}
\end{deluxetable}

\begin{deluxetable}{lccccc}
\tablecolumns{6}
\tablewidth{0pc}
\tabletypesize{\footnotesize}
\tablecaption{Predicted Excesses for Disks Around 1 \msun\ Stars\tablenotemark{a, b, c}}
\tablehead{
  \colhead{log $t$ (yr)} &
  \colhead{log $L_d/L_{\star}$} &
  \colhead{log $F_{24}/F_{24,0}$} &
  \colhead{log $F_{70}/F_{70,0}$} &
  \colhead{log $F_{160}/F_{160,0}$} &
  \colhead{log $F_{850}/F_{850,0}$}
}
\startdata
\cutinhead{$x_m$ = 0.01}
5.05 & -5.20 & 0.000 & 0.012 & 0.039 & 0.024 \\
5.15 & -5.21 & 0.000 & 0.012 & 0.039 & 0.024 \\
5.25 & -5.21 & 0.000 & 0.012 & 0.039 & 0.024 \\
5.35 & -5.22 & 0.000 & 0.012 & 0.039 & 0.024 \\
5.45 & -5.22 & 0.000 & 0.012 & 0.038 & 0.024 \\
\cutinhead{$x_m$ = 0.10}
5.05 & -4.28 & 0.000 & 0.077 & 0.222 & 0.154 \\
5.15 & -4.32 & 0.000 & 0.077 & 0.222 & 0.154 \\
5.25 & -4.35 & 0.000 & 0.077 & 0.222 & 0.154 \\
5.35 & -4.39 & 0.000 & 0.077 & 0.221 & 0.153 \\
5.45 & -4.40 & 0.000 & 0.073 & 0.214 & 0.148 \\
\cutinhead{$x_m$ = 1.00}
5.05 & -3.98 & 0.001 & 0.166 & 0.407 & 0.301 \\
5.15 & -4.00 & 0.001 & 0.159 & 0.388 & 0.285 \\
5.25 & -4.02 & 0.001 & 0.157 & 0.383 & 0.279 \\
5.35 & -4.03 & 0.001 & 0.155 & 0.378 & 0.273 \\
5.45 & -4.07 & 0.001 & 0.149 & 0.360 & 0.257 \\
\enddata
\label{tab:mod-1p0-a1}
\tablenotetext{a}{Table \ref{tab:mod-1p0-a1} is published in its entirety in the
electronic version of the {\it Astrophysical Journal Supplement.} A portion is shown here
for guidance regarding its form and content.}
\tablenotetext{b}{For calculations with $\Sigma \propto a^{-1}$,
planetesimals with initial radii of 1 m to 1~km only, and the 
$f_S$ fragmentation parameters}
\tablenotetext{c}{The typical inter-quartile range in the predicted fluxes
is $\pm$5\% to 10\% at 24 $\mu$m and $\pm$5\% at 70--850 $\mu$m.}
\end{deluxetable}
\clearpage

\begin{deluxetable}{lccccc}
\tablecolumns{6}
\tablewidth{0pc}
\tabletypesize{\footnotesize}
\tablecaption{Predicted Excesses for Disks Around 1.5 \msun\ Stars\tablenotemark{a, b, c}}
\tablehead{
  \colhead{log $t$ (yr)} &
  \colhead{log $L_d/L_{\star}$} &
  \colhead{log $F_{24}/F_{24,0}$} &
  \colhead{log $F_{70}/F_{70,0}$} &
  \colhead{log $F_{160}/F_{160,0}$} &
  \colhead{log $F_{850}/F_{850,0}$}
}
\startdata
\cutinhead{$x_m$ = 0.01}
5.05 & -5.03 & 0.000 & 0.023 & 0.054 & 0.028 \\
5.15 & -5.04 & 0.000 & 0.023 & 0.054 & 0.028 \\
5.25 & -5.05 & 0.000 & 0.023 & 0.054 & 0.028 \\
5.35 & -5.06 & 0.000 & 0.023 & 0.054 & 0.028 \\
5.45 & -5.07 & 0.000 & 0.023 & 0.053 & 0.028 \\
\cutinhead{$x_m$ = 0.10}
5.05 & -5.55 & 0.000 & 0.011 & 0.029 & 0.017 \\
5.15 & -5.50 & 0.000 & 0.011 & 0.029 & 0.017 \\
5.25 & -5.45 & 0.000 & 0.011 & 0.029 & 0.017 \\
5.35 & -5.39 & 0.000 & 0.011 & 0.029 & 0.017 \\
5.45 & -5.35 & 0.000 & 0.012 & 0.033 & 0.018 \\
\cutinhead{$x_m$ = 1.00}
5.05 & -3.89 & 0.003 & 0.224 & 0.426 & 0.273 \\
5.15 & -3.99 & 0.003 & 0.214 & 0.405 & 0.256 \\
5.25 & -3.98 & 0.003 & 0.202 & 0.378 & 0.234 \\
5.35 & -4.02 & 0.003 & 0.189 & 0.349 & 0.211 \\
5.45 & -4.06 & 0.003 & 0.175 & 0.314 & 0.182 \\
\enddata
\label{tab:mod-1p5-a1}
\tablenotetext{a}{Table \ref{tab:mod-1p5-a1} is published in its entirety in the
electronic version of the {\it Astrophysical Journal Supplement.} A portion is shown here
for guidance regarding its form and content.}
\tablenotetext{b}{For calculations with $\Sigma \propto a^{-1}$,
planetesimals with initial radii of 1 m to 1~km only, and the 
$f_S$ fragmentation parameters.}
tablenotetext{c}{The typical inter-quartile range in the predicted fluxes
is $\pm$5\% to 10\% at 24 $\mu$m and $\pm$5\% at 70--850 $\mu$m.}
\end{deluxetable}
\clearpage

\begin{deluxetable}{lccccc}
\tablecolumns{6}
\tablewidth{0pc}
\tabletypesize{\footnotesize}
\tablecaption{Predicted Excesses for Disks Around 2 \msun\ Stars\tablenotemark{a, b, c}}
\tablehead{
  \colhead{log $t$ (yr)} &
  \colhead{log $L_d/L_{\star}$} &
  \colhead{log $F_{24}/F_{24,0}$} &
  \colhead{log $F_{70}/F_{70,0}$} &
  \colhead{log $F_{160}/F_{160,0}$} &
  \colhead{log $F_{850}/F_{850,0}$}
}
\startdata
\cutinhead{$x_m$ = 0.01}
5.05 & -4.91 & 0.001 & 0.034 & 0.065 & 0.031 \\
5.15 & -4.92 & 0.001 & 0.034 & 0.065 & 0.031 \\
5.25 & -4.94 & 0.001 & 0.034 & 0.065 & 0.031 \\
5.35 & -4.95 & 0.001 & 0.034 & 0.065 & 0.031 \\
5.45 & -4.96 & 0.001 & 0.033 & 0.063 & 0.030 \\
\cutinhead{$x_m$ = 0.10}
5.05 & -3.99 & 0.003 & 0.186 & 0.334 & 0.185 \\
5.15 & -4.08 & 0.003 & 0.181 & 0.324 & 0.180 \\
5.25 & -4.16 & 0.003 & 0.176 & 0.314 & 0.175 \\
5.35 & -4.18 & 0.002 & 0.161 & 0.293 & 0.162 \\
5.45 & -4.21 & 0.002 & 0.155 & 0.285 & 0.158 \\
\cutinhead{$x_m$ = 1.00}
5.05 & -3.86 & 0.006 & 0.254 & 0.415 & 0.241 \\
5.15 & -3.97 & 0.006 & 0.243 & 0.394 & 0.225 \\
5.25 & -4.00 & 0.006 & 0.226 & 0.358 & 0.198 \\
5.35 & -4.04 & 0.006 & 0.213 & 0.332 & 0.178 \\
5.45 & -3.88 & 0.021 & 0.363 & 0.443 & 0.220 \\
\enddata
\label{tab:mod-2p0-a1}
\tablenotetext{a}{Table \ref{tab:mod-2p0-a1} is published in its entirety in the
electronic version of the {\it Astrophysical Journal Supplement.} A portion is shown here
for guidance regarding its form and content.}
\tablenotetext{b}{For calculations with $\Sigma \propto a^{-1}$,
planetesimals with initial radii of 1 m to 1~km only, and the 
$f_S$ fragmentation parameters.}
tablenotetext{c}{The typical inter-quartile range in the predicted fluxes
is $\pm$5\% to 10\% at 24 $\mu$m and $\pm$5\% at 70--850 $\mu$m.}
\end{deluxetable}
\clearpage

\begin{deluxetable}{lccccc}
\tablecolumns{6}
\tablewidth{0pc}
\tabletypesize{\footnotesize}
\tablecaption{Predicted Excesses for Disks Around 3 \msun\ Stars\tablenotemark{a, b, c}}
\tablehead{
  \colhead{log $t$ (yr)} &
  \colhead{log $L_d/L_{\star}$} &
  \colhead{log $F_{24}/F_{24,0}$} &
  \colhead{log $F_{70}/F_{70,0}$} &
  \colhead{log $F_{160}/F_{160,0}$} &
  \colhead{log $F_{850}/F_{850,0}$}
}
\startdata
\cutinhead{$x_m$ = 0.01}
5.05 & -4.38 & 0.006 & 0.107 & 0.132 & 0.051 \\
5.15 & -4.40 & 0.006 & 0.106 & 0.130 & 0.050 \\
5.25 & -4.42 & 0.006 & 0.105 & 0.128 & 0.049 \\
5.35 & -4.44 & 0.006 & 0.101 & 0.124 & 0.047 \\
5.45 & -4.46 & 0.006 & 0.097 & 0.120 & 0.046 \\
\cutinhead{$x_m$ = 0.10}
5.05 & -3.79 & 0.015 & 0.267 & 0.341 & 0.160 \\
5.15 & -4.03 & 0.013 & 0.243 & 0.319 & 0.148 \\
5.25 & -4.06 & 0.012 & 0.229 & 0.303 & 0.139 \\
5.35 & -4.09 & 0.011 & 0.217 & 0.290 & 0.132 \\
5.45 & -4.12 & 0.010 & 0.203 & 0.272 & 0.123 \\
\cutinhead{$x_m$ = 1.00}
5.05 & -3.85 & 0.024 & 0.301 & 0.373 & 0.174 \\
5.15 & -3.87 & 0.040 & 0.341 & 0.391 & 0.179 \\
5.25 & -3.46 & 0.099 & 0.508 & 0.520 & 0.238 \\
5.35 & -3.29 & 0.139 & 0.597 & 0.586 & 0.268 \\
5.45 & -3.26 & 0.153 & 0.618 & 0.594 & 0.268 \\
\enddata
\label{tab:mod-3p0-a1}
\tablenotetext{a}{Table \ref{tab:mod-3p0-a1} is published in its entirety in the
electronic version of the {\it Astrophysical Journal Supplement.} A portion is shown here
for guidance regarding its form and content.}
\tablenotetext{b}{For calculations with $\Sigma \propto a^{-1}$,
planetesimals with initial radii of 1 m to 1~km only, and the 
$f_S$ fragmentation parameters.}
tablenotetext{c}{The typical inter-quartile range in the predicted fluxes
is $\pm$5\% to 10\% at 24 $\mu$m and $\pm$5\% at 70--850 $\mu$m.}
\end{deluxetable}
\clearpage

\begin{deluxetable}{lccccc}
\tablecolumns{6}
\tablewidth{0pc}
\tabletypesize{\footnotesize}
\tablecaption{Predicted Excesses for Disks Around 1 \msun\ Stars\tablenotemark{a, b, c}}
\tablehead{
  \colhead{log $t$ (yr)} &
  \colhead{log $L_d/L_{\star}$} &
  \colhead{log $F_{24}/F_{24,0}$} &
  \colhead{log $F_{70}/F_{70,0}$} &
  \colhead{log $F_{160}/F_{160,0}$} &
  \colhead{log $F_{850}/F_{850,0}$}
}
\startdata
\cutinhead{$x_m$ = 0.01}
5.05 & -6.05 & 0.000 & 0.002 & 0.005 & 0.003 \\
5.15 & -6.06 & 0.000 & 0.002 & 0.005 & 0.003 \\
5.25 & -6.06 & 0.000 & 0.002 & 0.005 & 0.003 \\
5.35 & -6.06 & 0.000 & 0.002 & 0.005 & 0.003 \\
5.45 & -6.06 & 0.000 & 0.002 & 0.005 & 0.003 \\
\cutinhead{$x_m$ = 0.10}
5.05 & -5.07 & 0.000 & 0.017 & 0.049 & 0.029 \\
5.15 & -5.08 & 0.000 & 0.017 & 0.049 & 0.029 \\
5.25 & -5.09 & 0.000 & 0.017 & 0.049 & 0.029 \\
5.35 & -5.10 & 0.000 & 0.016 & 0.058 & 0.029 \\
5.45 & -5.10 & 0.000 & 0.016 & 0.047 & 0.028 \\
\cutinhead{$x_m$ = 1.00}
5.05 & -4.18 & 0.000 & 0.093 & 0.250 & 0.162 \\
5.15 & -4.22 & 0.000 & 0.091 & 0.245 & 0.160 \\
5.25 & -4.27 & 0.000 & 0.089 & 0.240 & 0.158 \\
5.35 & -4.31 & 0.000 & 0.087 & 0.235 & 0.156 \\
5.45 & -4.34 & 0.000 & 0.085 & 0.230 & 0.154 \\
\enddata
\label{tab:mod-1p0-a1p5-w}
\tablenotetext{a}{Table \ref{tab:mod-1p0-a1p5-w} is published in its entirety in the
electronic version of the {\it Astrophysical Journal Supplement.} A portion is shown here
for guidance regarding its form and content.}
\tablenotetext{b}{For calculations with $\Sigma \propto a^{-3/2}$,
planetesimals with initial radii of 1 m to 1~km only, and the 
$f_W$ fragmentation parameters.}
\tablenotetext{c}{The typical inter-quartile range in the predicted fluxes
is $\pm$5\% to 10\% at 24 $\mu$m and $\pm$5\% at 70--850 $\mu$m.}
\end{deluxetable}
\clearpage

\begin{deluxetable}{lccccc}
\tablecolumns{6}
\tablewidth{0pc}
\tabletypesize{\footnotesize}
\tablecaption{Predicted Excesses for Disks Around 1.5 \msun\ Stars\tablenotemark{a, b, c}}
\tablehead{
  \colhead{log $t$ (yr)} &
  \colhead{log $L_d/L_{\star}$} &
  \colhead{log $F_{24}/F_{24,0}$} &
  \colhead{log $F_{70}/F_{70,0}$} &
  \colhead{log $F_{160}/F_{160,0}$} &
  \colhead{log $F_{850}/F_{850,0}$}
}
\startdata
\cutinhead{$x_m$ = 0.01}
5.05 & -5.88 & 0.000 & 0.004 & 0.008 & 0.004 \\
5.15 & -5.88 & 0.000 & 0.004 & 0.008 & 0.004 \\
5.25 & -5.88 & 0.000 & 0.004 & 0.008 & 0.004 \\
5.35 & -5.89 & 0.000 & 0.004 & 0.008 & 0.004 \\
5.45 & -5.89 & 0.000 & 0.004 & 0.008 & 0.004 \\
\cutinhead{$x_m$ = 0.10}
5.05 & -4.89 & 0.000 & 0.031 & 0.065 & 0.033 \\
5.15 & -4.91 & 0.000 & 0.031 & 0.065 & 0.033 \\
5.25 & -4.93 & 0.000 & 0.031 & 0.065 & 0.033 \\
5.35 & -4.95 & 0.000 & 0.030 & 0.064 & 0.032 \\
5.45 & -4.96 & 0.000 & 0.029 & 0.062 & 0.032 \\
\cutinhead{$x_m$ = 1.00}
5.05 & -4.04 & 0.001 & 0.141 & 0.285 & 0.168 \\
5.15 & -4.13 & 0.001 & 0.141 & 0.285 & 0.168 \\
5.25 & -4.21 & 0.001 & 0.141 & 0.285 & 0.168 \\
5.35 & -4.23 & 0.001 & 0.133 & 0.273 & 0.161 \\
5.45 & -4.25 & 0.001 & 0.125 & 0.261 & 0.154 \\
\enddata
\label{tab:mod-1p5-a1p5-w}
\tablenotetext{a}{Table \ref{tab:mod-1p5-a1p5-w} is published in its entirety in the
electronic version of the {\it Astrophysical Journal Supplement.} A portion is shown here
for guidance regarding its form and content.}
\tablenotetext{b}{For calculations with $\Sigma \propto a^{-3/2}$,
planetesimals with initial radii of 1 m to 1~km only, and the 
$f_W$ fragmentation parameters.}
\tablenotetext{c}{The typical inter-quartile range in the predicted fluxes
is $\pm$5\% to 10\% at 24 $\mu$m and $\pm$5\% at 70--850 $\mu$m.}
\end{deluxetable}
\clearpage

\begin{deluxetable}{lccccc}
\tablecolumns{6}
\tablewidth{0pc}
\tabletypesize{\footnotesize}
\tablecaption{Predicted Excesses for Disks Around 2 \msun\ Stars\tablenotemark{a, b, c}}
\tablehead{
  \colhead{log $t$ (yr)} &
  \colhead{log $L_d/L_{\star}$} &
  \colhead{log $F_{24}/F_{24,0}$} &
  \colhead{log $F_{70}/F_{70,0}$} &
  \colhead{log $F_{160}/F_{160,0}$} &
  \colhead{log $F_{850}/F_{850,0}$}
}
\startdata
\cutinhead{$x_m$ = 0.01}
5.05 & -4.91 & 0.001 & 0.034 & 0.065 & 0.031 \\
5.15 & -4.92 & 0.001 & 0.034 & 0.065 & 0.031 \\
5.25 & -4.94 & 0.001 & 0.034 & 0.065 & 0.031 \\
5.35 & -4.95 & 0.001 & 0.034 & 0.065 & 0.031 \\
5.45 & -4.96 & 0.001 & 0.033 & 0.063 & 0.030 \\
\cutinhead{$x_m$ = 0.10}
5.05 & -3.99 & 0.003 & 0.176 & 0.314 & 0.175 \\
5.15 & -4.08 & 0.003 & 0.176 & 0.314 & 0.175 \\
5.25 & -4.16 & 0.003 & 0.176 & 0.314 & 0.175 \\
5.35 & -4.18 & 0.003 & 0.161 & 0.293 & 0.162 \\
5.45 & -4.21 & 0.003 & 0.155 & 0.284 & 0.158 \\
\cutinhead{$x_m$ = 1.00}
5.05 & -3.86 & 0.006 & 0.254 & 0.415 & 0.241 \\
5.15 & -3.97 & 0.006 & 0.243 & 0.394 & 0.225 \\
5.25 & -4.00 & 0.006 & 0.226 & 0.358 & 0.198 \\
5.35 & -4.04 & 0.006 & 0.213 & 0.332 & 0.178 \\
5.45 & -3.88 & 0.021 & 0.363 & 0.443 & 0.220 \\
\enddata
\label{tab:mod-2p0-a1p5-w}
\tablenotetext{a}{Table \ref{tab:mod-2p0-a1p5-w} is published in its entirety in the
electronic version of the {\it Astrophysical Journal Supplement.} A portion is shown here
for guidance regarding its form and content.}
\tablenotetext{b}{For calculations with $\Sigma \propto a^{-3/2}$,
planetesimals with initial radii of 1 m to 1~km only, and the 
$f_W$ fragmentation parameters.}
\tablenotetext{c}{The typical inter-quartile range in the predicted fluxes
is $\pm$5\% to 10\% at 24 $\mu$m and $\pm$5\% at 70--850 $\mu$m.}
\end{deluxetable}
\clearpage

\begin{deluxetable}{lccccc}
\tablecolumns{6}
\tablewidth{0pc}
\tabletypesize{\footnotesize}
\tablecaption{Predicted Excesses for Disks Around 3 \msun\ Stars\tablenotemark{a, b, c}}
\tablehead{
  \colhead{log $t$ (yr)} &
  \colhead{log $L_d/L_{\star}$} &
  \colhead{log $F_{24}/F_{24,0}$} &
  \colhead{log $F_{70}/F_{70,0}$} &
  \colhead{log $F_{160}/F_{160,0}$} &
  \colhead{log $F_{850}/F_{850,0}$}
}
\startdata
\cutinhead{$x_m$ = 0.01}
5.05 & -4.38 & 0.005 & 0.113 & 0.136 & 0.053 \\
5.15 & -4.40 & 0.005 & 0.109 & 0.132 & 0.051 \\
5.25 & -4.42 & 0.005 & 0.105 & 0.128 & 0.049 \\
5.35 & -4.44 & 0.005 & 0.101 & 0.124 & 0.047 \\
5.45 & -4.46 & 0.005 & 0.097 & 0.120 & 0.045 \\
\cutinhead{$x_m$ = 0.10}
5.05 & -4.00 & 0.014 & 0.257 & 0.335 & 0.157 \\
5.15 & -4.03 & 0.013 & 0.243 & 0.319 & 0.148 \\
5.25 & -4.06 & 0.012 & 0.229 & 0.303 & 0.139 \\
5.35 & -4.09 & 0.011 & 0.217 & 0.290 & 0.132 \\
5.45 & -4.12 & 0.010 & 0.203 & 0.272 & 0.123 \\
\cutinhead{$x_m$ = 1.00}
5.05 & -3.85 & 0.024 & 0.301 & 0.373 & 0.174 \\
5.15 & -3.87 & 0.040 & 0.341 & 0.391 & 0.179 \\
5.25 & -3.46 & 0.099 & 0.508 & 0.520 & 0.238 \\
5.35 & -3.29 & 0.139 & 0.597 & 0.586 & 0.268 \\
5.45 & -3.26 & 0.153 & 0.618 & 0.594 & 0.268 \\
\enddata
\label{tab:mod-3p0-a1p5-w}
\tablenotetext{a}{Table \ref{tab:mod-3p0-a1p5-w} is published in its entirety in the
electronic version of the {\it Astrophysical Journal Supplement.} A portion is shown here
for guidance regarding its form and content.}
\tablenotetext{b}{For calculations with $\Sigma \propto a^{-3/2}$,
planetesimals with initial radii of 1 m to 1~km only, and the 
$f_W$ fragmentation parameters.}
\tablenotetext{c}{The typical inter-quartile range in the predicted fluxes
is $\pm$5\% to 10\% at 24 $\mu$m and $\pm$5\% at 70--850 $\mu$m.}
\end{deluxetable}
\clearpage

\begin{deluxetable}{lccccc}
\tablecolumns{6}
\tablewidth{0pc}
\tabletypesize{\footnotesize}
\tablecaption{Predicted Excesses for Disks Around 1 \msun\ Stars\tablenotemark{a, b, c}}
\tablehead{
  \colhead{log $t$ (yr)} &
  \colhead{log $L_d/L_{\star}$} &
  \colhead{log $F_{24}/F_{24,0}$} &
  \colhead{log $F_{70}/F_{70,0}$} &
  \colhead{log $F_{160}/F_{160,0}$} &
  \colhead{log $F_{850}/F_{850,0}$}
}
\startdata
\cutinhead{$x_m$ = 0.01}
5.05 & -6.05 & 0.000 & 0.002 & 0.005 & 0.003 \\
5.15 & -6.06 & 0.000 & 0.002 & 0.005 & 0.003 \\
5.25 & -6.06 & 0.000 & 0.002 & 0.005 & 0.003 \\
5.35 & -6.06 & 0.000 & 0.002 & 0.005 & 0.003 \\
5.45 & -6.06 & 0.000 & 0.002 & 0.005 & 0.003 \\
\cutinhead{$x_m$ = 0.10}
5.05 & -5.06 & 0.000 & 0.017 & 0.049 & 0.029 \\
5.15 & -5.08 & 0.000 & 0.017 & 0.049 & 0.029 \\
5.25 & -5.09 & 0.000 & 0.017 & 0.049 & 0.029 \\
5.35 & -5.10 & 0.000 & 0.017 & 0.048 & 0.029 \\
5.45 & -5.10 & 0.000 & 0.016 & 0.047 & 0.028 \\
\enddata
\label{tab:mod-1p0-a15-s}
\tablenotetext{a}{Table \ref{tab:mod-1p0-a15-s} is published in its entirety in the
electronic version of the {\it Astrophysical Journal Supplement.} A portion is shown here
for guidance regarding its form and content.}
\tablenotetext{b}{For calculations with $\Sigma \propto a^{-3/2}$,
planetesimals with initial radii of 1 m to 1~km only, and the 
$f_S$ fragmentation parameters.}
\tablenotetext{c}{The typical inter-quartile range in the predicted fluxes
is $\pm$5\% to 10\% at 24 $\mu$m and $\pm$5\% at 70--850 $\mu$m.}
\end{deluxetable}
\clearpage

\begin{deluxetable}{lccccc}
\tablecolumns{6}
\tablewidth{0pc}
\tabletypesize{\footnotesize}
\tablecaption{Predicted Excesses for Disks Around 1.5 \msun\ Stars\tablenotemark{a, b, c}}
\tablehead{
  \colhead{log $t$ (yr)} &
  \colhead{log $L_d/L_{\star}$} &
  \colhead{log $F_{24}/F_{24,0}$} &
  \colhead{log $F_{70}/F_{70,0}$} &
  \colhead{log $F_{160}/F_{160,0}$} &
  \colhead{log $F_{850}/F_{850,0}$}
}
\startdata
\cutinhead{$x_m$ = 0.01}
5.05 & -5.88 & 0.000 & 0.004 & 0.008 & 0.004 \\
5.15 & -5.88 & 0.000 & 0.004 & 0.008 & 0.004 \\
5.25 & -5.88 & 0.000 & 0.004 & 0.008 & 0.004 \\
5.35 & -5.88 & 0.000 & 0.004 & 0.008 & 0.004 \\
5.45 & -5.89 & 0.000 & 0.004 & 0.008 & 0.004 \\
\cutinhead{$x_m$ = 0.10}
5.05 & -4.89 & 0.000 & 0.031 & 0.065 & 0.033 \\
5.15 & -4.90 & 0.000 & 0.031 & 0.065 & 0.033 \\
5.25 & -4.92 & 0.000 & 0.031 & 0.065 & 0.033 \\
5.35 & -4.94 & 0.000 & 0.031 & 0.065 & 0.033 \\
5.45 & -4.95 & 0.000 & 0.030 & 0.063 & 0.032 \\
\enddata
\label{tab:mod-1p5-a15-s}
\tablenotetext{a}{Table \ref{tab:mod-1p5-a15-s} is published in its entirety in the
electronic version of the {\it Astrophysical Journal Supplement.} A portion is shown here
for guidance regarding its form and content.}
\tablenotetext{b}{For calculations with $\Sigma \propto a^{-3/2}$,
planetesimals with initial radii of 1 m to 1~km only, and the 
$f_S$ fragmentation parameters.}
\tablenotetext{c}{The typical inter-quartile range in the predicted fluxes
is $\pm$5\% to 10\% at 24 $\mu$m and $\pm$5\% at 70--850 $\mu$m.}
\end{deluxetable}
\clearpage

\begin{deluxetable}{lccccc}
\tablecolumns{6}
\tablewidth{0pc}
\tabletypesize{\footnotesize}
\tablecaption{Predicted Excesses for Disks Around 2 \msun\ Stars\tablenotemark{a, b, c}}
\tablehead{
  \colhead{log $t$ (yr)} &
  \colhead{log $L_d/L_{\star}$} &
  \colhead{log $F_{24}/F_{24,0}$} &
  \colhead{log $F_{70}/F_{70,0}$} &
  \colhead{log $F_{160}/F_{160,0}$} &
  \colhead{log $F_{850}/F_{850,0}$}
}
\startdata
\cutinhead{$x_m$ = 0.01}
5.05 & -5.75 & 0.000 & 0.005 & 0.010 & 0.004 \\
5.15 & -5.76 & 0.000 & 0.005 & 0.010 & 0.004 \\
5.25 & -5.76 & 0.000 & 0.005 & 0.010 & 0.004 \\
5.35 & -5.76 & 0.000 & 0.005 & 0.010 & 0.004 \\
5.45 & -5.77 & 0.000 & 0.005 & 0.010 & 0.004 \\
\cutinhead{$x_m$ = 0.10}
5.05 & -4.77 & 0.001 & 0.044 & 0.079 & 0.036 \\
5.15 & -4.80 & 0.001 & 0.044 & 0.079 & 0.036 \\
5.25 & -4.83 & 0.001 & 0.044 & 0.079 & 0.036 \\
5.35 & -4.84 & 0.001 & 0.042 & 0.077 & 0.035 \\
5.45 & -4.86 & 0.001 & 0.040 & 0.074 & 0.034 \\
\enddata
\label{tab:mod-2p0-a15-s}
\tablenotetext{a}{Table \ref{tab:mod-2p0-a15-s} is published in its entirety in the
electronic version of the {\it Astrophysical Journal Supplement.} A portion is shown here
for guidance regarding its form and content.}
\tablenotetext{b}{For calculations with $\Sigma \propto a^{-3/2}$,
planetesimals with initial radii of 1 m to 1~km only, and the 
$f_S$ fragmentation parameters.}
\tablenotetext{c}{The typical inter-quartile range in the predicted fluxes
is $\pm$5\% to 10\% at 24 $\mu$m and $\pm$5\% at 70--850 $\mu$m.}
\end{deluxetable}
\clearpage

\begin{deluxetable}{lccccc}
\tablecolumns{6}
\tablewidth{0pc}
\tabletypesize{\footnotesize}
\tablecaption{Predicted Excesses for Disks Around 3 \msun\ Stars\tablenotemark{a, b, c}}
\tablehead{
  \colhead{log $t$ (yr)} &
  \colhead{log $L_d/L_{\star}$} &
  \colhead{log $F_{24}/F_{24,0}$} &
  \colhead{log $F_{70}/F_{70,0}$} &
  \colhead{log $F_{160}/F_{160,0}$} &
  \colhead{log $F_{850}/F_{850,0}$}
}
\startdata
\cutinhead{$x_m$ = 0.01}
5.05 & -5.58 & 0.000 & 0.008 & 0.010 & 0.004 \\
5.15 & -5.58 & 0.000 & 0.008 & 0.010 & 0.004 \\
5.25 & -5.59 & 0.000 & 0.008 & 0.010 & 0.004 \\
5.35 & -5.59 & 0.000 & 0.008 & 0.010 & 0.004 \\
5.45 & -5.60 & 0.000 & 0.008 & 0.010 & 0.004 \\
\cutinhead{$x_m$ = 0.10}
5.05 & -4.59 & 0.003 & 0.067 & 0.085 & 0.034 \\
5.15 & -4.62 & 0.003 & 0.065 & 0.083 & 0.033 \\
5.25 & -4.66 & 0.003 & 0.063 & 0.081 & 0.032 \\
5.35 & -4.70 & 0.003 & 0.061 & 0.079 & 0.030 \\
5.45 & -4.72 & 0.003 & 0.059 & 0.076 & 0.029 \\
\enddata
\label{tab:mod-3p0-a15-s}
\tablenotetext{a}{Table \ref{tab:mod-3p0-a15-s} is published in its entirety in the
electronic version of the {\it Astrophysical Journal Supplement.} A portion is shown here
for guidance regarding its form and content.}
\tablenotetext{b}{For calculations with $\Sigma \propto a^{-3/2}$,
planetesimals with initial radii of 1 m to 1~km only, and the 
$f_S$ fragmentation parameters.}
\tablenotetext{c}{The typical inter-quartile range in the predicted fluxes
is $\pm$5\% to 10\% at 24 $\mu$m and $\pm$5\% at 70--850 $\mu$m.}
\end{deluxetable}
\clearpage

\begin{deluxetable}{lcccc}
\tablecolumns{6}
\tablewidth{0pc}
%\tabletypesize{\small}
\tablecaption{Scaling Relations for Debris Disks\tablenotemark{(a,b)}}
\tablehead{
  \colhead{Parameter} & 
  \colhead{(1)} &
  \colhead{(2)} &
  \colhead{(3)} &
  \colhead{Notes}
}
\startdata
Fragmentation ($f_i$) & $f_S$ & $f_S$ & $f_W$ & \\
\\
$\Sigma$ (g cm$^{-2}$) & 30 $x_m ~  M_{\star,2} ~ a^{-1}$ & 30 $x_m ~  M_{\star,2} ~ a^{-3/2}$ & 30 $x_m ~ M_{\star,2} ~ a^{-3/2}$ & (c) \\
\\
$t_{d,max}$ (Myr) & 7.5~$x_m^{-1}$ & 25 $x_m^{-2/3}~M_{\star,2}$ & 10~$x_m^{-1}~M_{\star,2}^2$  & \\
\\
$L_{d,max} / L_{\star}$ & $10^{-2}~x_m$ & $2 \times 10^{-3} x_m$ & $2 \times 10^{-3} ~ x_m~M_{\star,2}^{1/3}$  & \\
\\
log $F_{24,max}/F_{24,0}$ &
$b + M_{\star,2}~{\rm log}~x_m$ &
$c + 0.5M_{\star,2}~{\rm log}~x_m$ &
$c + 0.5M_{\star,2}~{\rm log}~x_m$ & (d) \\
\\
$F_{70,max}/F_{70,0}$ & 400 $x_m^{0.9} M_{\star,2}^2$ & 55 $x_m^{0.9} ~ M_{\star,2}$ & 110 $x_m^d ~ M_{\star,2}$ & (e) \\
\\
$F_{160,max}/F_{160,0}$ & 600 $x_m^{0.9} M_{\star,2}^2$ & 65 $x_m^{0.9}$ & 80 $x_m^{0.9}$  & \\
\\
$F_{850,max}/F_{850,0}$ & 200 $x_m$ & 40 $x_m^{0.9}$ & 40 $x_m^{0.75}$  & (f) \\
$F_{850,max}/F_{850,0}$ & 200 $x_m$ &  25 $x_m^{0.9}$ & 25 $x_m^{0.9}$ & (g) \\
\enddata
\tablenotetext{a}{For planetesimals with initial radii of 1 m to 1~km}
\tablenotetext{b}{The inter-quartile ranges for the coefficients are
$\pm$5\% for $t_{d,max}$ and $L_{d,max}$ and $\pm$5\% to 10\% for the
maximum excesses at 24--850 $\mu$m.} 
\tablenotetext{c}{$M_{\star,2} = M_{\star} / 2~M_\odot$}
\tablenotetext{d}{$b = 0.05 + (M_\star - 1~M_\odot)$; $c = 0.74 (M_\star - 1~M_\odot)$}
\tablenotetext{e}{$d = 0.55 + (M_\star / 1~M_\odot)$}
\tablenotetext{f}{For disks around 1 \msun\ stars}
\tablenotetext{g}{For disks around 1.5--3 \msun\ stars}
\label{tab:eqdebris1}
\end{deluxetable}
\clearpage

\begin{deluxetable}{lcccc}
\tablecolumns{6}
\tablewidth{0pc}
%\tabletypesize{\small}
\tablecaption{Scaling Relations for Debris Disks\tablenotemark{a,b}}
\tablehead{
  \colhead{Parameter} & 
  \colhead{1~km only } &
  \colhead{10~km only} &
  \colhead{100~km only} &
  \colhead{Notes}
}
\startdata
$t_{d,max}$ (Myr) & 9~$x_m^{-1}~M_{\star,2}^{b}$ & 50~$x_m^{-1}~M_{\star,2}^{c}$ & 325~$x_m^{-1}~M_{\star,2}^{-2}$ & (c) \\
\\
$L_{d,max} / L_{\star}$ & $2 \times 10^{-3}~x_m^{1.25}$ & $5 \times 10^{-4} x_m^{1.5}$ & $5 \times 10^{-5} ~ x_m^2$  & \\
\\
log $F_{24,max}/F_{24,0}$ &
$c + M_{\star,2}~{\rm log}~x_m$ &
$d + 0.5M_{\star,2}~{\rm log}~x_m$ &
$e + 0.5M_{\star,2}~{\rm log}~x_m$ & (d) \\
\\
$F_{70,max}/F_{70,0}$ & 80 $x_m^{1.2} M_{\star,2}^{1/2}$ & 15 $x_m^{1.1} ~ M_{\star,2}^{1/2}$ & 2 $x_m^{1.5} ~ M_{\star,2}^{1/2}$ & \\
\\
$F_{160,max}/F_{160,0}$ & 65 $x_m^{1.2}$ & 15 $x_m^{1.2}$ & 2 $x_m^{1.25}$  & \\
\\
$F_{850,max}/F_{850,0}$ & 30 $x_m$ & 10 $x_m^{1.2}$ & 2 $x_m^{0.75}$  & (e) \\
$F_{850,max}/F_{850,0}$ & 20 $x_m^{1.2}$ & 5 $x_m^{1.2}$ & 1.25 $x_m$ & (f) \\
\enddata
\tablenotetext{a}{For calculations with $\Sigma \propto a^{-3/2}$,
planetesimals with a single initial radius, and the $f_S$ fragmentation parameters}
\tablenotetext{b}{The inter-quartile ranges for the coefficients are
$\pm$5\% to 10\% for $t_{d,max}$ and $L_{d,max}$ and $\pm$10\% to 20\% for the
maximum excesses at 24--850 $\mu$m.} 
\tablenotetext{c}{$b = 2 - 2~H(2)$, $c = 2 - 0.5~H(2)$; $H(x)$ = 0 for x $\le$ 2 \msun, 
$H(x)$ = 1 for x $>$ 2 \msun}
\tablenotetext{d}{$c = 0.68~(M_\star - 1~M_\odot)$; $d = 0.35~(M_\star - 1~M_\odot)$; 
$e = 0.05~(M_\star - 1~M_\odot)$}
\tablenotetext{e}{For disks around 1 \msun\ stars}
\tablenotetext{f}{For disks around 1.5--3 \msun\ stars}
\label{tab:eqdebris2}
\end{deluxetable}
\clearpage

%
%\centerline{\bf FIGURE CAPTIONS}
%\vskip 4ex
%

\clearpage

\clearpage

\clearpage

\clearpage

\clearpage

\clearpage

\clearpage
\clearpage

\clearpage

\clearpage

\clearpage

\clearpage

\clearpage

\clearpage

\clearpage

\clearpage

\clearpage

\clearpage

\clearpage

\clearpage

\clearpage

\clearpage

\clearpage

\clearpage

\clearpage

\clearpage

\clearpage

\clearpage

\clearpage

\clearpage

\clearpage

\clearpage

\clearpage

\clearpage

\end{document}